\documentclass[manuscript]{acmart}

\AtBeginDocument{%
  \providecommand\BibTeX{{%
    \normalfont B\kern-0.5em{\scshape i\kern-0.25em b}\kern-0.8em\TeX}}}

\acmBooktitle{Woodstock '18: ACM Symposium on Neural Gaze Detection,
  June 03--05, 2018, Woodstock, NY}
\acmPrice{15.00}
\acmISBN{978-1-4503-XXXX-X/18/06}

\usepackage{cleveref}
\graphicspath{{imgs/}}
\usepackage{overpic}
\usepackage{multirow}

\begin{document}

\title{Solid NURBS Conforming Scaffolding for Isogeometric Analysis}


\author{Stefano Moriconi}
\affiliation{%
  \institution{King's College London}
  \streetaddress{Strand}
  \city{London}
  \country{United Kingdom}}
\email{stefano.moriconi@kcl.ac.uk}

\author{Parashkev Nachev}
\affiliation{%
  \institution{University College London}
  \streetaddress{Gower St.}
  \city{London}
  \country{United Kingdom}}
\email{p.nachev@ucl.ac.uk}

\author{S\'ebastien Ourselin}
\affiliation{%
  \institution{King's College London}
  \streetaddress{Strand}
  \city{London}
  \country{United Kingdom}}
\email{sebastien.ourselin@kcl.ac.uk}

\author{M. Jorge Cardoso}
\affiliation{%
  \institution{King's College London}
  \streetaddress{Strand}
  \city{London}
  \country{United Kingdom}}
\email{m.jorge.cardoso@kcl.ac.uk}


% \renewcommand{\shortauthors}{Anonymous Author, et al.}

\begin{abstract}
  This work introduces a scaffolding framework to compactly parametrise solid structures with conforming NURBS elements for isogeometric analysis. A novel formulation introduces a topological, geometrical and parametric subdivision of the space in a minimal plurality of conforming vectorial elements. These determine a multi-compartmental scaffolding for arbitrary branching patterns. A solid smoothing paradigm is devised for the conforming scaffolding achieving higher than positional geometrical and parametric continuity. Results are shown for synthetic shapes of varying complexity, for modular CAD geometries, for branching structures from tessellated meshes and for organic biological structures from imaging data. Representative simulations demonstrate the validity of the introduced scaffolding framework with scalable performance and groundbreaking applications for isogeometric analysis.
\end{abstract}




\maketitle

\begin{figure}[h!]
    \centering
    \includegraphics[width=.99\textwidth]{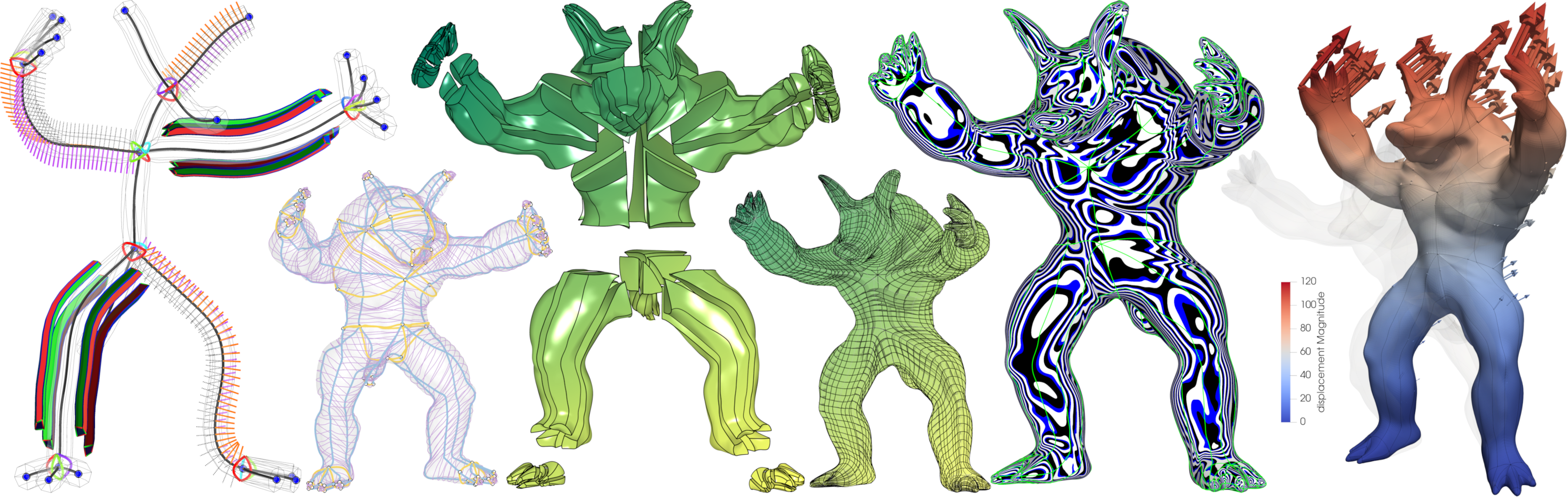}
    \caption*{Graphical Abstract}
    \label{fig:GraphAbstract}
\end{figure}

\section{Introduction}

Devising an end-to-end framework to trace parametric geometries fully compatible with continuous-domain computational simulations is an extremely complex task, and currently stands as an open challenge.
The design of geometrical structures traditionally employs computer-graphics techniques based on modelling raster surfaces, which often require high polygonal density to accurately represent smooth and irregular shapes \cite{mortenson1997geometric,russo2006polygonal}.
In computer-aided design (CAD) vectorial and piece-wise continuous entities, i.e. Non-Uniform Rational B-Spline (NURBS) patches, leverage similar polygonal subdivision schemes in a more compact and resolution-free parametrisation by representing the underlying geometry exactly \cite{Piegl1991,piegl1996nurbs}.
Nonetheless, most of the commercially available CAD tools exclusively focus on surface geometries and do not integrate parametric simulations or a solid finite-element analysis framework in the same vectorial domain.
Solid and volumetric meshing is predominantly addressed in a raster fashion, where discrete polyhedral subdivision schemes fill with an unstructured plurality of solid elements the hollow space delimited by the superficial (closed) boundary \cite{ho1988finite}.
%In this case, conventional finite-elements methods employ such discrete polyhedral tessellation to ultimately perform computational simulations, approximating partial differential equations (PDEs) for a wide range of applications \cite{rao2017finite}.
In this case, conventional finite-elements methods ultimately approximate partial differential equations (PDEs) solutions over such discrete solid tessellations \cite{rao2017finite}.

Isogeometric analysis (IGA) \cite{hughes2005isogeometric,cottrell2009isogeometric} is a relatively recent computational approach that directly integrates finite-element analysis with vectorial NURBS-based geometries.
By means of highly compact shape parametrisations, IGA has the advantage of high-performance and high-throughput continuous simulations while reducing geometrical approximation errors and providing higher numerical stability for the simulation solution profiles.
Raster polygonal or polyhedral meshes, however, are not compatible with IGA.
Conversely, a consistent NURBS-based geometrical domain is required to satisfy stringent conforming parametric conditions.
The lack of a parametric framework to compactly trace solid shapes in IGA-compatible domains currently limited the development and application of Isogeometric Analysis to simple 2D/3D toy examples and few case studies.

With the aim of bridging the technological gap, this work first proposes a novel solid scaffolding framework jointly leveraging a topological, geometrical and parametric formulation.
A solid scaffolding of minimal parameters and vectorial elements is sought to model the shape of the underlying structure by enforcing a conforming condition and preserving smoothness properties of a solid continuum, while optimising for the computational complexity of downstream simulations.
The introduced scaffolding framework paves the way towards novel approaches to rationalising space for industrial design, CAD modelling, organic structures representation, and compact digital shape tracing with a particular focus on scalable, accurate and high-performance finite-elements analysis using IGA.

\subsection{Related Work}
\paragraph{NURBS Geometries} The pioneering work of \cite{Piegl1991,piegl1996nurbs} first introduced a parametrisation framework to exactly represent 3D geometries with vectorial elements as opposed to raster finite-resolution meshes employed in polygonal modelling.
Curvilinear and surface NURBS elements have \textit{de-facto} become the standard for computed-aided design (CAD), manufacturing and engineering in a number of commercial and industrial applications.
Beyond the established standards, the NURBS formulation generalises also for the parametrisation of higher-dimensional entities \cite{chen2016nurbs,xu2014high}.
Among those, solid NURBS build on extending and combining linear NURBS operators on univariate primitives, thus constituting a free-form cuboid in the most simple form.
Recent research toolboxes \cite{de2011geopdes,vazquez2016new,bingol2019geomdl,PetIGA,juttler2014geometry} have developed a multi-platform formulation for high-dimensional vectorial representations, including hierarchical NURBS \cite{garau2018algorithms,bracco2018refinement} and IGA-compatible frameworks \cite{hesch2016hierarchical,bazilevs2010isogeometric} for the solution of PDEs based on splines.
However, the design and construction of higher-dimensional geometrical domains has only been addressed for simple toy examples or individual case studies \cite{hughes2005isogeometric,cottrell2009isogeometric}.

\paragraph{Scaffolding Construction} 
The design and construction of geometrical partitioning domains was initially addressed in \cite{suarez2017scaffolding,suarez2018scaffolding,panotopoulou2018scaffolding} with a scaffolding approach.
In particular, a quadrilateral mesh is first introduced as the coarsest shape representation around a one-dimensional skeleton of high genus. Proof is provided for the existence of such coarse representation based on Voronoi diagrams and the problem of constructing an optimal scaffold is formalised as an integer linear program \cite{suarez2018scaffolding}.
In these cases, affinities between the scaffolding and tensor product splines are mentioned, yet a structural subdivision and a parametric formulation specific for NURBS elements and their arrangement are not addressed.
Also, strategies to recover the geometrical embedding of the underlying solid structure from the scaffolding are limited to coarse polygonal refinement schemes.

\paragraph{Shape Estimation}
In \cite{livesu2016skeleton} a solid meshing approach builds from a skeleton topology and integrates an adaptive refinement strategy to locally recover the shape of a structure delimited by a surface mesh.
In a similar fashion, \cite{LPPSC20} proposed a hexahedral meshing method which optimises for cuboid elements leveraging a cascade of loop-wise cuts based on a surface-field aware block decomposition of a generic mesh.
Alternative solid meshing techniques \cite{tarini2004polycube,livesu2013polycut,gao2017robust,gregson2011all,si2015tetgen,wang2008polycube} approximate the geometry of the tessellation by locally sampling the boundary of the structure and further subdivide the solid mesh in smaller polyhedra or finer inner partitions with adaptive or pre-defined resolution.
Shape and curvature features as well as geometrical cues are often retrieved from the boundary mesh with directional fields \cite{panozzo2014frame,pietroni2016tracing} and geodesic \cite{kimmel1998computing,martinez2004geodesic,surazhsky2005fast,crane2020survey} techniques, by first partitioning and segmenting superficial patches of the geometry and ultimately fitting the 3D coordinates of the solid polyhedral mesh.
Dense solid meshes recover accurately the underlying geometry of the structure, often in detriment to a compact representation, where the resulting unorganised plurality of polyhedra exhibits a piece-wise linear and faceted lattice with pure positional continuity among neighbouring elements.

\paragraph{Smoothness and Continuity} 
A number of studies \cite{catmull1978recursively,doo1978behaviour,loop1987smooth,stam1998evaluation} have proposed methods and strategies to recover a smoother lattice and higher degrees of continuity for associated limit surfaces focusing on polygonal meshes.
In general, they build on refining first the mesh with a denser lattice and subsequently adjusting the coordinates of the geometry based on the local neighbourhood.
In \cite{stam2001subdivision} a new class of subdivision surfaces is introduced by bridging the gap between polygonal meshes and uniform B-spline surfaces of arbitrary degree.
In \cite{stam1998exact}, an exact solution is provided for similar spline-based surfaces, and a closed form is devised for patches joining at extraordinary vertices of arbitrary valence.
Irregular surface structures exactly recover the typically organic smoothness and higher degree of parametric continuity in their differential forms, which provided accurate simulation profiles for surface geometries using IGA \cite{pan2016isogeometric}.
Rectification routines and extended subdivisions are devised for convex polyhedral meshes \cite{livesu2015practical,burkhart2010iso}, where the coordinates of the inner lattice are constrained by the local connected neighbourhood.
Yet, a solid smoothing strategy for higher degrees of continuity in a conforming solid NURBS scaffolding for IGA has not been investigated, and currently remains an open challenge.

\paragraph{IGA Applications}
An increasing number of studies \cite{bazilevs2006isogeometric,bazilevs2008isogeometric,cottrell2009isogeometric,urick2019review,carraturo2019suitably,bucelli2021multipatch} are investigating and employing finite-element analysis striving for a unified framework embedding vectorial geometries with continuous-domain accurate and fast simulations.
Among those, the pioneering work by \cite{hughes2005isogeometric,bazilevs2006isogeometric} first introduced a CAD-compatible integration and a monolithic formulation of computational analyses with IGA, where spline-basis functions solve for linear elastic problems and (bio)-mechanical characterisations, as opposed to conventional polynomial finite-element methods \cite{rao2017finite}.
In early medical applications, \cite{zhang2007patient,bazilevs2008isogeometric} presented a subject-specific approach for tracing vessels with a set of conforming solid NURBS geometries.
In a similar fashion, \cite{de2011geopdes,vazquez2016new,bingol2019geomdl,juttler2014geometry} developed vectorial tools for modelling complex interactions between composite and multi-compartmental structures in a compact, general-purpose and light-weight manner.
Quantitative IGA analyses proved higher accuracy and numerical stability \cite{bazilevs2006isogeometric} compared with conventional approaches, as the solution profile of PDEs eliminates the propagation of geometrical approximation errors.
Most works, however, have focused on a limited set of geometries, addressing feasibility studies and proof-of-concept applications.
This is mainly due to the lack of a principled scaffolding framework able to capture the underlying structure of more complex geometries and generalise for compatible real case scenarios on large scale \cite{bazilevs2010isogeometric,urick2019review}.

\subsection{Outline}

Aiming to tackle the aforementioned open problems in a unified manner, a conforming solid scaffolding for a generic 3D structure is described in the following sections.
In \cref{GraphConfigBranchStruct}, a graph configuration is first considered as the underlying skeleton of the solid structure.
In \cref{ConformSolidScaff}, the generic form of a solid scaffolding is defined as a structured set of adjacent vectorial elements, and the conforming condition is introduced.
In \cref{LumScaff} the luminal solid scaffolding construction is introduced, where the minimal number of vectorial elements, the associated spatial arrangement and the conforming parametric configurations are detailed. A straightforward extension, i.e. the wall scaffolding, is addressed in \cref{WallScaff} and alternative formulations are mentioned in \cref{Othercaff}.
An interfacing configuration is introduced in \cref{QuadJuncSimplex} for vectorial elements joining at a generic junction with arbitrary branching pattern.
In \cref{GeomEmbedding} the geometrical embedding of the scaffolding elements is described, including a data-driven fitting scheme.
Lastly, in \cref{SmoothPrdgm}, a smoothing paradigm is devised for the scaffolding, achieving higher degrees of geometrical and parametric continuity, as in an organic medium in the form of a continuum.

%Lastly, in \cref{IGAConfig} a representative set of boundary conditions are presented for an IGA simulation employing the conforming solid scaffolding.

\section{Graph Configuration of the Structure}
\label{GraphConfigBranchStruct}
\begin{figure}
    \centering
    \begin{overpic}[width=.99\textwidth]{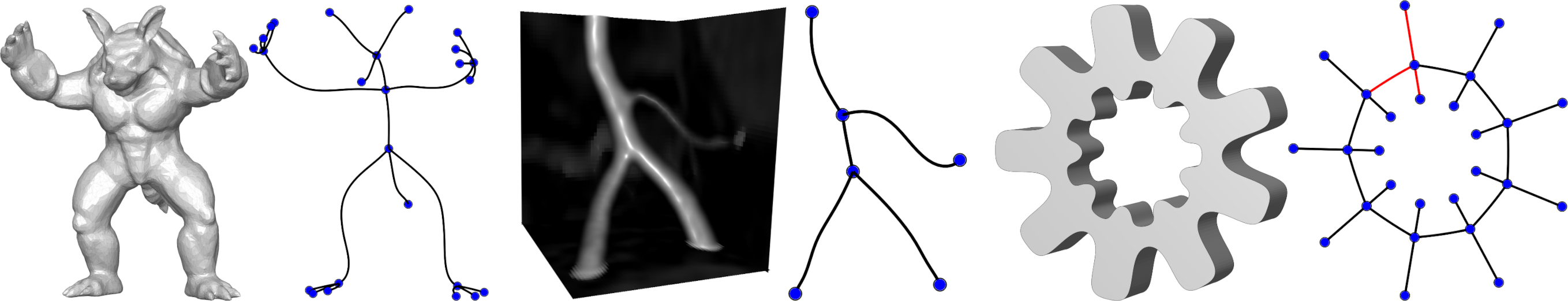}
            \put(0,19){\small{(a)}}
            \put(31,19){\small{(b)}}
            \put(63,19){\small{(c)}}
    \end{overpic}

    \caption{Graph configuration of the structure: (a) Skeleton underlying a 3D surface mesh: \textit{Armadillo}; (b) Connecting graph of a biological structure from imaging data: \textit{Phantom}; (c) Skeleton underlying the modular (red) CAD geometry: \textit{Gear} by construction.}
    \label{fig:GraphStructure}
\end{figure}

A solid structure is often characterised by one or more branches, as elongated regions protruding from its main body.
Complex structures may exhibit holes or concave regions, others have the form of laminar thin or thick shells.
Convoluted and irregular organic structures, as well as composite CAD geometries, may show several of the above components, which can be locally decomposed in sub-blocks and modules \cite{chuang2000skeletonisation,lu2017evaluation,wang2017sheet,takayama2019dual}.
% The extent of such branching structure can be determined e.g. by delineating or segmenting 3D volumetric images, as in clinical applications; alternatively by considering the delimiting closed boundary of the structure as a polygonal surface mesh.
In any case, a graph can be determined as the skeleton underlying a branching portion of the structure by means of a number of (semi-)automatic computational routines and methods for different data types \cite{saha2016survey,saha2017skeletonization}. Such graph embeds both spatial and topological configurations, where generally the nodes correspond to junctions or terminal end-points, and the edges correspond to the elongated connecting regions underlying each branch \cref{fig:GraphStructure}.

\section{Conforming Solid Scaffolding}
\label{ConformSolidScaff}
\begin{figure}
    \centering
    \begin{tabular}{c}
         \begin{overpic}[width=.99\textwidth]{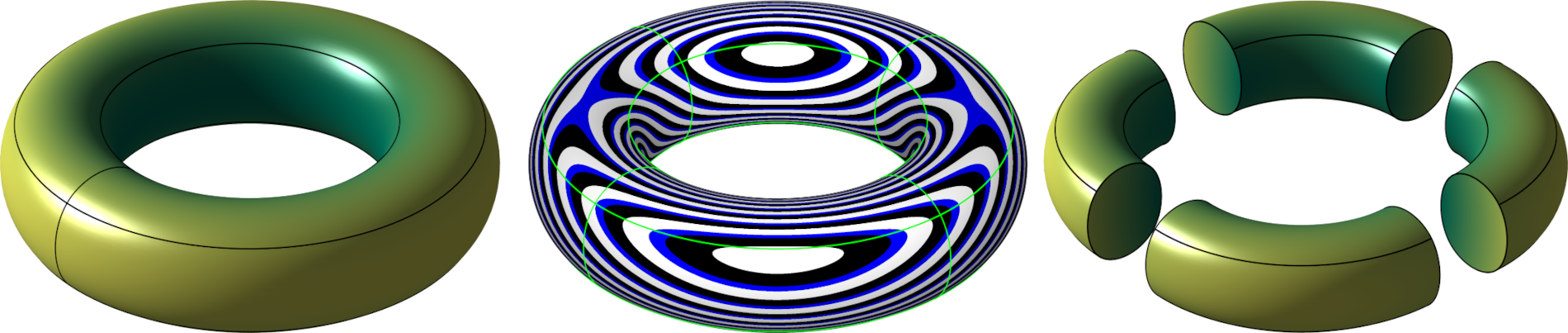}
            \put(1,20){\small{(a)}}
            \put(35,20){\small{(b)}}
            \put(68,20){\small{(c)}}
         \end{overpic}\\\\
         \begin{overpic}[width=.99\textwidth]{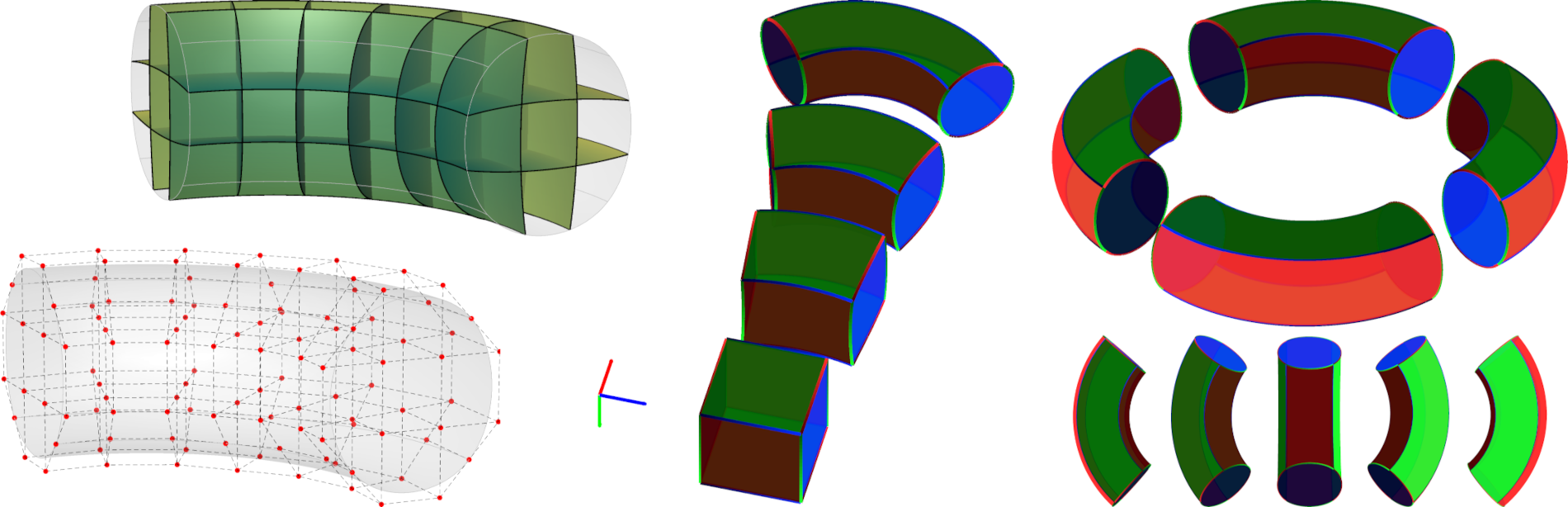}
            \put(3,30){\small{(d)}}
            \put(43,30){\small{(e)}}
            \put(68,30){\small{(f)}}
            \put(39,10){\small{$u$}}
            \put(38,3.5){\small{$v$}}
            \put(41,7){\small{$w$}}
         \end{overpic}
    \end{tabular}
    \caption{Solid Scaffolding of a \textit{Torus}: (a) Minimal element, (b) Equivalent multi-patch (4 elements) with reflection lines; (c) Multi-patch explosion view; (d) Individual element: control points lattice and inner solid sectioning; (e) Parametric embedding: individual element isomorphic to a cube; (f) Multi-patch sides (edges) facing (along) directions $(u,v,w)$, respectively: dark/light red for $u=\{0,1\}$, dark/light green for $v=\{0,1\}$, and dark/light blue for $w=\{0,1\}$ - conforming interfaces facing $w$; boundaries facing $u$ and $v$.}
    \label{fig:Torus}
\end{figure}

The scaffolding builds on the graph configuration of the branching structure in \cref{GraphConfigBranchStruct}, as it subdivides each connecting branch in a pre-defined minimal number of vectorial elements.
A rational partitioning of the structure's space is introduced, and the scaffolding configures a set of conforming elements being jointly defined over high-dimensional, geometrical and parametric domains.
In particular, the scaffolding comprises an organised and structured lattice of vectorial elements, each conforming to the neighbouring ones, and constitutes a conforming solid NURBS multi-patch \cref{fig:Torus}.
The conforming condition is necessary for an IGA-compatible domain, and it represents the key parametric constraint and the structural criterion at the base of the solid scaffolding.

\subsection{Vectorial Element} A vectorial element is the atomic component of the scaffolding and consists of a 3D free-form solid patch determined by NURBS primitives \cref{fig:Torus}. 
As introduced in \cite{piegl1996nurbs}, the vectorial element $\textrm{H}(u,v,w)$ is defined as a trivariate tensor product of NURBS curves along three independent parametric directions $(u,v,w)$, and has the form
\begin{equation}
\label{eq_NURBScuboid}
\begin{matrix}
\textrm{H}(u,v,w) = \sum_{i=1}^{l_u} \sum_{j=1}^{l_v} \sum_{k=1}^{l_w} R_{i,j,k}(u,v,w) \mathbf{P}_{i,j,k}, & \textrm{with}
\end{matrix}
\end{equation}
\begin{equation}
\label{eq_NURBSrationalBF}
    R_{i,j,k}(u,v,w) = \frac{ N_{i,d_u}(u) N_{j,d_v}(v) N_{k,d_w}(w) \textrm{w}_{i,j,k} }{\sum_{\hat{i}}^{l_u} \sum_{\hat{j}}^{l_v} \sum_{\hat{k}}^{l_w} N_{\hat{i},d_u}(u) N_{\hat{j},d_v}(v) N_{\hat{k},d_w}(w) \textrm{w}_{\hat{i},\hat{j},\hat{k}}},
\end{equation}
where $R_{i,j,k}(u,v,w)$ are the rational basis functions; $\mathbf{P}_{i,j,k} \in \mathbb{R}^3$ are the control points of cardinality $l_u$, $l_v$, and $l_w$ in each independent parametric direction; $N_{i,d_u}(u)$, $N_{j,d_v}(v)$, and $N_{k,d_w}(w)$ are the univariate B-spline basis functions of degree $d_u$, $d_v$, and $d_w$ respectively defined on the three independent parametric directions; and $\textrm{w}_{i,j,k}$ are the strictly positive weights of the rational basis functions.
The span of the basis functions' parametric domain is sampled in each independent direction by a set of intervals defined by the knot vectors $\mathbf{k}_u$, $\mathbf{k}_v$ and $\mathbf{k}_w$ respectively. The generic knot vector $\mathbf{k}$ comprises a set of real-valued knots ${k_1,...,k_r} \in [0,1]$ and is defined as
\begin{equation}
\label{eq_knotVect}
\begin{matrix}
\mathbf{k} = \{~\underbrace{k_1 , \dots , k_1}_{\textrm{m}} ~,~ k_2 , \dots , k_{r-1} ~,~ \underbrace{k_r , ... , k_r}_{\textrm{m}} ~\}, & \textrm{with} & r = l + d + 1.
\end{matrix}
    %\mathbf{k} = \{~\underbrace{k_1 , \dots , k_1}_{\scriptsize{\textrm{m}}} ~,~ k_2 , \dots , k_{r-1} ~,~ \underbrace{k_r , ... , k_r}_{\scriptsize{\textrm{m}}} ~\}, ~~~ \textrm{with} ~~~ r = l + d + 1.
\end{equation}
The knot vector is assumed open, i.e. $k_1 = 0$, $k_r = 1$ and the multiplicity $\textrm{m} = d + 1$, with $r$ being the total cardinality of the knots, $l$ being the number of control points in the generic parametric direction, and $d$ being the arbitrary degree of the associated univariate B-spline basis function.

The parametrisation in \cref{eq_NURBScuboid} maps a solid tensor grid defined on the parametric domain to a physical-space, where each vectorial element represents a solid free-form cuboid.
The free-form shape is modulated by the 3D coordinates of the control points $\mathbf{P}_{i,j,k}$ and by the weights $\textrm{w}_{i,j,k}$ of the rational basis functions.
The boundary of the element comprises 6 sides of quadrilateral profile, which are oriented, in opposite pairs, to face the directions of the parametric domain.
% Differently from a discrete hexahedral cuboid defined by a polygonal surface mesh, the vectorial element underlies a solid portion of the space in the form of a continuum. 

\subsubsection{Conforming Condition}
\label{ConformCond}
A pair of vectorial elements is said to be \textit{adjacent} when there is at least one side in common for each element. Such shared (or adjacent) sides are referred as \textit{interfaces}, whereas the free-end unshared sides are referred as \textit{boundaries}.
The same pair of adjacent elements is also said to be \textit{conforming} when all the interfaces are conforming, i.e. the parametrisation of each pair of adjacent sides is matching.
In other words, for each interface, all the control points, as well as, all the knot vectors and all the weights of the rational basis functions and the degrees of the associated univariate B-spline basis functions defining the sides of the adjacent sides must coincide, up to a different orientation of the elements in the parametric domain \cref{fig:Torus}.

A conforming solid scaffolding requires \textit{all} adjacent elements in the lattice jointly meet the conforming condition.

\subsection{Luminal Scaffolding Construction}
\label{LumScaff}
\begin{figure}
    \centering
    \begin{overpic}[width=.99\textwidth]{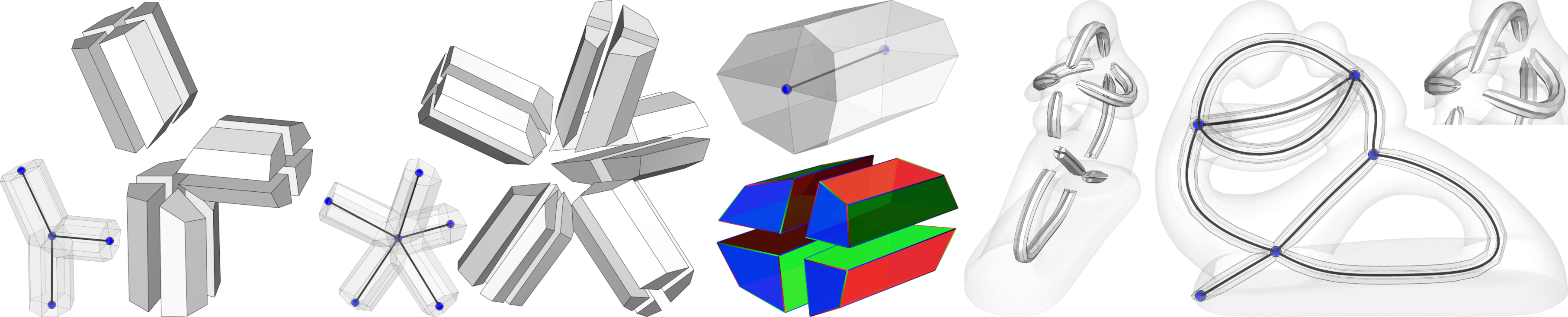}
            \put(1,18){\small{(a)}}
            \put(22,18){\small{(b)}}
            \put(44,18){\small{(c)}}
            \put(63,18){\small{(d)}}
         \end{overpic}
    \caption{Luminal Scaffolding Construction from Graph: (a) Bifurcation; (b) 5-way-junction: 4 elements required per branch; (c) Cross-sectional quadrants arrangement and elements adjacency per branch - note the conforming interfaces of adjacent elements couple pairs of sides facing $u$ and $v$; (d) Scaffolding construction for a 3D mesh structure (\textit{Fertility}) with multiple junctions and elongated portions. For simplicity, all scaffolding elements are shown as piece-wise linear segments.}
    \label{fig:LuminalScaff}
\end{figure}

%\label{LumWallScaff}
A luminal scaffolding builds on the branching structure graph in \cref{GraphConfigBranchStruct}, where higher density of branches and greater complexity of the branching pattern require an increasing number of elements to minimally parametrise the underlying spatial regions. 
%In the followings, a luminal scaffolding is first introduced, then the respective minimal number of elements and the associated arrangement is generalised considering a generic branching structure with arbitrary branching pattern.
%sections, two types of scaffolding are introduced, namely the luminal scaffolding and the wall scaffolding.
%For each type, the minimal number of vectorial elements is generalised considering a generic branching structure, whose graph represents a network with arbitrary branching pattern.
%Then, the arrangement of the elements and the conforming constraints are provided for the formulation of both the luminal and wall scaffoldings.
% \subsubsection{Luminal Scaffolding}
% \label{LumScaff}
The luminal scaffolding comprises the organised set of vectorial elements filling the innermost space of a branching structure, e.g. the fluid region as in a set of tubes or connected chambers, or the volumetric space occupied by a solid object. Leveraging the free-form shape of each element, the luminal scaffolding models the branching structure as a set of connecting portions that elongate for each structure protrusion encoded in the graph (\cref{fig:LuminalScaff}).

\paragraph{Minimal Elements} The scaffolding subdivides the branching structure in a pre-defined number of vectorial elements based on the topology of the graph. 
In general, the luminal scaffolding requires 4 vectorial elements per branch, assuming branches arbitrarily connect with generic \mbox{$n$-way-junctions} (\cref{fig:LuminalScaff}). 
%By way of example and independently from the geometric regularity of the luminal structure, the associated luminal scaffolding of a solid cylinder or an ellipsoid requires at least one vectorial element; in this case the graph consists of a two terminal nodes connected by one edge.
%For simple graph topologies, e.g. a bifurcating tube with a `Y'- or `T'-shape branching pattern, the luminal scaffolding may exceptionally reduce to 2 elements per branch. %; in this case the graph is an unrooted, undirected binary tree with 3 edges and 4 nodes (3 terminal end-points and 1 junction).
%
%In the general case, the number of vectorial elements of the luminal scaffolding is directly proportional to the number of branches.

\paragraph{Arrangement and Adjacency} For a generic graph with arbitrary branching pattern, the luminal scaffolding arranges each branch in 4 organised cross-sectional \textit{quadrants}, each accounting for an individual element.
This results in 4 adjacent vectorial elements elongated along the longitudinal direction of the branch, either joining a junction with a terminal endpoint, alternatively connecting two junctions.
Assuming the parametric direction $w$ mapping the longitudinal direction along the branch, each element is adjacent to other two elements of the same branch along one of its cross-sectional sides.
In particular, two interfaces are determined for each vectorial element in the correspondence of the interior part of the quadrant: one interface facing $u$ and the other interface facing $v$ respectively (dark red for $u$ and light green for $v$ in \cref{fig:LuminalScaff}). The remaining pair of sides, i.e. the boundaries, determines the exterior part of the quadrant: respectively, one boundary side facing $u$ (light red side) and another boundary side facing $v$ (dark green side), opposite to the interfaces.
The boundaries %do not share any side in common with any other element of the branch and of the whole luminal scaffolding, as they 
define a pseudo-circumferential profile relative to the elongation of the branch.

At any terminal branch, each element of the scaffolding terminates with a boundary side facing the longitudinal direction $w$.
At junctions, each element determines an interface with another adjacent element of a neighbouring incident branch.
The scaffolding configuration at the junction interfaces is defined by a \textit{quadrilateral junction simplex}.

\paragraph{Quadrilateral Junction Simplex}
\label{QuadJuncSimplex}

\begin{figure}
    \centering
    \begin{overpic}[width=.99\textwidth]{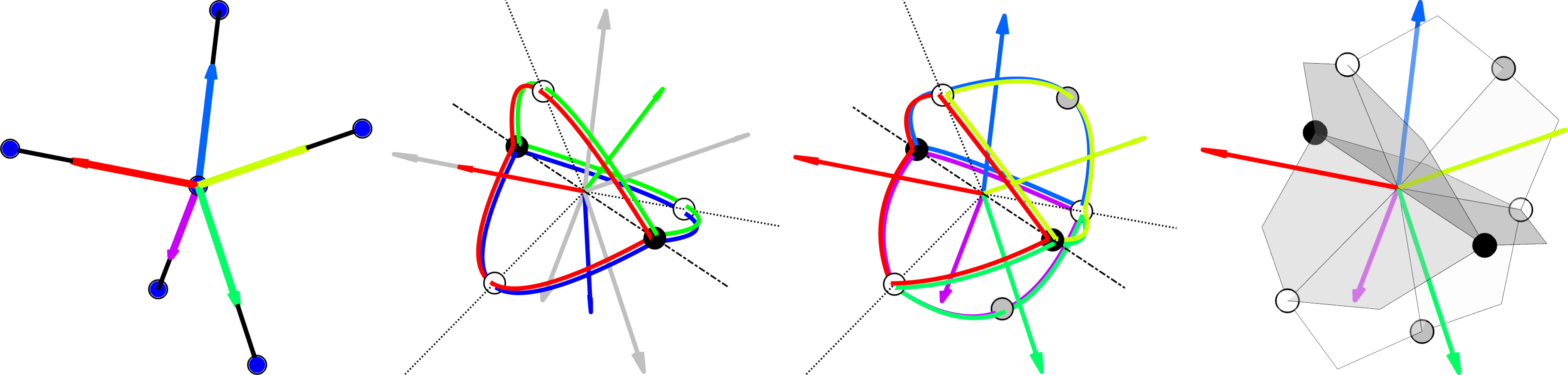}
            \put(2,21){\small{(a)}}
            \put(27,21){\small{(b)}}
            \put(52,21){\small{(c)}}
            \put(77,21){\small{(d)}}
            \put(4,15){\small{$\hat{\mathbf{n}}^{(5)}_{1}$}}
            \put(10,17){\small{$\hat{\mathbf{n}}^{(5)}_{2}$}}
            \put(16,15){\small{$\hat{\mathbf{n}}^{(5)}_{3}$}}
            \put(12,2){\small{$\hat{\mathbf{n}}^{(5)}_{4}$}}
            \put(8,8){\small{$\hat{\mathbf{n}}^{(5)}_{5}$}}
            \put(28,11){\small{$\hat{\mathbf{n}}_{1}$}}
            \put(43,18){\small{$\hat{\mathbf{n}}_{2}$}}
            \put(37,2){\small{$\hat{\mathbf{n}}_{3}$}}
            \put(45,4){\small{$\hat{\mathbf{v}}$}}
            \put(34,20){\small{$\hat{\mathbf{n}}_{1,2}^{\perp}$}}
            \put(47,11){\small{$\hat{\mathbf{n}}_{2,3}^{\perp}$}}
            \put(30,3){\small{$\hat{\mathbf{n}}_{3,1}^{\perp}$}}
            \put(70,4){\small{$\hat{\mathbf{v}}$}}
         \end{overpic}
    \caption{Quadrilateral junction simplex: (a) Set of incident directions $\hat{\mathbf{n}}^{(n)}$ of a generic \mbox{$n$-way-junction}, here $n$ = 5. (b) Associated atomic simplex $\mathcal{Q}^{\textrm{atom}}$ from three clustered principal directions: white nodes (original extraordinary vertices in black), colour-coded quadrant arcs surrounding the initial directional partition with a quadrilateral profile. (c) Resulting simplex $\mathcal{Q}$ after $n$-3 iterative bisections: new extraordinary vertices in white. (d) Quadrilateral and quadrant-based interfaces all facing $w$ constructed on $\mathcal{Q}$.}
    \label{fig:QuadJuncsimplex}
\end{figure}

In a generic junction, all the elements of each incident branch adjacently meet in the neighbourhood of the junction node.
This determines a structured set of interfaces, all facing the $w$ parametric direction, where each interface (as any other side) of the elements are configured with a quadrilateral profile.

These structural requirements are met by introducing the quadrilateral junction simplex $\mathcal{Q}$ (\cref{fig:QuadJuncsimplex}).
The simplex is a rational partitioning of the junction's space and determines an interfacing scaffolding configuration, whose atomic form underlies a bifurcation.

The topology of the atomic simplex $\mathcal{Q}^{\textrm{atom}}$ accounts for a total of 5 nodes and 6 edges being configured in 3 loops, each comprising 4 nodes and 4 edges, which are jointly connected by 2 common nodes, i.e. the initial \textit{extraordinary} vertices of the atomic simplex.
The associated spatial embedding of the atomic simplex delimits the interfacing cross-sectional areas of three incident branches, where each loop of $\mathcal{Q}^{\textrm{atom}}$ surrounds the incidental direction of the branch with 4 connecting arcs.
In this form, each loop defines a quadrilateral profile, and each arc of each loop underlies a quadrant.

The geometrical embedding of an atomic simplex is formulated by considering the directions of the incident branches as three non-coincident unit vectors $\hat{\mathbf{n}}_i \in \mathbb{R}^3$, with $i = \{1,2,3\}$, centred at the origin of a local reference system.
The initial extraordinary vertices lie along a common axis in opposite directions, where the common axis $\hat{\mathbf{v}}$ is defined as
\begin{equation}
\hat{\mathbf{v}} = \frac{ \sum_{i=1}^{3} \hat{\mathbf{n}}_{i} \times \hat{\mathbf{n}}_{j} }{| \sum_{i=1}^{3} \hat{\mathbf{n}}_{i} \times \hat{\mathbf{n}}_{j} |}.
\end{equation}
The unit vector $\hat{\mathbf{v}}$ is determined as the average of the cross-products of the incidental directions considered in pairs, in circular shift, i.e. \mbox{$j=i~(\textrm{mod}~3)+1$}.
The remaining three independent nodes of the atomic simplex $\mathcal{Q}^{\textrm{atom}}$ lie along the intermediate directions $\hat{\mathbf{n}}_{i,j}^{\perp}$, which are orthogonal to the common axis $\hat{\mathbf{v}}$.
The intermediate directions are defined as
\begin{equation}
\begin{matrix}
\hat{\mathbf{n}}_{i,j}^{\perp} = f_{i,j} ~ \frac{ \hat{\mathbf{n}}_{i,j} - (\hat{\mathbf{n}}_{i,j} \cdot \hat{\mathbf{v}})\hat{\mathbf{v}} }{| \hat{\mathbf{n}}_{i,j} - (\hat{\mathbf{n}}_{i,j} \cdot \hat{\mathbf{v}})\hat{\mathbf{v}} |} , & \textrm{with} & \hat{\mathbf{n}}_{i,j} = \frac{ \hat{\mathbf{n}}_{i} + \hat{\mathbf{n}}_{j} }{| \hat{\mathbf{n}}_{i} + \hat{\mathbf{n}}_{j} |},
\end{matrix}
    %\hat{\mathbf{n}}_{i,j}^{\perp} = f_{i,j} ~ \frac{ \hat{\mathbf{n}}_{i,j} - (\hat{\mathbf{n}}_{i,j} \cdot \hat{\mathbf{v}})\hat{\mathbf{v}} }{| \hat{\mathbf{n}}_{i,j} - (\hat{\mathbf{n}}_{i,j} \cdot \hat{\mathbf{v}})\hat{\mathbf{v}} |} , ~~~ \textrm{with} ~~~ \hat{\mathbf{n}}_{i,j} = \frac{ \hat{\mathbf{n}}_{i} + \hat{\mathbf{n}}_{j} }{| \hat{\mathbf{n}}_{i} + \hat{\mathbf{n}}_{j} |},
\end{equation}
being $\hat{\mathbf{n}}_{i,j}$ the bisecting unit vector between each pair of incident directions.
The flip factor $f_{i,j}$ determines whether the associated intermediate direction are considered in opposite sign.
This avoids inconsistent partitions of the space for particularly close and narrow sets of incidental directions.
Each flip factor $f_{i,j}$ is defined as
\begin{equation}
\begin{matrix}
f_{i,j} = \bigg{\{} \begin{matrix}
    -1~~ & \textrm{when}~~~p_{i,j} > 0 ~ \land ~ p_{j,k} \leq 0 ~ \land ~ p_{k,i} \leq 0\\
    1~~ & \textrm{otherwise}
    \end{matrix}, & \textrm{where}
\end{matrix}
    %f_{i,j} = \bigg{\{} \begin{matrix}
    %-1~~ & \textrm{when}~~~p_{i,j} > 0 ~ \land ~ p_{j,k} \leq 0 ~ \land ~ p_{k,i} \leq 0\\
    %1~~ & \textrm{otherwise}
    %\end{matrix}, ~~~ \textrm{where}
\end{equation}
\begin{equation}
\begin{matrix}
p_{i,j} = \hat{\mathbf{n}}_{i,j} \cdot \hat{\mathbf{n}}_{k}^{\perp}, & \textrm{and} & \hat{\mathbf{n}}_{i}^{\perp} = \frac{ \hat{\mathbf{n}}_{i} - (\hat{\mathbf{n}}_{i} \cdot \hat{\mathbf{v}})\hat{\mathbf{v}} }{| \hat{\mathbf{n}}_{i} - (\hat{\mathbf{n}}_{i} \cdot \hat{\mathbf{v}})\hat{\mathbf{v}} |}.
\end{matrix}
    %p_{i,j} = \hat{\mathbf{n}}_{i,j} \cdot \hat{\mathbf{n}}_{k}^{\perp}, ~~~ \textrm{and} ~~~
    %\hat{\mathbf{n}}_{i}^{\perp} = \frac{ \hat{\mathbf{n}}_{i} - (\hat{\mathbf{n}}_{i} \cdot \hat{\mathbf{v}})\hat{\mathbf{v}} }{| \hat{\mathbf{n}}_{i} - (\hat{\mathbf{n}}_{i} \cdot \hat{\mathbf{v}})\hat{\mathbf{v}} |}.
\end{equation}
The projection $p_{i,j}$ is computed as scalar-product between the bisecting unit vector $\hat{\mathbf{n}}_{i,j}$ and $\hat{\mathbf{n}}_{k}^{\perp}$, being the latter the unit vector associated to the $k$-th incident direction and orthogonal to the common axis $\hat{\mathbf{v}}$, with  $k=(i+1~(\textrm{mod}~3))+1$.

In the general form, the quadrilateral junction simplex $\mathcal{Q}$ of a junction with $n$ incident branches topologically consists in an adjacent stack of atomic simplexes, % i.e. \mbox{$\mathcal{Q} = \{\mathcal{Q}^{\textrm{atom}}_{1},\mathcal{Q}^{\textrm{atom}}_{2},\dots,\mathcal{Q}^{\textrm{atom}}_{n-2}\}$},
where each pair mutually shares a common loop.

In a physical-space embedding, such topological composition is equivalent to a cascade of spatial bisections of an initial atomic simplex, which first considers three principal incidental directions clustered from the \mbox{$n$-way-junction} (\cref{fig:QuadJuncsimplex}).
In this case, the total number of required bisections equals to $n-3$, for $n$ incident branches.
At each bisection, the initial \textit{parent} loop is split into a pair of \textit{child} loops following a binary spatial partition by connecting any alternating pair of vertices of the parent loop with an arc, and by splitting the connecting arc at the midpoint.
This binary spatial partition is iterated for the newly generated child loops until convergence, where each child loop surrounds every incidental direction of the \mbox{$n$-way-junction}.
At convergence, each loop of the simplex inherits a quadrilateral profile.

%However, two possible sectioning configurations arise: namely the (Z)- and the (E)-sectioning configurations\footnote{The E-Z notation loosely recalls the stereochemical isomerism, where E stands for \textit{entgegen}, and Z for \textit{zusammen}, both German words for \textit{opposite} and \textit{together}, respectively.}.
%They differ in choosing which pair of alternating vertices is considered to further bisect the parent loop.

%For a (Z)-junction, all the child loops of the resulting simplex $\mathcal{Q}$ keep sharing \textit{together} the initial pair of extraordinary vertices of the initial atomic simplex.
%Conversely, for a (E)-junction, at least one child loop of the resulting simplex $\mathcal{Q}$ does not share the initial pair of extraordinary vertices of the initial atomic simplex.
%Instead, new pairs of local extraordinary vertices are generated at each bisection, these being \textit{opposite} to the extraordinary vertices of the respective parent loop.

The resulting spatial configuration of $\mathcal{Q}$ is data-driven and is determined upon the incidental directions and sizes of the branches at the junction. Each bisection is based on maximising the volumetric partition for each incident branch.

A set of interfacing quadrants is obtained for each loop in $\mathcal{Q}$, by first connecting all the vertices of the simplex with its centroid, then by splitting each arc of each loop at the midpoint (\cref{fig:QuadJuncsimplex}).
This determines a consistent quadrant-based configuration and an initial geometrical embedding of the incident elements.
In particular, for each interface facing $w$, the subset of vertices belonging to the quadrant-based simplex coincides with the a subset of control points in \cref{eq_NURBScuboid}.

\paragraph{Conforming Constraints}
\label{LuminScaffConformConstr}
The conforming condition of the considered scaffolding introduces a certain dependency among the parameters of each vectorial element in the lattice.
In line with the construction, the parametric components along $u$ and $v$ directions are equivalent, i.e. the the univariate basis functions degrees $d_u = d_v$ in \cref{eq_NURBSrationalBF}, as well as, the knot vectors $\mathbf{k}_u = \mathbf{k}_v$, and the cardinality of control points $l_u = l_v$ as in \cref{eq_NURBScuboid} and \cref{eq_NURBSrationalBF}, respectively for each vectorial element.
Geometrically, the control points coordinates associated to each adjacent side of any pair of conforming vectorial elements coincide as well.
Each element of the luminal scaffolding has independent parametric components along $w$, i.e. arbitrary degree $d_w$, arbitrary spacing of the knot vector $\mathbf{k}_w$ and arbitrary cardinality of control points $l_w$.

\section{Geometrical Embedding}
\label{GeomEmbedding}

The conforming scaffolding associates a 3D geometrical embedding to the lattice of control points in \cref{eq_NURBScuboid}.
Each free-form element of the scaffolding recovers both shape and size of the underlying portion of the structure, similarly to \cite{krishnamurthy1996fitting}.
The directional formulation of the interfacing quadrilateral junction simplex in \cref{QuadJuncSimplex} provides an initial directional embedding, whereas the geometrical extent and configuration of each branch of the scaffolding is detailed in the following sections, by viable fitting strategies for different input data.

\subsection{Curvilinear Primitives and Lattice}
The geometrical embedding of each vectorial element is determined by a solid profiling operator.
This is controlled by a cross-sectional quadrilateral loop of varying shape sweeping along the branch-wise skeletal curvilinear primitive.
In particular, the quadrant cross-sectional quadrilateral loop comprises the curvilinear profiles at the interfacing simplexes and the outlines estimated from the superficial boundary of the structure and the underlying skeleton, at the extremities and along each branch, respectively.
Internal control points of the solid lattice are derived with a composition of blended interpolations from such curvilinear primitives \cite{piegl1996nurbs}.
% For simplicity, the luminal scaffolding is considered in the following formulation, as the geometrical embedding of a wall scaffolding is a conforming extension.

\subsubsection{Skeletal Primitives}

\begin{figure}
    \centering
    \begin{overpic}[width=.99\textwidth]{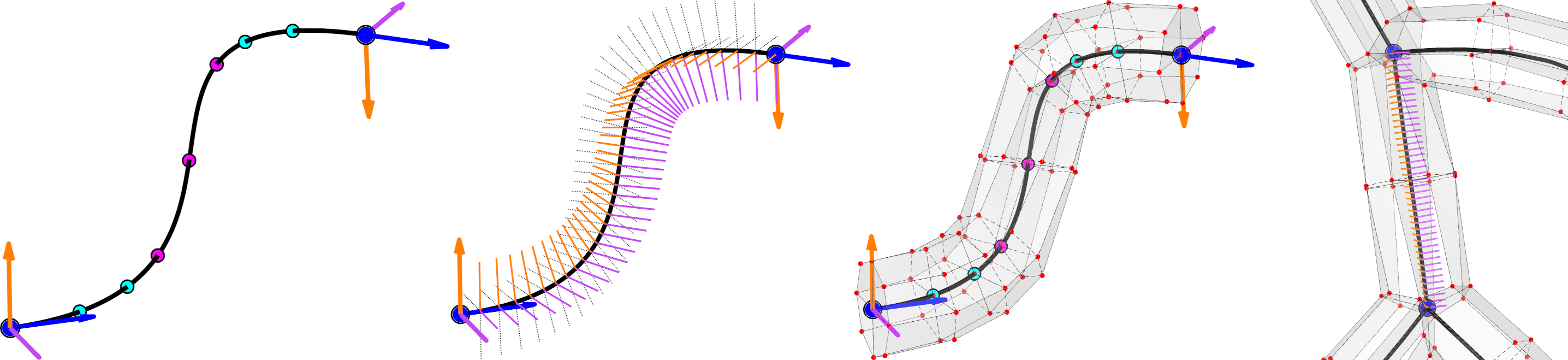}
        \put(2,21){\small{(a)}}
        \put(32,21){\small{(b)}}
        \put(57,21){\small{(c)}}
        \put(81,21){\small{(d)}}
        \put(12,8){\small{$\textrm{C}_{\textrm{skel}}(w)$}}
        \put(1,8){\small{$\hat{\mathbf{e}}_i^{(a)}$}}
        \put(3,0){\small{$\hat{\mathbf{f}}_i^{(a)}$}}
        \put(7,2){\small{$\hat{\mathbf{n}}_i^{(a)}$}}
        \put(22,13){\small{$\hat{\mathbf{e}}_i^{(b)}$}}
        \put(26.5,21.5){\small{$\hat{\mathbf{f}}_i^{(b)}$}}
        \put(27,17.5){\small{$\hat{\mathbf{n}}_i^{(b)}$}}
    \end{overpic}
    \caption{Curvilinear primitive: (a) Skeletal curve $\textrm{C}_{\textrm{skel}}(w)$: curvature-preserving (magenta) and uniformly spaced (cyan) samples, with terminal orthogonal bases; (b) Normal $\hat{\mathbf{e}}$ (orange) and binormal $\hat{\mathbf{f}}$ (purple) vectors matching the terminal quadrant bases for $\theta$; (c) Minimal torsion scaffolding and control points; (d) Scaffolding between two junctions (\textit{Phantom}) matching the quadrant bases for $\theta$.}
    \label{fig:SkelPrimitiveBranchMinTwist}
\end{figure}

The underlying skeletal centerline of the $i$-th branch is parametrised as a NURBS curve $\textrm{C}_{\textrm{skel}}(w)$, conforming to the univariate formulation in \cref{eq_NURBScuboid} along the longitudinal direction $w$.
The control points and the associated knot vector are estimated following \cite{piegl1996nurbs,ma1998nurbs,pagani2018curvature}, considering an adaptive sampling scheme accounting for uniform-, or non-uniform and curvature-preserving samples, up to a desired spatial resolution (\cref{fig:SkelPrimitiveBranchMinTwist}).
The curvilinear primitive $\textrm{C}_{\textrm{skel}}(w)$ links either a pair of junctions, or a junction with a terminal node, and each point of $\textrm{C}_{\textrm{skel}}(w)$ is associated with a set of orthogonal bases, comprising a tangential unit vector $\hat{\mathbf{n}}_i$ and a pair of normal $\hat{\mathbf{e}}_i$ and binormal $\hat{\mathbf{f}}_i$ unit vectors, as in a Frenet–Serret apparatus \cite{hanson1994quaternion}.
Both normal and binormal bases along the skeletal primitive determine the directional embedding of the branch-wise scaffolding quadrants.

At a junction simplex $\mathcal{Q}$, the normal $\hat{\mathbf{e}}_i$ and binormal $\hat{\mathbf{f}}_i$ unit vectors are defined for the $i$-th branch as
\begin{equation}
\label{eq_normbinorm}
\begin{matrix}
\hat{\mathbf{e}}_i = \frac{\hat{\mathbf{x}}_i + \hat{\mathbf{x}}_i^{\perp}}{|\hat{\mathbf{x}}_i + \hat{\mathbf{x}}_i^{\perp}|} & \textrm{and} & \hat{\mathbf{f}}_i = \frac{\hat{\mathbf{y}}_i + \hat{\mathbf{y}}_i^{\perp}}{|\hat{\mathbf{y}}_i + \hat{\mathbf{y}}_i^{\perp}|}, & \textrm{with}
\end{matrix}
    %\hat{\mathbf{e}}_i = \frac{\hat{\mathbf{x}}_i + \hat{\mathbf{x}}_i^{\perp}}{|\hat{\mathbf{x}}_i + \hat{\mathbf{x}}_i^{\perp}|} ~~~ \textrm{and} ~~~ \hat{\mathbf{f}}_i = \frac{\hat{\mathbf{y}}_i + \hat{\mathbf{y}}_i^{\perp}}{|\hat{\mathbf{y}}_i + \hat{\mathbf{y}}_i^{\perp}|}, ~~~ \textrm{with}
\end{equation}
\begin{equation}
\begin{matrix}
\hat{\mathbf{x}}_i^{\perp} = \frac{\hat{\mathbf{x}}_i - ( \hat{\mathbf{x}}_i \cdot \hat{\mathbf{y}}_i ) \hat{\mathbf{y}}_i} {|\hat{\mathbf{x}}_i - ( \hat{\mathbf{x}}_i \cdot \hat{\mathbf{y}}_i ) \hat{\mathbf{y}}_i|} & \textrm{and} & \hat{\mathbf{y}}_i^{\perp} = \frac{\hat{\mathbf{y}}_i - ( \hat{\mathbf{x}}_i \cdot \hat{\mathbf{y}}_i ) \hat{\mathbf{x}}_i} {|\hat{\mathbf{y}}_i - ( \hat{\mathbf{x}}_i \cdot \hat{\mathbf{y}}_i ) \hat{\mathbf{x}}_i|},
\end{matrix}
    %\hat{\mathbf{x}}_i^{\perp} = \frac{\hat{\mathbf{x}}_i - ( \hat{\mathbf{x}}_i \cdot \hat{\mathbf{y}}_i ) \hat{\mathbf{y}}_i} {|\hat{\mathbf{x}}_i - ( \hat{\mathbf{x}}_i \cdot \hat{\mathbf{y}}_i ) \hat{\mathbf{y}}_i|} ~~~ \textrm{and} ~~~ \hat{\mathbf{y}}_i^{\perp} = \frac{\hat{\mathbf{y}}_i - ( \hat{\mathbf{x}}_i \cdot \hat{\mathbf{y}}_i ) \hat{\mathbf{x}}_i} {|\hat{\mathbf{y}}_i - ( \hat{\mathbf{x}}_i \cdot \hat{\mathbf{y}}_i ) \hat{\mathbf{x}}_i|},
\end{equation}
where, $\hat{\mathbf{x}}_i$ and $\hat{\mathbf{y}}_i$ are the unit vectors orthogonal to the tangential $\hat{\mathbf{n}}_i$, respectively joining the local pair of extraordinary vertices of the quadrilateral loop and the opposite pair of loop's midpoints.
The normal and binormal bases of a terminal node follow the Frenet–Serret formulas \cite{hanson1994quaternion} starting from the bases at the junction simplex as in \cref{eq_normbinorm}, and osculate with minimal intrinsic torsion.

By directly matching the the normal and binormal bases of two connected junctions, the parametrisation of the $i$-th branch results in an additive structural twist.
To reduce such structural twist, the bases are quadrant-wise matched by minimising a torsional angle and by diluting it along the skeletal primitive (\cref{fig:SkelPrimitiveBranchMinTwist}).

Assuming $\hat{\mathbf{n}}_i^{(a)}$, $\hat{\mathbf{e}}_i^{(a)}$, $\hat{\mathbf{f}}_i^{(a)}$ and $\hat{\mathbf{n}}_i^{(b)}$, $\hat{\mathbf{e}}_i^{(b)}$, $\hat{\mathbf{f}}_i^{(b)}$ being the orthogonal bases at the endpoints of $\textrm{C}_{\textrm{skel}}(w)$ underlying the $i$-th connecting branch, at the corresponding junctions $a$ and $b$, the torsional angle $\theta$ is given by
\begin{equation}
\label{tortionangle}
\begin{matrix}
\theta = \min \Bigg( \arccos{ \Bigg( \hat{\mathbf{e}}_i^{(a)\perp} \cdot \underbrace{ \left( \mathrm{R}_{\hat{\mathbf{n}}_i^{(b)}}\left(\frac{j\pi}{2}\right) \hat{\mathbf{e}}_i^{(b)} \right) }_{ \big\{ \hat{\mathbf{e}}_i^{(b)} , \hat{\mathbf{f}}_i^{(b)} , -\hat{\mathbf{e}}_i^{(b)} , -\hat{\mathbf{f}}_i^{(b)} \big\} }  \Bigg) } \Bigg), & \textrm{for} & j = \{0,1,2,3\},
\end{matrix}
    %\theta = \min \Bigg( \arccos{ \Bigg( \hat{\mathbf{e}}_i^{(a)\perp} \cdot \underbrace{ \left( \mathrm{R}_{\hat{\mathbf{n}}_i^{(b)}}\left(\frac{j\pi}{2}\right) \hat{\mathbf{e}}_i^{(b)} \right) }_{ \big\{ \hat{\mathbf{e}}_i^{(b)} , \hat{\mathbf{f}}_i^{(b)} , -\hat{\mathbf{e}}_i^{(b)} , -\hat{\mathbf{f}}_i^{(b)} \big\} }  \Bigg) } \Bigg), ~~~ \textrm{for} ~~~ j = 0,1,2,3, 
\end{equation}
where $\hat{\mathbf{e}}_i^{(a)\perp}$ is the orthogonal projection of $\hat{\mathbf{e}}_i^{(a)}$ on $\hat{\mathbf{n}}_i^{(b)}$, and the rotation matrix $\mathrm{R}_{\hat{\mathbf{n}}_i^{(b)}}(\frac{j\pi}{2})$ accounts for quadrant rotations of the bases $\hat{\mathbf{e}}_i^{(b)}$, $\hat{\mathbf{f}}_i^{(b)}$ around $\hat{\mathbf{n}}_i^{(b)}$.

Leveraging the symmetry of the branch-wise quadrants, the maximal structural torsion along the skeletal primitive corresponds to $\theta_{\max} = \frac{\pi}{4}$.

\subsubsection{Cross-Sectional Quadrilateral Loops}

\begin{figure}
    \centering
    \begin{overpic}[width=.99\textwidth]{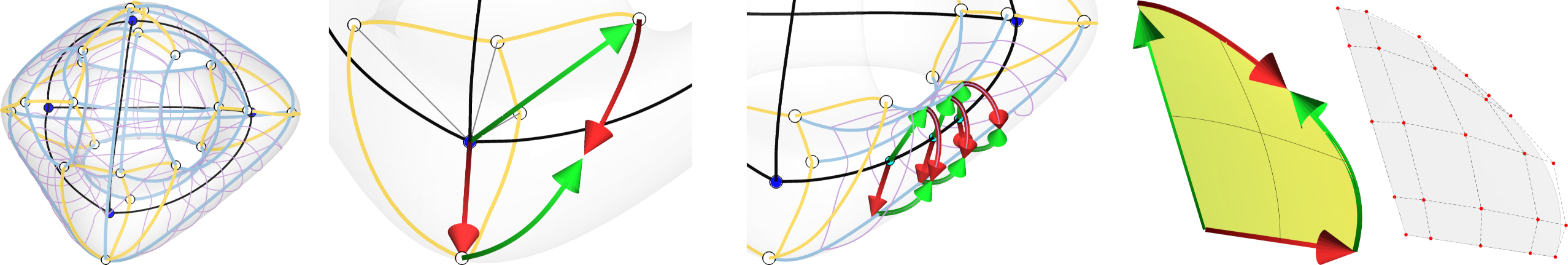}
        \put(1,16){\small{(a)}}
        \put(18,16){\small{(b)}}
        \put(45,16){\small{(c)}}
        \put(81,16){\small{(d)}}
        \put(24.5,4){\small{$\textrm{C}_0(u)$}}
        \put(33,14){\small{$\textrm{C}_0(v)$}}
        \put(39,6.5){\small{$\textrm{C}_1(u)$}}
        \put(34,0.5){\small{$\textrm{C}_1(v)$}}
    \end{overpic}
    \caption{Cross-Sectional Quadrilateral Loops: (a) Projected simplex(es) - gold arcs, white nodes on the surface of \textit{Genus3}; (b) Qruadrant-wise oriented parametrisation of boundary curves at the junction: $\textrm{C}_0(u)$ and $\textrm{C}_0(v)$ in a pseudo-radial direction, and $\textrm{C}_1(u)$ and $\textrm{C}_1(v)$ in a pseudo-circumferential direction, respectively; (c) Qruadrant-wise parametrisation of boundary curves (lilac) for the elongation of the branch delimited by longitudinal seam-cuts (cerulean). (d) Bilinearly blended Coons patch delimited by the boundary curves: the interpolated cross-sectional control points determine the internal lattice - splines with cubic degree(s).}
    \label{fig:CrossSectQuadLoops}
\end{figure}

The projected image of a junction simplex $\mathcal{Q}$ on the superficial boundary of the structure determines the curvilinear profiles of the cross-sectional quadrilateral loops in each respective quadrant.
In particular, each pair of extraordinary vertices and loops' midpoints of $\mathcal{Q}$ is first mapped following the directional formulation in \cref{QuadJuncSimplex}.
Then, each arc of the simplex loops is delineated with geodesics \cite{kimmel1998computing,martinez2004geodesic,surazhsky2005fast,crane2020survey}.
%The outer part of the cross-sectional quadrilateral loop of each quadrant at junctions is given by the image of $\mathcal{Q}$ on the superficial boundary of the structure (\cref{fig:CrossSectQuadLoops}).
%Following the directional formulation in \cref{QuadJuncSimplex}, each pair of extraordinary vertices and each pair of loop's midpoints in the simplex $\mathcal{Q}$ are first mapped and projected onto the superficial boundary of the structure.
%Then, each arc of the simplex loops is delineated over the superficial boundary of the structure with geodesics \cite{kimmel1998computing,martinez2004geodesic,surazhsky2005fast,crane2020survey}.

The parametrisation of each geodesic arc is constrained by the conforming conditions along the parametric directions $u$ and $v$.
For a luminal scaffolding, each geodesic arc of a quadrant is first split in two segments at the midpoint, and each segment is parametrised along $u$ and $v$ respectively, in an alternating and opposite scheme (\cref{fig:CrossSectQuadLoops}).
This determines a pair of associated curvilinear primitives $\textrm{C}_1(u)$ and $\textrm{C}_1(v)$ for each arc of the quadrant, following the pseudo-circumferential arrangement as in \cref{LumScaff}.
%The inner part of the cross-sectional quadrilateral loop of each quadrant comprises
The inner part comprises pairs of curvilinear primitives along $u$ and $v$ fitting the segments connecting each pair of extraordinary vertices and each pair of loop's midpoints to the junction node itself.
This determines another pair of curvilinear primitives, namely $\textrm{C}_0(u)$ and $\textrm{C}_0(v)$, originating from the junction node, each underlying a pseudo-radial direction.
For each quadrant, the 4-tuple of curvilinear primitives $\textrm{C}_0(u)$, $\textrm{C}_0(v)$, $\textrm{C}_1(u)$ and $\textrm{C}_1(v)$ constitute the sides of a quadrilateral loop on $(u,v)$ as in \cref{fig:CrossSectQuadLoops}.

In a similar manner, cross-sectional quadrilateral loops are determined for the elongation of each branch (\cref{fig:CrossSectQuadLoops}), where consecutive 4-tuple of curvilinear primitives are sampled orthogonal to the longitudinal direction $w$.
Seam-cuts conforming to the skeletal primitive $\textrm{C}_{\textrm{skel}}(w)$ connect the projected images of the simplexes along the branch as in the underlying graph (\cref{fig:CrossSectQuadLoops}).
At the outer part of the quadrant along the branch, similar curvilinear geodesic arcs are determined by leveraging a linear combination of the normal $\hat{\mathbf{e}}_i$ and binormal $\hat{\mathbf{f}}_i$ directions along $\textrm{C}_{\textrm{skel}}(w)$.
The orthogonal bases are (non-)linearly mapped onto the superficial boundary of the structure, and the associated projections constrain the local geodesics outlines delineation.
%The curvilinear geodesic arcs are parametrised with conforming curvilinear primitives along $u$ and $v$ similarly to the junction simplex case.
%Also, at the inner part of the quadrant, pairs of curvilinear primitives complete the 4-tuple of cruvilinear primitives at each conforming sample along $\textrm{C}_{\textrm{skel}}(w)$, by fitting the segments joining the skeletal primitive to the images of the normal and binormal projections.

\subsubsection{Internal lattice}
The internal lattice of each quadrant, and of each vectorial element, is determined with a composition of blended interpolations from the cross-sectional quadrilateral loops along the skeletal primitive.
In particular, each 4-tuple of curvilinear primitives $\textrm{C}_0(u)$, $\textrm{C}_0(v)$, $\textrm{C}_1(u)$ and $\textrm{C}_1(v)$ encloses a bilinearly blended Coons patch \cite{farin1999discrete}, whose conforming parametrisation on $(u,v)$ accounts for the interpolated inner control points coordinates of the lattice (\cref{fig:CrossSectQuadLoops}).
The conforming parametrisation of the internal lattice is given by orderly stacking and further interpolating the cross-sectional Coons patches along $\textrm{C}_{\textrm{skel}}(w)$, with a profiling operator as in \cite{Piegl1991,piegl1996nurbs}.

\subsection{Geodesic Fitting Strategies}
Curvilinear arcs are determined using geodesics by connecting two end-points on the superficial boundary of the branching structure.
Fitting and parametrising the curvilinear primitives follows \cite{piegl1996nurbs,ma1998nurbs,pagani2018curvature}, in keeping with the conforming constraints.
%In the following sections two representative geodesic approaches are mentioned to extract minimal paths on different input data, including surface meshes of polygonal tessellation, and anatomical structures captured by volumetric imaging.

\paragraph{Polygonal Surface Meshes}

\begin{figure}
    \centering
    \includegraphics[width=.99\textwidth]{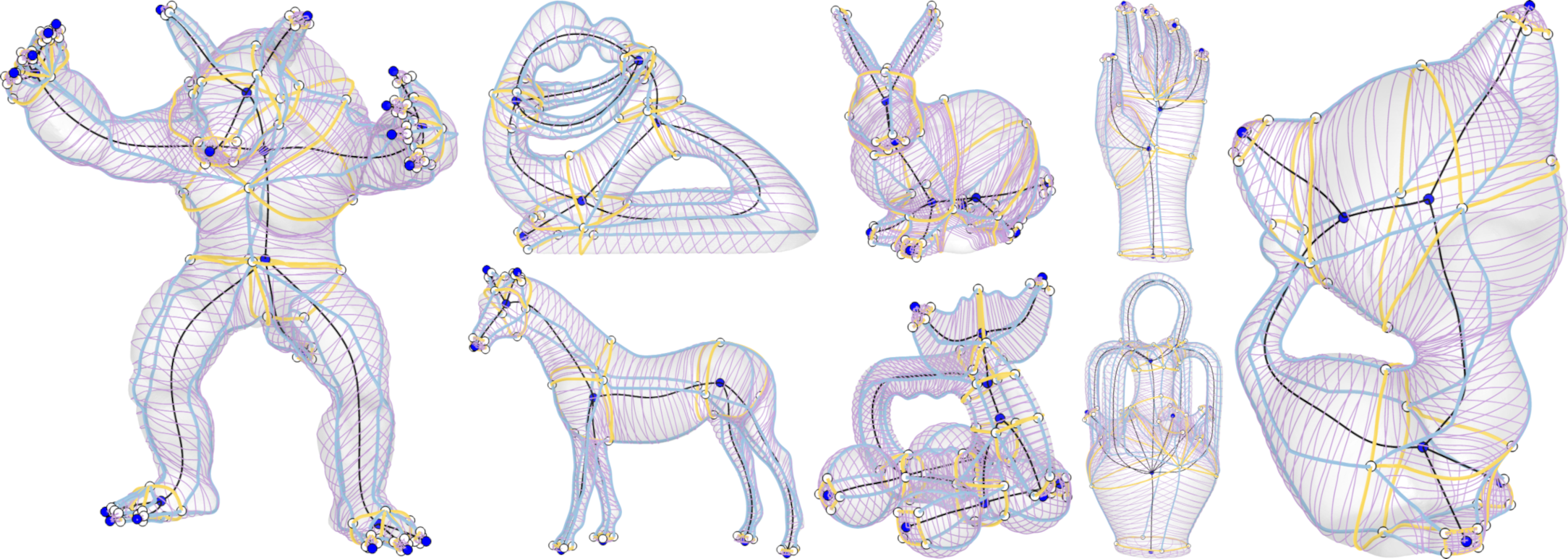}
    \caption{Geodesic Fitting -- Gallery of Polygonal Surface Meshes: mapping the geometrical embedding of the scaffolding in quad-based segmentation seam-cuts. Junction simplexes and terminal nodes: (gold), elongated branches (cerulean) and sampling outlines (lilac).}
    \label{fig:FittingMeshGallery}
\end{figure}

Superficial outlines underlie pseudo-circumferential and longitudinal profiles along each branch.
Geodesic curves are extracted with state-of-the-art methods as in \cite{surazhsky2005fast,crane2020survey} leveraging exact or approximated level-sets on a tessellated polygonal mesh.
The directional embedding of the scaffolding is first projected on the polygonal surface mesh by means of a non-linear \textit{inverse} skteleton-to-mesh mapping as in \cite{au2008skeleton}, or by semi-automatic user adjustments.
The projected images of each junction simplex nodes constitute the geodesic end-points. 
Along each branch, the orthogonal bases of the Frenet–Serret apparatus elicit preferential diffusion directions on the mesh, which constrain the geodesics along directional fields \cite{panozzo2014frame,pietroni2016tracing} in quad-regularised segmentation seam-cuts and outlines (\cref{fig:FittingMeshGallery}).

\paragraph{Volumetric Imaging}

\begin{figure}
    \centering
    \begin{overpic}[width=.9\textwidth]{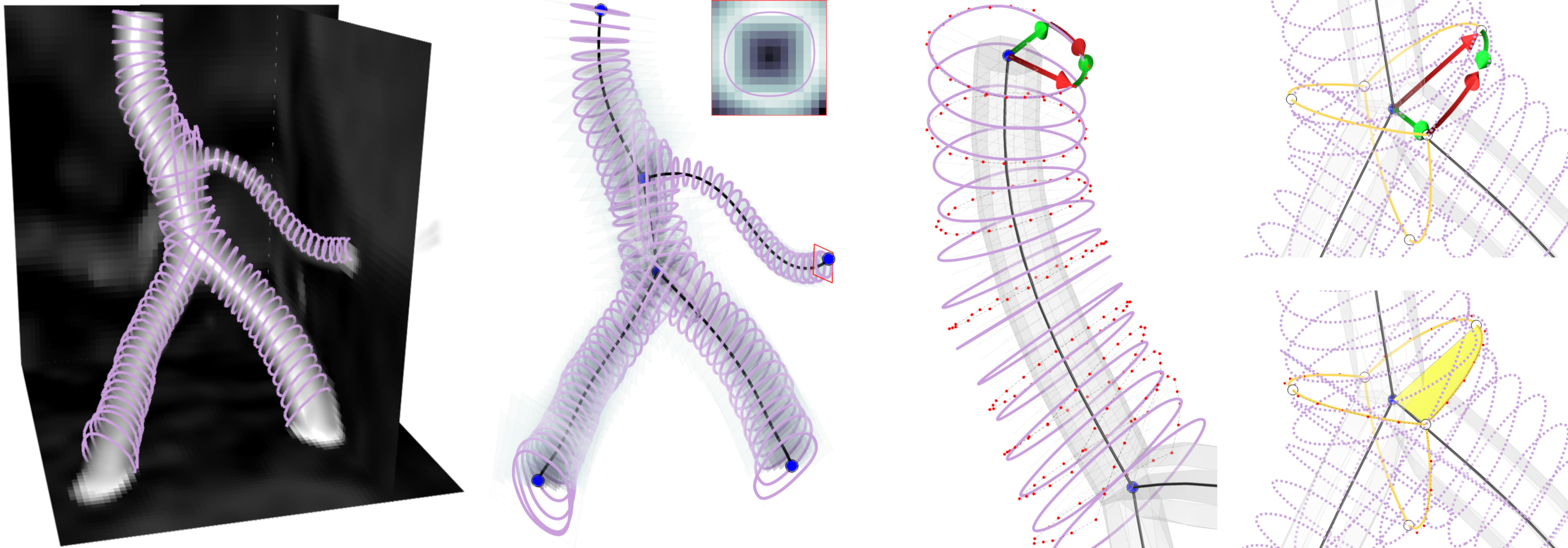}
        \put(-4,32){\small{(a)}}
        \put(31.5,32){\small{(b)}}
        \put(55.5,32){\small{(c)}}
        \put(77,32){\small{(d)}}
    \end{overpic}
    \caption{Geodesic Fitting -- Anatomical Structures (\textit{Phantom}): mapping the geometrical embedding of the scaffolding on snakes. (a) Volumetric image segmentation with snakes (lilac); (b) Cross-sectional curve-to-planar projections along each branch and contour delineation; (c) Scaffolding branch fitting on cross-sectional snakes; (d) Junction simplex fitting on overlapping snakes point-cloud.}
    \label{fig:FittingVesselsGallery}
\end{figure}

Geodesic active contours \cite{cheng2015accurate}, i.e. snakes, delineate the superficial boundary of a structure from 3D volumetric imaging.
In particular, snakes contour pseudo-circumferential profiles along the longitudinal axis of each branch (\cref{fig:FittingVesselsGallery}).
%Delineating the superficial boundary of a structure from 3D volumetric imaging is obtained with geodesic active contours (snakes) \cite{cheng2015accurate}.
%A set of snakes contour the pseudo-circumferential profile of the structure along the longitudinal axis of each branch (\cref{fig:FittingVesselsGallery}).
The volumetric image is first cross-sectionally resampled with a series of curve-to-planar projections \cite{Kanitsar2002} forming a branch-wise stack of slices orthogonal to the centerline.
Snakes evolve on each slice minimising a level-set potential \cite{sethian1996fast,sethian1999level} in the correspondence of the boundary of the structure.
For the elongation of the branch, the directional embedding of the scaffolding is linearly projected onto the snakes coordinates, directly fitting the curvilinear arcs (\cref{fig:FittingVesselsGallery}).
At the junctions, the directional embedding of the simplexes fits a 3D point-cloud determined by the overlapping snakes with least-squares (\cref{fig:FittingVesselsGallery}).
%Conversely, the directional embedding of the simplexes fits a 3D point-cloud determined by the overlapping snakes at the junctions with least-squares approximation (\cref{fig:FittingVesselsGallery}).

\section{Smoothing Paradigm}
\label{SmoothPrdgm}

The parametric and geometrical continuity of each scaffolding element depends on the degrees of the basis functions and on the multiplicity of each knot in the associated knot vector.
In particular, referring to \cref{eq_knotVect} and considering a generic parametric direction of a vectorial element, the number of continuous derivatives at the knot $k$ equals to $d - \textrm{m}_{(k)}$, with $d$ and $\textrm{m}_{(k)}$ being the univariate B-spline degree and the multiplicity of the knot $k$, respectively.

Considering open knot vectors as in \cref{eq_knotVect}, the parametric and geometrical continuity of any pair of conforming elements of the scaffolding is only \textit{positional}, i.e. the conforming elements are mutually $\textrm{G}^{0}$.

%First-order derivatives along $u$, $v$ and $w$ have arbitrary directions and magnitudes in the correspondence of interfacing and contiguous boundary sides of any adjacent pair of elements in the scaffolding.
%Similarly, high-order partial derivatives exhibit arbitrary values at the boundary of each vectorial element.
This may result in discontinuous curvature, geometrical creases and visible sharp edges, as well as in differential discontinuities of computational simulation profiles at the contiguous boundaries and across interfaces of the scaffolding.
%Although higher continuity is not strictly required for IGA, a solid smoothing paradigm is devised for the scaffolding, achieving higher smoothness, typical of an organic structure as a continuum.
Similarly to isoparametric Hermite elements \cite{petera1994isoparametric}, a solid smoothing paradigm is devised for the scaffolding, achieving higher smoothness and continuity, typical of an organic structure as a continuum.

The solid smoothing paradigm first builds on a topological subdivision of the scaffolding control points into a set of nested sub-lattices, i.e. \textit{medial shells}.
Then, the geometrical embedding of each sub-lattice is adjusted with a smoothing scheme leveraging concepts from state-of-the-art smoothing algorithms employed in subdivision surface modelling.

\subsection{Medial Shells Subdivision}
\label{MedialShellsSubdv}

\begin{figure}
    \centering
    \begin{overpic}[width=.99\textwidth]{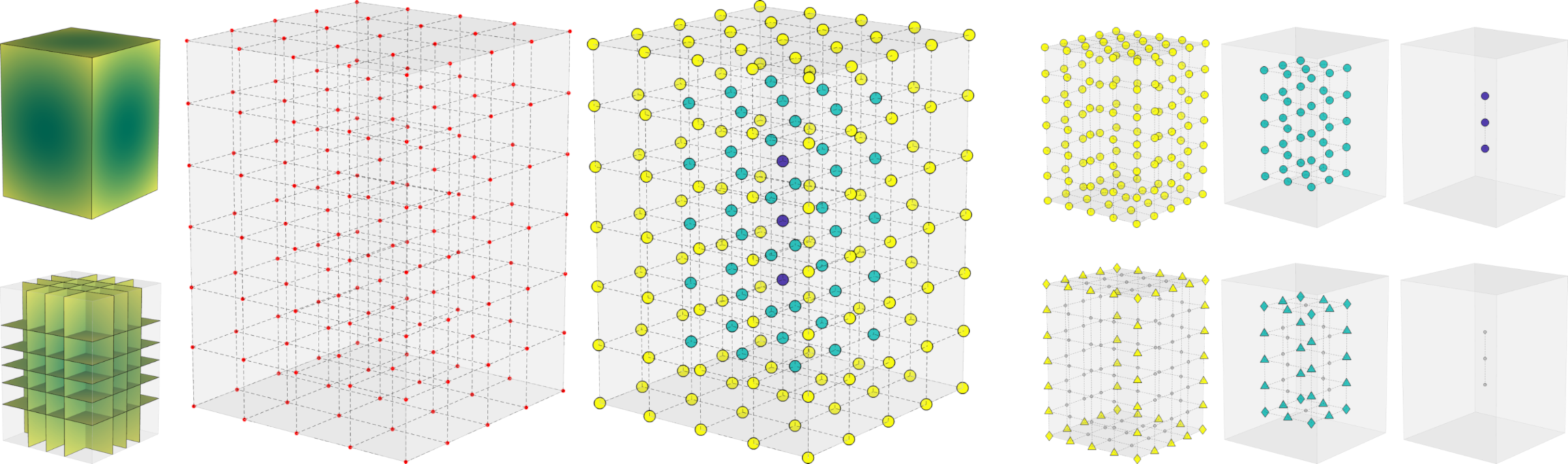}
        \put(10,29){\small{(a)}}
        \put(63.5,29){\small{(b)}}
        \put(63.5,13.5){\small{(c)}}
        \put(70,28.5){\small{$\mathcal{M}_{2}$}}
        \put(81.5,28.5){\small{$\mathcal{M}_{1}$}}
        \put(91,28.5){\small{$\mathcal{M}_{0} = \mathcal{M}_{\tilde{\tau}}$}}
        \put(70,13.5){\small{$\mathcal{F}_{(\mathcal{M}_{2})}$}}
        \put(81,13.5){\small{$\mathcal{F}_{(\mathcal{M}_{1})}$}}
        \put(93,13.5){\small{$\mathcal{F}_{(\mathcal{M}_{\tilde{\tau}})}$}}
    \end{overpic}
    \caption{Medial Shells Subdivision -- Single Cuboid Element: (a) Depth-wise colour-coded lattice of control points for each \textit{closed} $\mathcal{M}_{\tau}$. (b) Individual medial shell decomposition down to the medial locus $\mathcal{M}_{\tilde{\tau}}$. (c) Minimal fringe $\mathcal{F}_{(\mathcal{M}_{\tau})}$ of each medial shell: corner-lattice vertices ($\Diamond$) and edge-lattice vertices ($\bigtriangleup$).}
    \label{fig:MedialShellsSinglePatch}
\end{figure}

\begin{figure}
    \centering
    \begin{overpic}[width=.99\textwidth]{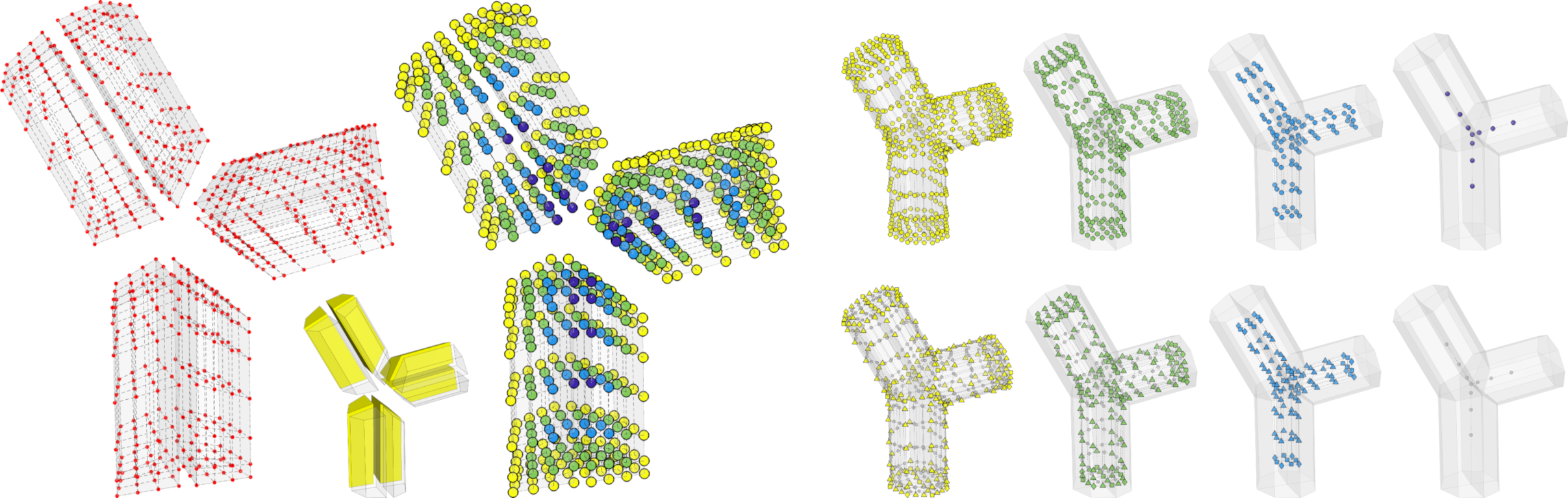}
        \put(20,30){\small{(a)}}
        \put(51,30){\small{(b)}}
        \put(51,14){\small{(c)}}
        \put(58.5,30){\small{$\mathcal{M}_{3}$}}
        \put(70,30){\small{$\mathcal{M}_{2}$}}
        \put(81.5,30){\small{$\mathcal{M}_{1}$}}
        \put(90.5,30){\small{$\mathcal{M}_{0} = \mathcal{M}_{\tilde{\tau}}$}}
        \put(57.5,13.5){\small{$\mathcal{F}_{(\mathcal{M}_{3})}$}}
        \put(69.5,13.5){\small{$\mathcal{F}_{(\mathcal{M}_{2})}$}}
        \put(81,13.5){\small{$\mathcal{F}_{(\mathcal{M}_{1})}$}}
        \put(92.5,13.5){\small{$\mathcal{F}_{(\mathcal{M}_{\tilde{\tau}})}$}}
    \end{overpic}
    \caption{Medial Shells Subdivision -- Multi-patch 3-way-junction: (a) Depth-wise colour-coded lattice of control points for a section of \textit{closed} $\mathcal{M}_{\tau}$ (exploded view). (b) Individual medial shell decomposition down to the medial locus $\mathcal{M}_{\tilde{\tau}}$. (c) Composite fringe $\mathcal{F}_{(\mathcal{M}_{\tau})}$ of each medial shell: corner-lattice vertices ($\Diamond$), edge-lattice vertices ($\bigtriangleup$), and face-lattice vertices ($\square$).}
    \label{fig:MedialShellsMultiPatch}
\end{figure}

A topological subdivision of the scaffolding control points determines a set of nested sub-lattices at different depths, i.e. a set of \textit{medial shells} (\cref{fig:MedialShellsSinglePatch,fig:MedialShellsMultiPatch}).
A medial shell $\mathcal{M}$ is a lattice graph of regular tiling, comprising a subset of control points $P$ and a subset of edges $E$ connecting points of the same shell in quadrilateral cells.
Such medial shell can also be regarded as a quadrilateral (open or closed) mesh.
A set of medial shells (\cref{fig:MedialShellsSinglePatch}) can be defined over the structured lattice of control points in an individual element as
\begin{equation}
\label{eq_MedialShells}
\begin{matrix}
\mathcal{M}_{\tau} = \{ (P,E) | P=\{\mathbf{P}_{i,j,k}|\mathrm{t}_{i,j,k} = \tau \} , E = \mathcal{N}_6(P) \} , & \mathrm{with} & \tau = 0,1,\dots,\bar{\tau}.
\end{matrix}
\end{equation}
The scalar value $\mathrm{t}_{i,j,k} \in \mathbb{N}$ is the depth tag of each control point $\mathbf{P}_{i,j,k}$ in the structured lattice. All control points with same depth tag $\mathrm{t}_{i,j,k} = \tau$ constitute the vertices of the associated medial shell $\mathcal{M}_{\tau}$, and the respective edges are defined by the 6-connected neighbourhood $\mathcal{N}_6$ pattern, as in the original structured lattice of the element.
The depth-wise medial shell subdivision follows a recursive definition of the tag $\mathrm{t}_{i,j,k}$. All tags are first initialised as \textit{infinite} values for the entirety of the structured lattice, whereas $0$ values are imposed for the boundary sides.
A recursive scheme assigns a \textit{finite} value to each tag in the lattice as
\begin{equation}
\label{eq_rcrsvShellTag}
    \textrm{t}_{i,j,k} = \min \left(~\textrm{t}_{i,j,k}~,~\min \left(~\textbf{t}_{\mathcal{N}_{26}(i,j,k)}~\right)+1~\right),
\end{equation}
where $\textbf{t}_{\mathcal{N}_{26}(i,j,k)}$ is the array of tag values associated to the 26-connected neighbourhood $\mathcal{N}_{26}$ relative to the index tuple $(i,j,k)$ in the structured lattice of each (multi-patch) element.

%The same medial shell subdivision applies to the lattice of a conforming (coupled) scaffolding, where the depth tags at the interfacing sides are initialised with infinite values and recursively determined as in \cref{eq_rcrsvShellTag}. 

Multiple nested shells are generated from the superficial boundary of the lattice, i.e at null depth, following an iterative integer-step shrinking of the exterior shell towards the innermost region of the structure.
In particular, each inner shell represents a discrete \textit{contraction} of the relative outer shell, up to an ultimate irreducible shell $\mathcal{M}_{\tilde{\tau}}$, i.e. the \textit{medial locus}, which lies in the neighbourhood of the structure skeleton and can be a degenerate lattice graph.
%For a generic scaffolding, the (odd or even) number of the control points in the original lattice and the extent of the scaffolding boundary sides affect the cardinality of medial shells and the topology of the associated medial locus.
% The latter may exhibit, in each branch, either a singular, or linear, or planar cell pattern, associated to an odd or even number of control points in the original lattice.
%In the physical domain, $\mathcal{M}_{\tau}$ approximates a stack of foliating layers decomposing the solid elements of the scaffolding and the underlying continuum.

\subsection{Medial Shells Smoothing}
\label{SmoothMedialShells}

\begin{figure}
    \centering
    \begin{overpic}[width=.99\textwidth]{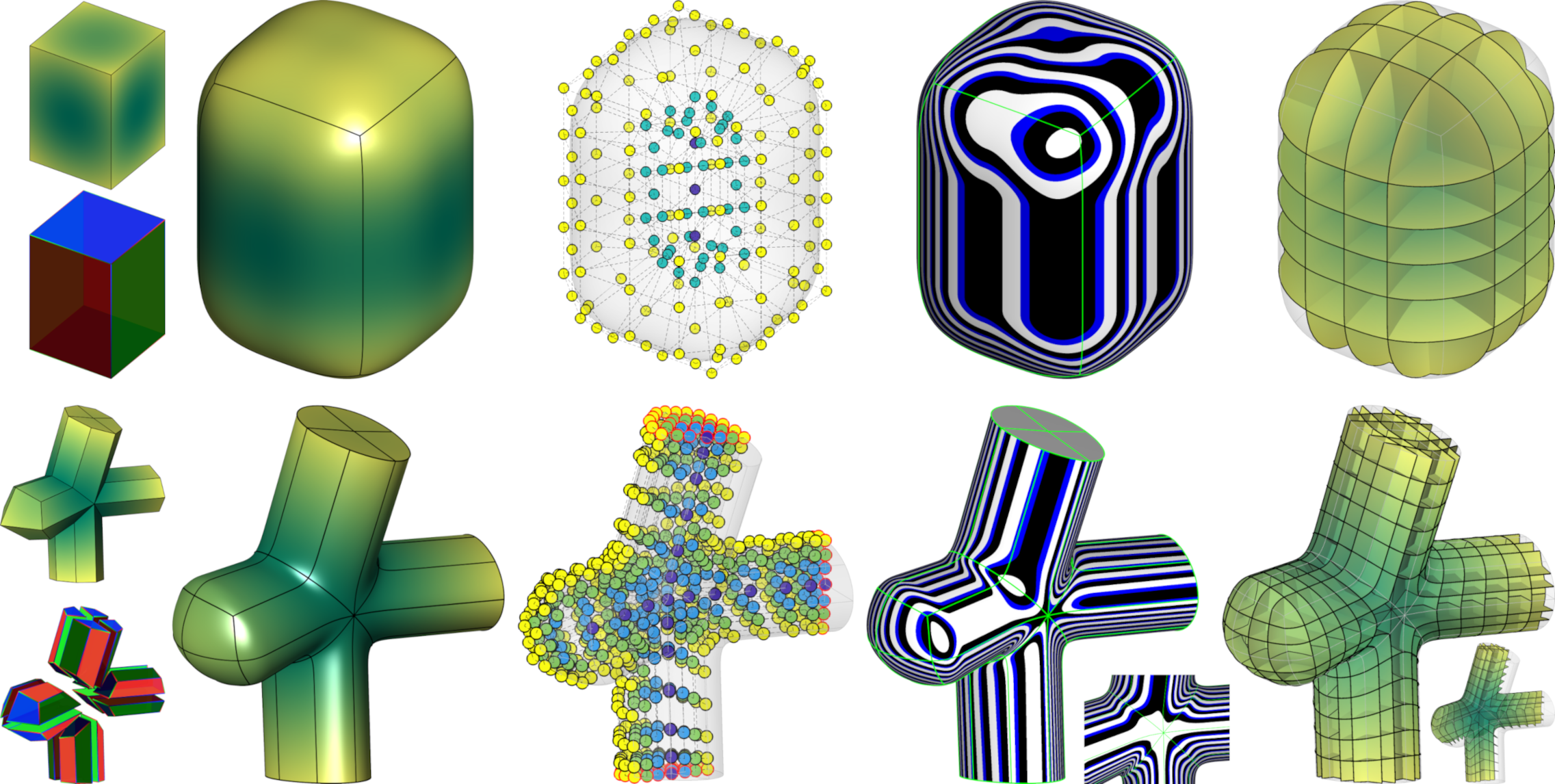}
        \put(0,49){\small{(a)}}
        \put(0,23){\small{(b)}}
    \end{overpic}
    \caption{Solid Smoothing Paradigm at Convergence: (a) Single cuboid; (b) Multi-patch 4-way-junction with locally mixed open and closed medial shells. Resulting geometry, depth-wise medial shells of control points, reflection lines and inner solid trabecular view.}
    \label{fig:SmoothingParadigmResults}
\end{figure}

Given the analogy between medial shells and quadrilateral meshes, the smoothing paradigm leverages concepts introduced by state-of-the-art subdivision surface algorithms \cite{catmull1978recursively,stam2001subdivision}.
Reference literature generally employs first a local refinement of the mesh with local subdivisions.
Then, vertices coordinates are adjusted according to a smoothing scheme based on the vertex \textit{valence} $\nu$, i.e. the number of edges that meet at each vertex.
In a similar manner, a lattice refinement by knot-insertion can be alternated to the following smoothing paradigm at each smoothing step, reaching the underlying limit geometry (\cref{fig:SmoothingParadigmResults}).
%The solid multi-patch arrangement and the conforming conditions limit the straightforward application of state-of-the-art algorithms, as a lattice refinement by knot-insertion implies a global increase of parameters propagating consistently over the entirety of the solid scaffolding.
%The smoothing paradigm below describes a geometrical adjusting scheme without altering the original density of the scaffolding lattice.

\subsubsection{Medial Shell Fringe}
\label{MedialShellFringe}

\begin{figure}
    \centering
    \begin{overpic}[width=.99\textwidth]{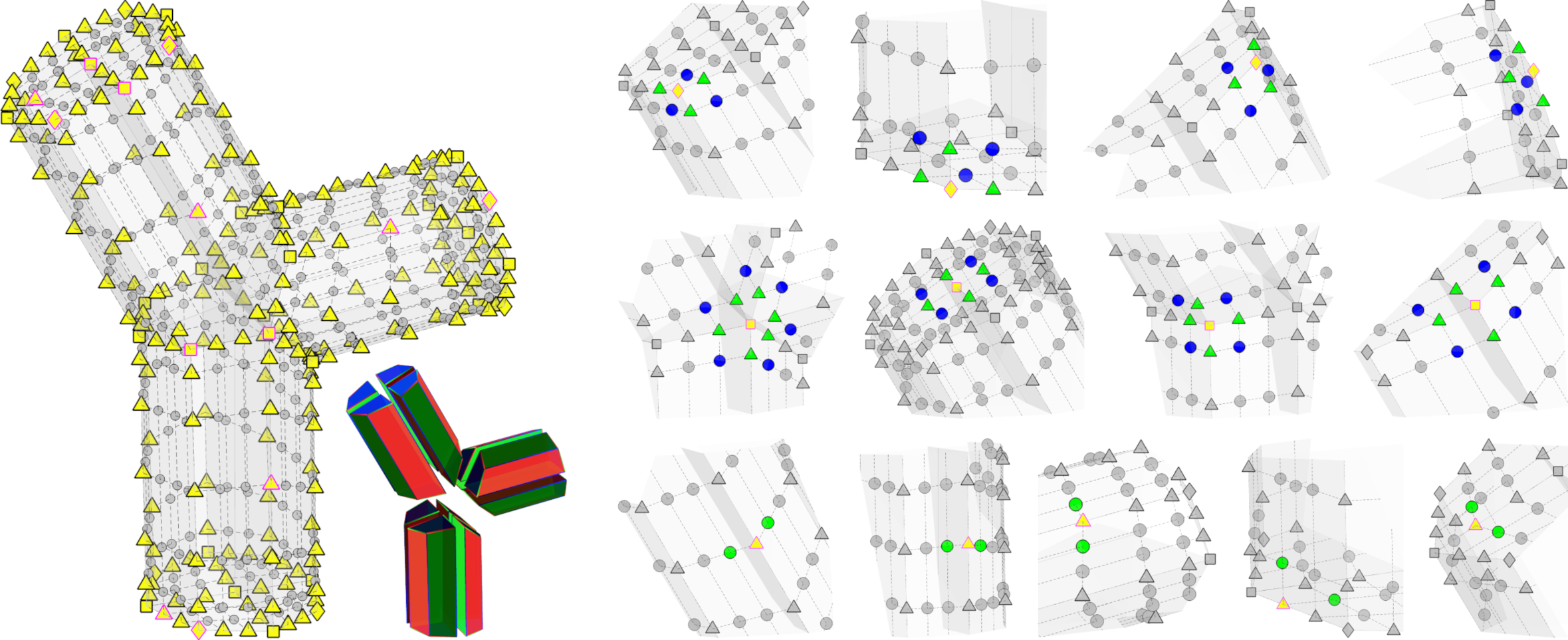}
        \put(19,39){\small{(a)}}
        \put(38,39){\small{(b)}}
        \put(38,25){\small{(c)}}
        \put(38,11){\small{(d)}}
    \end{overpic}
    \caption{Closed Medial Shell Fringe -- Multi-patch 3-way-junction: (a) Parametric embedding and exterior shell fringe - selected vertices (magenta outlines); (b) Neighbourhoods of corner-lattice vertices ($\Diamond$): edge-connected neighbours (green), face-connected neighbours (blue); (c) Neighbourhoods of face-lattice vertices ($\square$): edge-connected neighbours (green) and face-connected neighbours (blue). (d) Neighbourhoods of edge-lattice vertices ($\bigtriangleup$): edge-connected neighbours as in \cref{eq_smthEdgeLatticeVertex} (green).}
    \label{fig:MultiPatchFringe}
\end{figure}

The smoothing paradigm operates on each medial shell \textit{fringe} $\mathcal{F}$ of the (sub-)lattice(s).
The fringe is defined as the union of the boundary rims of a lattice within a certain outskirt of influence.
By way of example, the outermost and minimal-influence fringe of a solid scaffolding coincides with the set of edges belonging to the boundary sides of each vectorial element. %and by increasing the outskirt of influence, the fringe is dilated towards the sides of the conforming elements.

Considering the topological subdivision in \cref{MedialShellsSubdv}, the definition of a shell-wise fringe $\mathcal{F}_{(\mathcal{M}_{\tau})}$ is applied to each contracted medial shell, in keeping with the parametric arrangement of the scaffolding elements.
In detail, the medial shell fringe $\mathcal{F}_{(\mathcal{M}_{\tau})}$ includes the vertices belonging to the \textit{boundaries} rims of the contracted sub-lattice.
These are labelled either as \textit{corner}-, or \textit{edge-lattice} vertices (or \textit{face-lattice} vertices, with increasing outskirt of influence).
Additional vertices at the rims of the scaffolding \textit{interfaces} are also included in the shell-wise fringe $\mathcal{F}_{(\mathcal{M}_{\tau})}$, these all being labelled as \textit{edge-lattice} vertices for a minimal outskirt of influence, or labelled as \textit{face-lattice} vertices when they are corners of the respective individual adjacent elements.
%The notation corner-, edge- or face-lattice vertex is relative to the parametric (sub-)lattice context of the conforming elements in the scaffolding.
Note that all the \textit{extraordinary} vertices of a simplex $\mathcal{Q}$ in \cref{QuadJuncSimplex} coincide with \textit{face-lattice} vertices of the fringe.

\paragraph{Smoothing Scheme}
\label{SmoothScheme}
The smoothing scheme determines a \textit{geometric dual} configuration for the vertices in each fringe $\mathcal{F}_{(\mathcal{M}_{\tau})}$. 
Similarly to \cite{stam2001subdivision}, in the simplest form, the smoothing scheme adjusts the coordinates of each vertex by adopting an iterative weighting scheme based on the vertex valence $\nu$ and on the label of the vertex. % at the fringe.

For each \textit{corner-} or \textit{face-lattice} vertex (\cref{fig:MultiPatchFringe}), the geometrical coordinates are iteratively adjusted as
\begin{equation}
\label{eq_smthCrnrFaceLatticeVertex}
    \mathbf{P}_{i+1} = \alpha(\nu)\mathbf{P}_{i} + \sum_{\nu}~\beta(\nu)\mathbf{P}_{i}^{(\mathrm{e}_{\nu})} + \sum_{\nu}~\gamma(\nu)\mathbf{P}_{i}^{(\mathrm{f}_{\nu})},
\end{equation}
where $\mathbf{P}_{i}$ represents the coordinates of the considered corner- or face-lattice vertex of the fringe $\mathcal{F}_{(\mathcal{M}_{\tau})}$, $\mathbf{P}_{i}^{(\mathrm{e}_{\nu})}$ represents the coordinates of the respective $\nu$-th edge-wise connected neighbour in $\mathcal{M}_{\tau}$, and $\mathbf{P}_{i}^{(\mathrm{f}_{\nu})}$ represents the coordinates of the respective $\nu$-th face-wise connected neighbour in $\mathcal{M}_{\tau}$, respectively at the $i$-th iteration.
The weights are defined by the valence $\nu$ of the fringe vertex as
\begin{align}
\label{eq_CatmullClarkStamWeights}
    \begin{matrix}
    ~~~\alpha(\nu) = \frac{\nu-3}{\nu}~~~ & ~~~\beta(\nu) = \frac{2}{\nu^2}~~~ & ~~~\gamma(\nu) = \frac{1}{\nu^2},~~~
    \end{matrix}
\end{align}
and satisfy an affine invariant scheme, i.e. $\alpha(\nu) + \nu~\beta(\nu) + \nu~\gamma(\nu) = 1$.

In this case, corner-lattice vertices of the fringe usually have valence $\nu=3$, whereas face-lattice vertices of the fringe are associated either to regular vertices ($\nu=4$), or to extraordinary vertices with arbitrary valence $\nu>4$.

Conversely, for each \textit{edge-lattice} vertex (\cref{fig:MultiPatchFringe}), the geometrical coordinates are iteratively adjusted as 
\begin{equation}
\label{eq_smthEdgeLatticeVertex}
    \mathbf{P}_{i+1} = \tilde{\alpha}\mathbf{P}_{i} + \sum_{\nu}~\tilde{\beta}\mathbf{P}_{i}^{(\mathrm{e}_{\nu}^{\perp})} ,
\end{equation}
where, at the $i$-th iteration, $\mathbf{P}_{i}$ represents the coordinates of the considered edge-lattice vertex of the fringe $\mathcal{F}_{(\mathcal{M}_{\tau})}$, and $\mathbf{P}_{i}^{(\mathrm{e}_{\nu}^{\perp})}$ represents the coordinates of the $\nu$-th edge-wise connected neighbour in $\mathcal{M}_{\tau}$, whose parametric direction in the (sub-)lattice lies \textit{perpendicularly} to the one of the considered fringe vertex.
In other words, each edge-wise connected neighbour in $\mathcal{M}_{\tau}$ belonging to the same rim in the (sub-)lattice is excluded.
The latter condition imposes each edge-lattice vertex of the fringe has valence $\nu=2$.
The weights, in this case, are fixed as $\tilde{\alpha} = \frac{1}{2}$ and $\tilde{\beta} = \frac{1}{4}$. % also satisfying an affine invariant scheme.

The iterative smoothing scheme ultimately converges for a machine error evaluated on the norm $|\mathbf{P}_{i+1} - \mathbf{P}_{i}|$.

\paragraph{Smoothing Scheme of Open Medial Shells}
\label{SmoothSchemeOpenMesh}

\begin{figure}
    \centering
    \begin{overpic}[width=.99\textwidth]{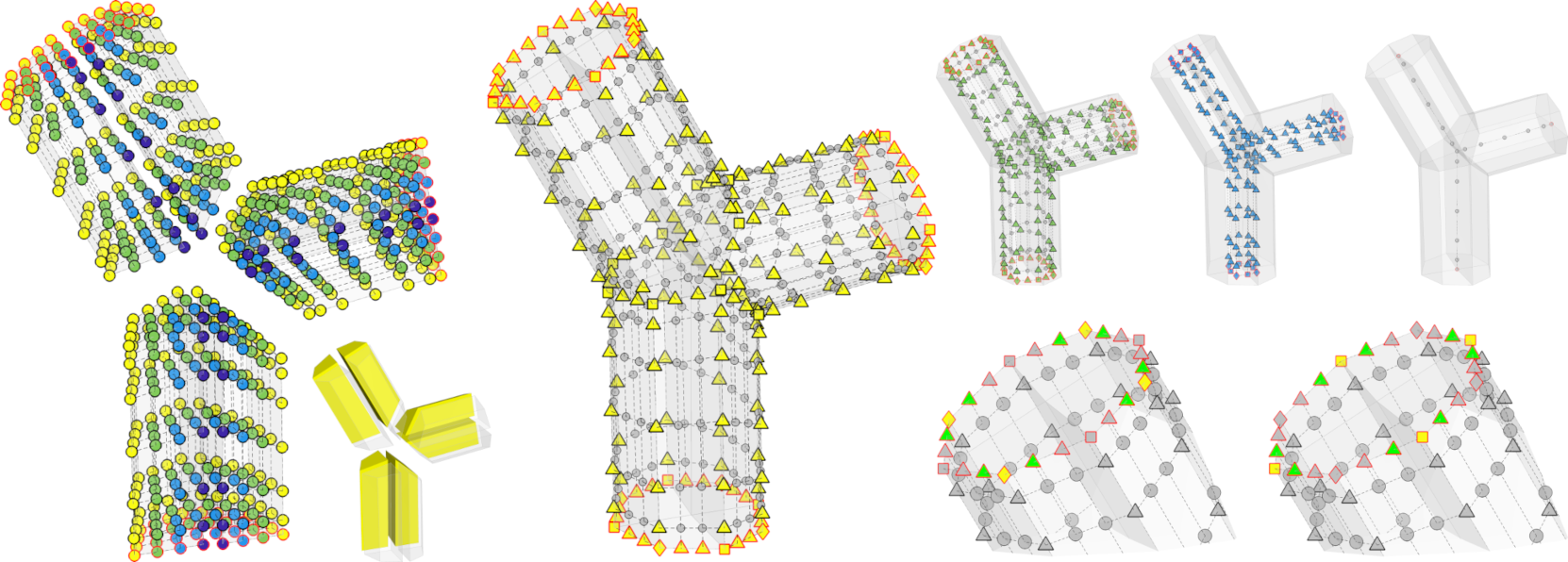}
        \put(0,35){\small{(a)}}
        \put(30,35){\small{(b)}}
        \put(60,14){\small{(c)}}
        \put(81,14){\small{(d)}}
        \put(15,33){\small{$\mathcal{M}_{\tau}$}}
        \put(44,33){\small{$\mathcal{F}_{(\mathcal{M}_{3})}$}}
        \put(65,33){\small{$\mathcal{F}_{(\mathcal{M}_{2})}$}}
        \put(79,33){\small{$\mathcal{F}_{(\mathcal{M}_{1})}$}}
        \put(92.5,33){\small{$\mathcal{F}_{(\mathcal{M}_{\bar{\tau}})}$}}
    \end{overpic}
    \caption{Open Medial Shells Fringe -- Multi-patch 3-way-Junction: (a) Depth-wise colour-coded lattice for an open $\mathcal{M}_{\tau}$ (section) - \textit{open-fan} (red); (b) Medial shell fringe decomposition; (c) Neighbourhood of open-fan corner-lattice vertices ($\Diamond$), and (d) face-lattice vertices ($\square$). Respective edge-connected neighbours as in \cref{eq_smthOpenFanEdgeLatticeVertex} (green). Edge-lattice vertices of the open-fan ($\bigtriangleup$) remain unaltered.}
    \label{fig:MultiPatchFringeOpen}
\end{figure}

Certain configurations of the medial shells result in vertices of the fringe $\mathcal{F}_{(\mathcal{M}_{\tau})}$ having an \textit{open-fan} in the quadrilateral tiling.
In this particular case, the smoothing scheme of the \textit{corner-}, \textit{edge-} and \textit{face-lattice} vertices belonging to the open-fan changes to account for the open boundary of the lattice.

Each \textit{edge-lattice} vertex of the open-fan is kept unaltered, whereas each \textit{corner-} and \textit{face-lattice} vertex of the open-fan is treated as a particular edge-lattice vertex of $\mathcal{M}_{\tau}$ (\cref{fig:MultiPatchFringeOpen}), and the geometrical coordinates are iteratively adjusted as
\begin{equation}
    \label{eq_smthOpenFanEdgeLatticeVertex}
    \mathbf{P}_{i+1} = \tilde{\alpha}\mathbf{P}_{i} + \sum_{\nu}~\tilde{\beta}\mathbf{P}_{i}^{(\mathrm{e}_{\nu}^{\parallel})}.
\end{equation}
Here, $\mathbf{P}_{i}^{(\mathrm{e}_{\nu}^{\parallel})}$ represents the coordinates of the $\nu$-th edge-wise connected neighbour in the open-fan of $\mathcal{M}_{\tau}$.
As in \cref{eq_smthEdgeLatticeVertex}, the resulting valence is $\nu=2$, and the weights coincide to $\tilde{\alpha} = \frac{1}{2}$ and $\tilde{\beta} = \frac{1}{4}$.

\paragraph{Smoothing Scheme of the Medial Locus}
\label{MedialLocusSmooth}
The medial locus $\mathcal{M}_{\tilde{\tau}}$ makes exception in the smoothing scheme, as the fringe of a degenerate lattice graph may not be defined.
For each vertex of the medial locus, the original edge-wise \mbox{6-connected} neighbourhood ($\mathcal{N}_6$) of the conforming lattice is considered, regardless of the medial shell of belonging.
The associated geometrical coordinates are iteratively adjusted as
\begin{equation}
    \label{eq_smthMedialLocusVertex}
    \mathbf{P}_{i+1} = \bar{\alpha}\mathbf{P}_{i} + \sum_{\nu}~\bar{\beta}(\nu)\mathbf{P}_{i}^{(\mathrm{e}_{\nu}^{\mathcal{N}_6})},
\end{equation}
where, at the $i$-th iteration, $\mathbf{P}_{i}$ represents the coordinates of the considered vertex of the medial locus $\mathcal{M}_{\tilde{\tau}}$, and $\mathbf{P}_{i}^{(\mathrm{e}_{\nu}^{\mathcal{N}_6})}$ represents the coordinates of the $\nu$-th edge-wise connected neighbour in $\mathcal{N}_6$ of the conforming lattice.
The weights are defined as $\bar{\alpha} = \frac{1}{2}$ and $\bar{\beta}(\nu) = \frac{1}{2\nu}$, satisfying an affine invariant scheme.

Open-fan vertices of the medial locus ($\nu<6$) are adjusted by considering only the neighbours in $\mathcal{N}_6$ belonging to the same boundary side of the conforming scaffolding.

\section{Results}
\label{ExperimResults}

\paragraph{Input Data}
\label{InputData}
A set of synthetic and CAD geometries with varying shape, complexity, composite graph structure and branching patterns were obtained by construction. Popular benchmarking geometries used in computer graphics applications have been considered form \cite{zhou2016thingi10k,attene2016web} as reference surface meshes for the scaffolding parametrisation. Also, representative multi-modal scans, i.e. computed tomography (CT) and magnetic resonance (MR), from available medical repositories \cite{piccinelli2009framework,BULLITT20051232} were considered for tracing anatomical structures from raster volumetric imaging.

\paragraph{Parametric and Implementation Details}
\label{ParamImplementDetails}

All parametrisations are considered for solid vectorial elements of cubic degree, i.e. $d_u=d_v=d_w=3$, with unitary weights of the rational basis functions.
Minimal number of control points $l_u,l_v,l_w\geq 4$ are adopted by enforcing an adaptive spatial resolution, based on \cite{pagani2018curvature}.
For simplicity, and in keeping with the conforming constraints, the parametrisations along $u$ and $v$ directions in the luminal scaffolding employed either uniformly-spaced, or non-uniformly symmetrically-spaced knot vectors ($\mathbf{k}_u$, $\mathbf{k}_v$).
The technical implementation of the introduced scaffolding framework builds on available tools from \cite{de2011geopdes,vazquez2016new,bingol2019geomdl}.% and is released as an evaluation tool-kit.

\paragraph{Parametrisation Evaluation}
\label{QuantEval}

Features of the parametrised structures account for the total number of scaffolding elements, the density of control points lattice, the computational time, and the accuracy for fitted geometries.

For synthetic and CAD structures, the compactness of the proposed approach is first evaluated by comparing our parametrisation with state-of-the-art solid meshing techniques employed in computational simulations.
For each structure, an optimal tetrahedral mesh is generated using \cite{si2015tetgen}, reporting the number of quadratic-element tetrahedra, total number of vertices and computational time.

Similarly, the scaffolding parametrisation is evaluated for polygonal meshes.
The fitting accuracy of the geometric embedding is reported considering the input surface mesh as reference.
High accuracy corresponds to minimal deviation from the reference mesh, in particular, the accuracy is evaluated as the complement of the percentage parametrisation error, i.e. $\text{Acc.} = (1 - \varepsilon)\cdot100 \%$, and the error $\varepsilon$ is computed as the Euclidean (or Hausdorff: $\varepsilon_{95}$) distance between the closest points on the boundary surface of the parametrised structure and on the reference surface using \cite{Aspert2002}, normalised by the local radius of the underlying inscribed sphere \cite{stolpner2011medial}.
For representative structures, the proposed parametrisation is also compared against the state-of-the-art linear-elements hexahedral meshing method \cite{LPPSC20}.

The parametric accuracy of fitted anatomical structures is evaluated on the normalised distance between the closest points on the parametrised boundary surface and the sampled geodesic active contours as reference.
A qualitative comparison with seminal work by \cite{zhang2007patient,urick2019review} is also provided.

\begin{table}[ht]
    \caption{Compactness of synthetic shapes and CAD geometries: Solid NURBS Scaffolding: $|\mathrm{H_3}|$ total cubic-degree elements; $|\mathbf{P}|$ total control points in the multi-patch lattice; the computational time (s). Equivalent Solid Tetrahedral Meshes \cite{si2015tetgen}: $|\mathrm{T_2}|$ total quadratic tetrahedra, $|\mathbf{V}|$ total vertices; the computational time (s). Coupled luminal ($L$) and wall ($W$) multi-compartments in subscript.}
    \label{tab:ResultsSynthCAD}
    \begin{minipage}{\columnwidth}
    \centering
    \begin{tabular}{c|ccc|ccc}
    %\toprule
        &\multicolumn{3}{c}{Solid NURBS Scaffolding}&\multicolumn{3}{|c}{Solid Tetrahedral Mesh}\\
        \toprule
        Structure & $|\mathrm{H_3}|$ & $|\mathbf{P}|$ & Time & $|\mathrm{T_2}|$ & $|\mathbf{V}|$ & Time \\
        \midrule
        3-Junct.$_L$ & 12 & 1.3k & 5.3 & 26.7k & 46.9k & 0.4 \\
        3-Junct.$_W$ & 24 & 2.7k & 5.3 & 40.0k & 80.4k & 0.7 \\
        5-Junct.$_L$ & 20 & 2.2k & 10.2 & 41.2k & 74.9k & 0.7 \\
        5-Junct.$_W$ & 40 & 4.5k & 10.2 & 68.2k & 135.5k & 1.3 \\
        7-Junct.$_L$ & 28 & 3.1k & 20.3 & 62k & 109.2k & 0.9 \\
        7-Junct.$_W$ & 56 & 6.3k & 20.3 & 97.6k & 191.8k & 1.9 \\
        9-Junct.$_L$ & 36 & 4.0k & 29.1 & 40.4k & 137.1k & 1.3 \\
        9-Junct.$_W$ & 72 & 8.0k & 29.1 & 128.0k & 249.2k & 2.8 \\
        %13-Junct.$_L$ & 52 & 5.8k & 44.8 & 111.2k & 198.7k & 1.9 \\
        %13-Junct.$_W$ & 104 & 11.6k & 44.8 & 188.6k & 363.6k & 4.2 \\
        Cross & 7 & 875 & 2.4 & 38.0k & 74.0k & 0.9 \\
        Egg$_L$ & 4 & 256 & 1.4 & 6.2k & 12.0k & 0.1 \\
        Egg$_W$ & 16 & 1.0k & 1.4 & 18.8k & 28.5k & 0.2 \\
        Frame & 8 & 1.0k & 2.7 & 40.0k & 78.5k & 1.0 \\
        Net 1 & 40 & 5.2k & 21.7 & 224.4k & 345.3k & 8.8 \\
        Net 2 & 72 & 8.4k & 58.5 & 317.6k & 569.9k & 21.9 \\
        Net 3 & 80 & 8.0k & 54.6 & 408.9k & 604.1k & 15.2 \\
        Pinocchio & 24 & 1.5k & 5.6 & 115k & 196.7k & 1.9 \\
        Quadball & 1 & 64 & 0.2 & 5.8k & 10.9k & 0.1 \\
        Torus & 4 & 448 & 0.3 & 10.5k & 21.1k & 0.2 \\
        Twist & 1 & 625 & 1.6 & 16.1k & 29.7k & 0.3 \\
        \midrule
        Air & 36 & 2.3k & 3.3 & 61.8k & 114.1k & 1.4 \\
        Blades & 48 & 3.1k & 3.3 & 64.3k & 125.3k & 1.3 \\
        Gear & 54 & 3.5k & 6.5 & 73.2k & 145.1k & 1.7 \\
        Hooks & 14 & 0.9k & 2.9 & 54.1k & 104.4k & 0.9 \\
        Plate & 72 & 4.6k & 5.1 & 73.1k & 145.7k & 1.6 \\
        Socket & 32 & 2.0k & 7.3 & 36.3k & 68.3k & 0.6 \\
        Stent & 120 & 11.5k & 29.7 & 580.8k & 789.9k & 10.1 \\
        T-pipe & 12 & 2.4k & 5.5 & 39.9k & 80.3k & 0.8 \\
        \bottomrule
    \end{tabular}
\end{minipage}
\end{table}

\subsection{Synthetic and CAD Geometries}
\label{SynthCADGeometries}

\begin{figure}
    \centering
    \includegraphics[width=.99\textwidth]{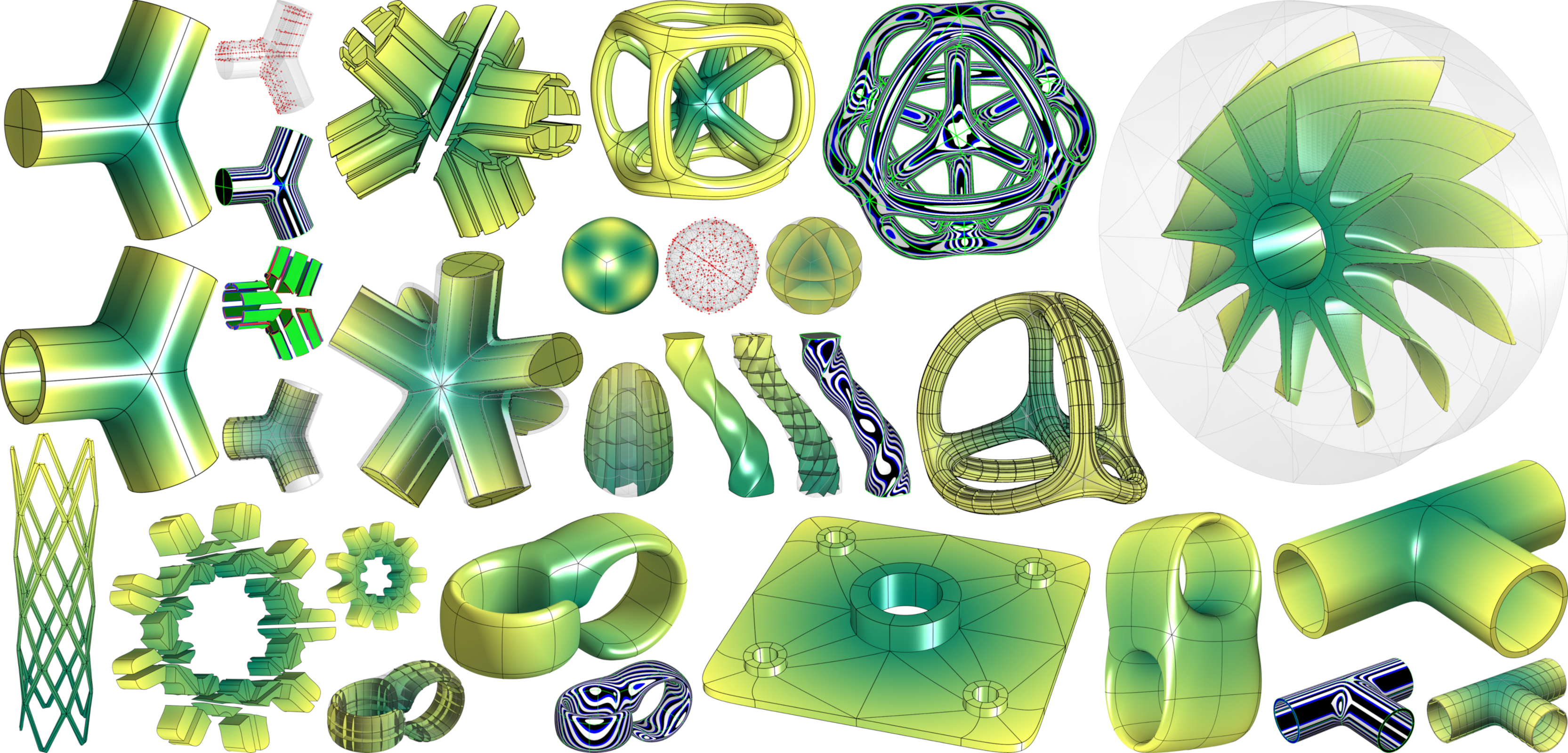}
    \caption{Gallery of Solid Synthetic and CAD Geometries -- Minimal elements scaffolding and modular decomposition in the explosion view; smooth and sharp features are preserved for different medial shells configurations. Control points lattice, parametric embedding, reflection lines and inner solid trabecular and sectioning views.}
    \label{fig:ResultsSynthCAD}
\end{figure}

\begin{figure}
    \centering
    \begin{overpic}[width=.99\textwidth]{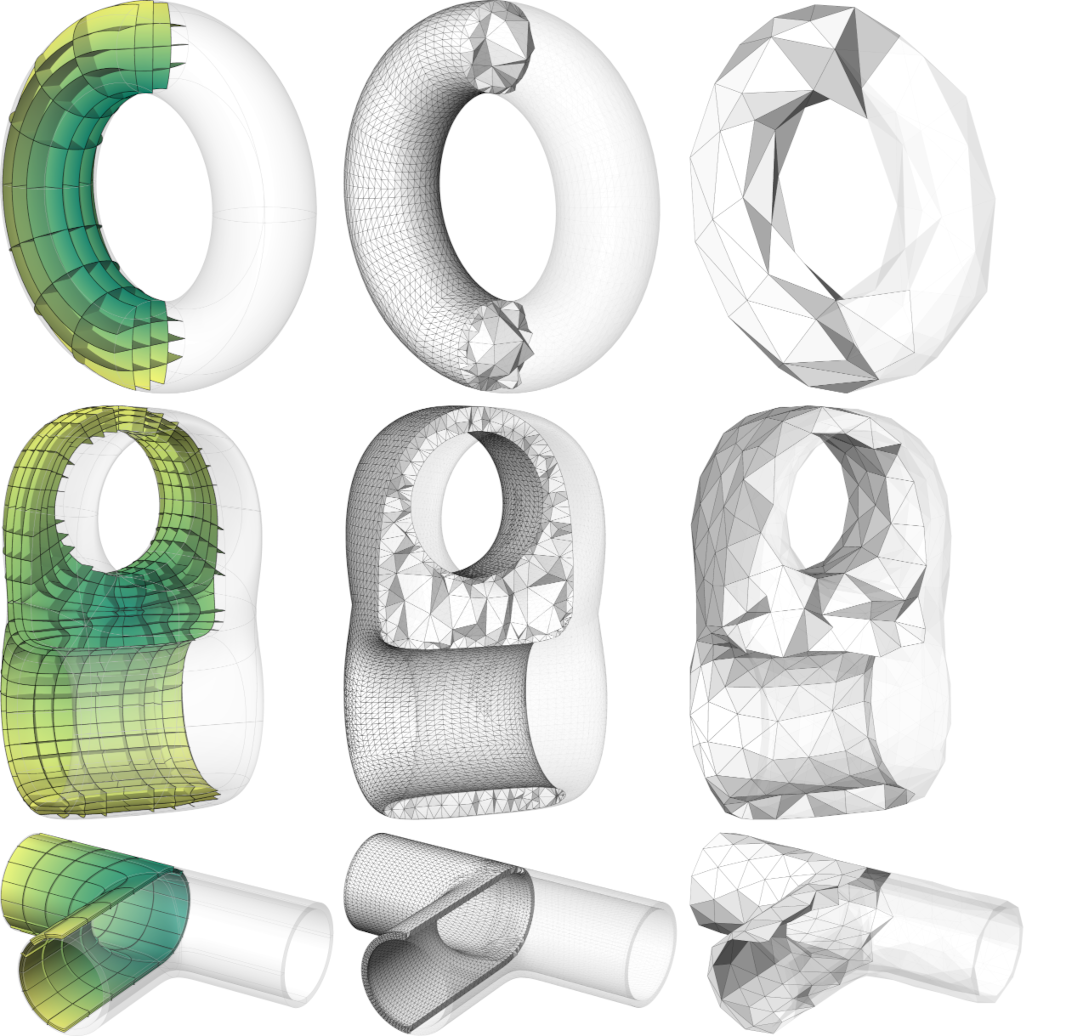}
        \put(2,94){\small{(a)}}
        \put(32,94){\small{(b)}}
        \put(66,94){\small{(c)}}
        % First Column
        \put(20.5,92){\small{$|\mathrm{H_3}|: 4$}}
        \put(20.5,90){\small{$|\mathbf{P}|: 448$}}
        \put(20.5,54){\small{$|\mathrm{H_3}|: 32$}}
        \put(20.5,52){\small{$|\mathbf{P}|: 2.0$k}}
        \put(20.5,16){\small{$|\mathrm{H_3}|: 12$}}
        \put(20.5,14){\small{$|\mathbf{P}|: 2.4$k}}
        % Second Column
        \put(53,92){\small{$|\mathrm{T_2}|: 10.5$k}}
        \put(53,90){\small{$|\mathbf{V}|: 21.1$k}}
        \put(53,88){\small{$\text{Acc.} \approx 100\%$}}
        \put(53,54){\small{$|\mathrm{T_2}|: 36.3$k}}
        \put(53,52){\small{$|\mathbf{V}|: 68.3$k}}
        \put(53,50){\small{$\text{Acc.} \approx 100\%$}}
        \put(53,16){\small{$|\mathrm{T_2}|: 39.9$k}}
        \put(53,14){\small{$|\mathbf{V}|: 80.3$k}}
        \put(53,12){\small{$\text{Acc.} \approx 100\%$}}
        % Third Column
        \put(86,92){\small{$|\mathrm{T_2}|: 204$}}
        \put(86,90){\small{$|\mathbf{V}|_{\mathrm{coarse}}: 427$}}
        \put(86,88){\small{$\text{Acc.}\, 74.2\pm17.4\%$}}
        \put(86,86){\small{$\text{Acc.}_{\varepsilon_{95}}\, 42.3\%$}}
        \put(86,54){\small{$|\mathrm{T_2}|: 1.5$k}}
        \put(86,52){\small{$|\mathbf{V}|_{\mathrm{coarse}}: 2.9$k}}
        \put(86,50){\small{$\text{Acc.}\, 87.6\pm10.3\%$}}
        \put(86,48){\small{$\text{Acc.}_{\varepsilon_{95}}\, 67.2\%$}}
        \put(86,16){\small{$|\mathrm{T_2}|: 1.1$k}}
        \put(86,14){\small{$|\mathbf{V}|_{\mathrm{coarse}}: 2.2$k}}
        \put(86,12){\small{$\text{Acc.}\, 88.4\pm10.3\%$}}
        \put(86,10){\small{$\text{Acc.}_{\varepsilon_{95}}\, 68.9\%$}}
    \end{overpic}
    \caption{Compactness of the Parametrisation -- representative examples: (a) Solid NURBS scaffolding: exact geometrical domain - sectioning inner trabecular view. (b) High-resolution equivalent tetrahedral mesh using \cite{si2015tetgen} with adaptive sampling and quadratic-elements. (c) Coarse quadratic-tetrahedral mesh optimised for $|\mathbf{V}|_{\textrm{coarse}} \approx |\mathbf{P}|$. Superficial degradation for increasing coarsening.}
    \label{fig:CoarseIsoParamMesh}
\end{figure}

Simple to complex shapes are modelled using the introduced scaffolding.
More complex geometries combine a composition of interfacing luminal and wall scaffoldings to model holes and complementary regions, leveraging modular and local blocks decomposition.
Individual elements are assessed in the explosion view (\cref{fig:ResultsSynthCAD}), where minimal multi-patches exactly recover the geometrical shape of large and complex geometries.
Smooth and sharp features of the structures are preserved by enforcing arbitrary smoothing configurations to the medial shells.
The continuum structure of the parametrised solids is shown in the trabecular view (\cref{fig:ResultsSynthCAD}), where free-form cuboid cells are sampled from the vectorial domain.
The compactness of the proposed solid parametrisation accounts for dozens to hundreds of cubic-degree elements, comprising few thousands of control points in the solid lattice for increasing structure complexity (\cref{tab:ResultsSynthCAD}).
The compactness is compared against adaptive quadratic-element tetrahedral meshing methods employed in computational simulations \cite{si2015tetgen}.
To accurately represent the same structure, tetrahedral meshes required a dense tessellation of elements and vertices, which differs by several orders of magnitude compared to the proposed parametrisation. 
Coarser tessellations are feasible (\cref{fig:CoarseIsoParamMesh}), yet higher deviation from the boundary structure is observed, introducing larger geometrical errors and approximations for computational simulations.
The structures parametrised with our approach exhibit higher degree of smoothness and parametric continuity. This is shown for the superficial boundary in the reflection lines \cite{balzer2010principles}, where nearly G\textsuperscript{2} smoothness is found for organic and smoothly blended portions of the structures (\cref{fig:ResultsSynthCAD}).
Also, structural parametric continuity is observed in the smooth profiles of the trabecular view, where inner sections and structural cells of the vectorial solid domains are arbitrarily sampled (\cref{fig:ResultsSynthCAD}).
Different smoothing configurations can be defined by altering the medial shells subdivision for mixed sharp and organically blended sub-portions, as in conventional smoothing groups.
The construction of fully-coupled geometries, e.g. \mbox{$n$-way-junctions}, accounting for both luminal and wall multi-compartments are simultaneously generated, determining a conforming multi-compartmental domain \cref{fig:ResultsSynthCAD}.

\subsection{Polygonal Surface Meshes}
\label{PolyMesh}

\begin{table}[ht]
    \caption{Fitting polygonal surface meshes: features of parametrised structures. Input surface geometry: $|\mathrm{T}|$ total triangles, $|\mathbf{V}|$ total vertices. Solid NURBS Scaffolding: $|\mathrm{H_3}|$ total cubic-degree elements; $|\mathbf{P}|$ total control points in the multi-patch lattice; the computational time (s); geometrical embedding fitting accuracy ($\%$) is reported mean$\pm$sd and relative to Hausdorff distance ($\varepsilon_{95}$).}
    \label{tab:ResultsMesh}
    \begin{minipage}{\columnwidth}
    \centering
    \begin{tabular}{c|cc|ccccc}
    %\toprule
        &\multicolumn{2}{|c|}{Surface Mesh}&\multicolumn{5}{c}{Solid NURBS Scaffolding}\\
        \toprule
        Structure & $|\mathrm{T}|$ & $|\mathbf{V}|$ & $|\mathrm{H_3}|$ & $|\mathbf{P}|$ & Time & $\text{Acc.~\%}$ & $\text{Acc.}_{\varepsilon_{95}}~\%$ \\
        \midrule
        Armadillo & 83.4k & 41.7k & 96 & 20.5k & 463.9 & 95.37$\pm$6.44 & 82.76 \\
        Botijo & 82.3k & 41.2k & 52 & 22.0k & 426.8 & 97.18$\pm$2.61 & 92.54 \\
        Bunny & 104.3k & 52.2k & 40 & 13.8k & 362.1 & 97.70$\pm$3.19 & 92.40 \\
        Elephant & 78.3k & 39.2 & 76 & 18.2k & 324.3 & 97.74$\pm$2.79 & 93.08 \\
        Elk & 108.3k & 54.1 & 48 & 17.2k & 401.4 & 98.05$\pm$2.53 & 93.19\\
        Fertility & 483.2k & 241.6k & 32 & 11.4k & 862.1 & 98.67$\pm$1.64 & 95.68 \\
        Genus3 & 13.3k & 6.7k & 24 & 4.5k & 15.9 & 96.87$\pm$3.29 & 90.63 \\
        Hand & 73.2k & 36.6k & 28 & 10.8k & 206.8 & 97.83$\pm$3.78 & 93.08 \\
        Horse & 61.3k & 30.6k & 36 & 14.2k & 239.8 & 97.21$\pm$4.36 & 91.42 \\
        Kitten & 110.1k & 55.4k & 28 & 10.0k & 249.3 & 98.67$\pm$1.80 & 95.67 \\
        \bottomrule
    \end{tabular}
\end{minipage}
\end{table}

\begin{figure}
    \centering
    \includegraphics[width=.99\textwidth]{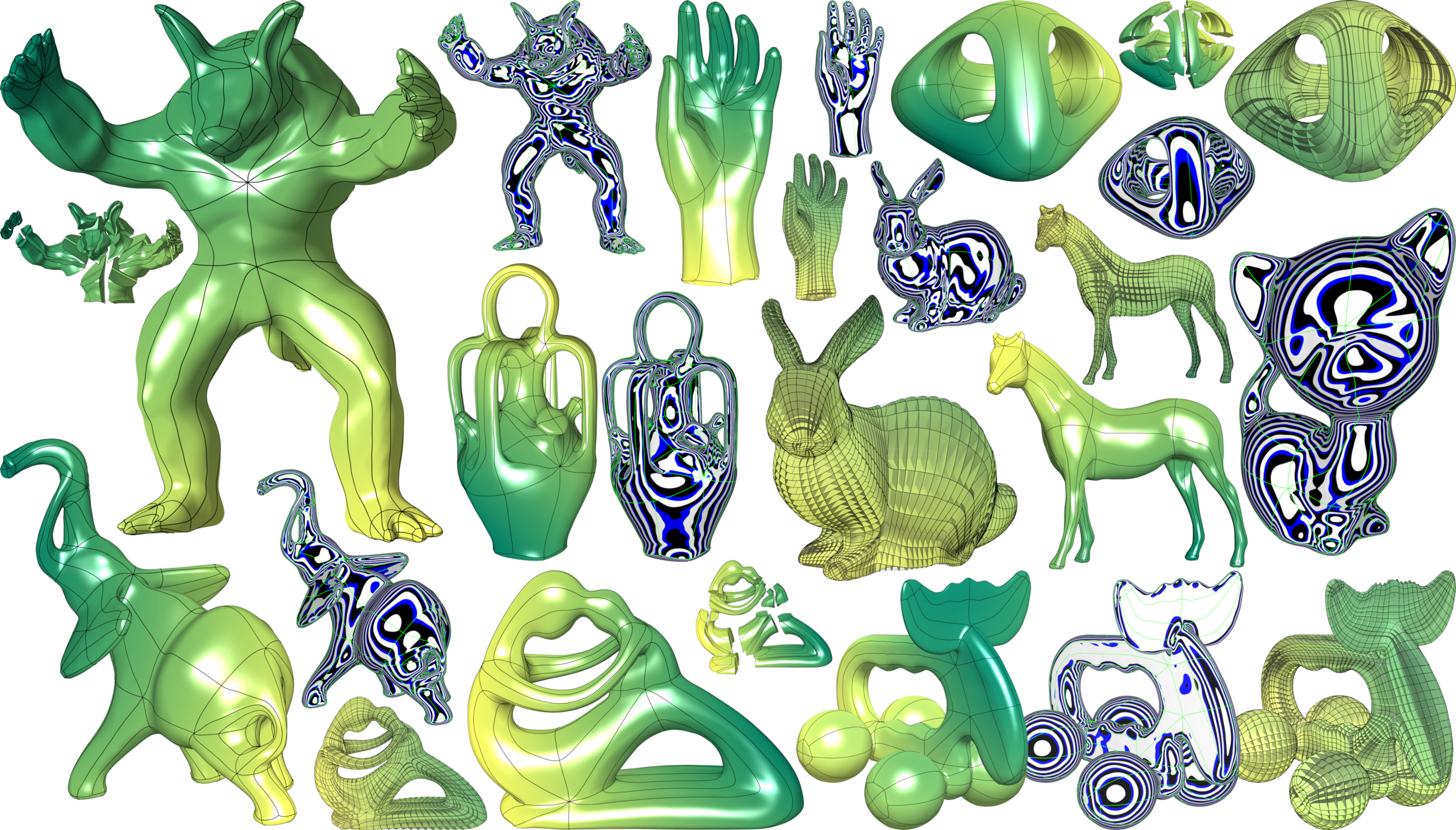}
    \caption{Gallery of Solid Parametrised Geometries from Polygonal Surface Meshes -- Explosion view, reflection lines and inner solid trabecular view. Nearly G\textsuperscript{2} smoothness and parametric continuity overall for boundaries and interfaces with organic profiles.}
    \label{fig:ResultsSurfMesh}
\end{figure}

\begin{figure}
    \centering
    \begin{overpic}[width=.99\textwidth]{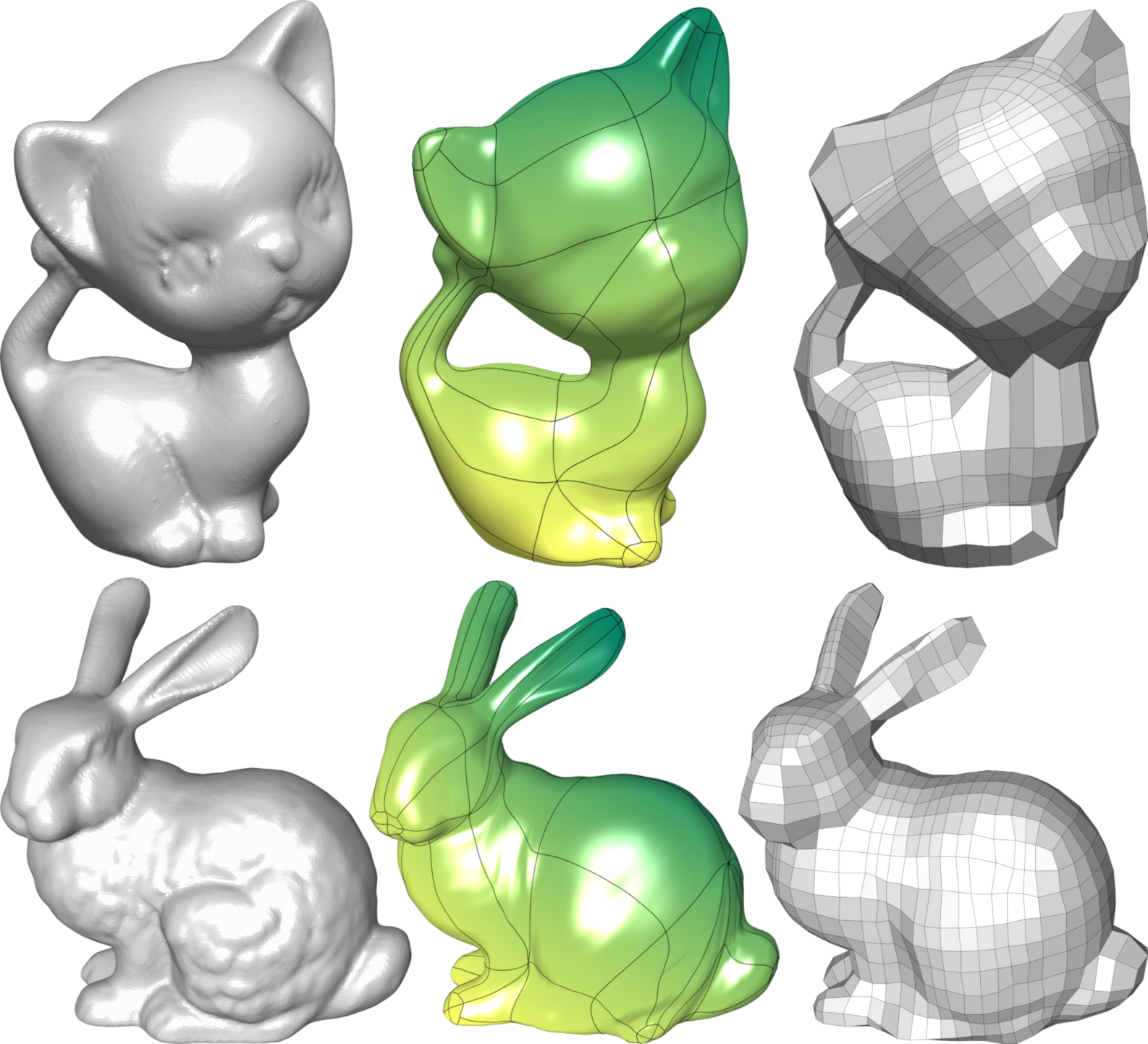}
        \put(0,90){\small{(a)}}
        \put(31,90){\small{(b)}}
        \put(65,90){\small{(c)}}
        % First Row
        \put(2.5,88){\small{$|\mathrm{T}|: 110.1$k}}
        \put(2.5,86){\small{$|\mathbf{V}|: 55.4$k}}
        \put(33.5,88){\small{$|\mathrm{H_3}|: 28$}}
        \put(33.5,86){\small{$|\mathbf{P}|: 10$k}}
        \put(33.5,84){\small{$\text{Acc.}\, 98.67\pm1.8 \%$}}
        \put(33.5,82){\small{$\text{Acc.}_{\varepsilon_{95}}\, 95.67\%$}}
        \put(67.5,88){\small{$|\mathrm{H_1}|: 1.5$k}}
        \put(67.5,86){\small{$|\mathbf{P}|: 1.9$k}}
        \put(67.5,84){\small{$\text{Acc.}\, 96.7\pm4.74 \%$}}
        \put(67.5,82){\small{$\text{Acc.}_{\varepsilon_{95}}\, 88.54\%$}}
        % Second Row
        \put(23,36){\small{$|\mathrm{T}|: 104.3$k}}
        \put(23,34){\small{$|\mathbf{V}|: 52.2$k}}
        \put(55,36){\small{$|\mathrm{H_3}|: 40$}}
        \put(55,34){\small{$|\mathbf{P}|: 13.8$k}}
        \put(55,32){\small{$\text{Acc.}\, 97.7\pm3.19 \%$}}
        \put(55,30){\small{$\text{Acc.}_{\varepsilon_{95}}\, 92.4\%$}}
        \put(86,36){\small{$|\mathrm{H_1}|: 2.2$k}}
        \put(86,34){\small{$|\mathbf{P}|: 2.8$k}}
        \put(86,32){\small{$\text{Acc.}\, 97.53\pm4.12 \%$}}
        \put(86,30){\small{$\text{Acc.}_{\varepsilon_{95}}\, 90.14\%$}}
    \end{overpic}
    \caption{Comparison with Hexahedral Meshing: (a) Reference triangle surface mesh; (b) Solid NURBS scaffolding; (c) Solid hexahedral mesh using \cite{LPPSC20}. Quantitative attributes reporting the compactness and the accuracy at the boundary. $|\mathrm{H}_{3}|$ and $|\mathrm{H}_{1}|$ respectively for cubic-degree and linear-degree total elements.}
    \label{fig:CompareHexahedralMesh}
\end{figure}

A set of benchmark polygonal meshes are parametrised as reference (\cref{fig:ResultsSurfMesh}).
The luminal scaffolding recovers the underlying shape by fitting the mesh geometries with high accuracy (\cref{tab:ResultsMesh}) with respect to the reference boundary surface.
Complex structures with multiple busy branching patterns are compactly parametrised with minimal element saffoldings.
Free-form cuboids of the conforming scaffolding recover the geometrical embedding of the underlying structure with localised deviations and minor distortions for high-curvature regions.
The spatial partitions of junctions are automatically determined for busy skeletal branch-points.
The automatic mapping of the scaffolding initially produced asymmetric quadrants and projected irregular quad-based segmentations on the boundary surface, especially for busy junctions with severe changes in size and shape.
The resulting smoothness and continuity of the parametrised solid structures is shown in the reflection lines for the outer boundary and in the trabecular view for the innermost region in \cref{fig:ResultsSurfMesh}.
Nearly G\textsuperscript{2} smoothness is achieved globally, and high structural regularity of the lattice is obtained in the form of an organic continuum.
In the neighbourhood of extraordinary vertices, the solid scaffolding exhibits G\textsuperscript{1} smoothness and continuity, whereas pure positional continuity (G\textsuperscript{0}) is observed exactly for the extraordinary vertices.
Representative structures are parametrised using state-of-the-art hexahedral meshing \cite{LPPSC20} in \cref{fig:CompareHexahedralMesh}.
Higher compactness and higher accuracy is observed for the proposed approach with respect to linear-element hexahedral meshes.
Note that a further NURBS-based parametrisation is required in \cite{LPPSC20} prior to employing the hexahedral mesh in IGA.

\subsection{Anatomical Structures from Clinical Imaging}
\label{VolumetricAngio}

\begin{figure}
    \centering
    \includegraphics[width=.99\textwidth]{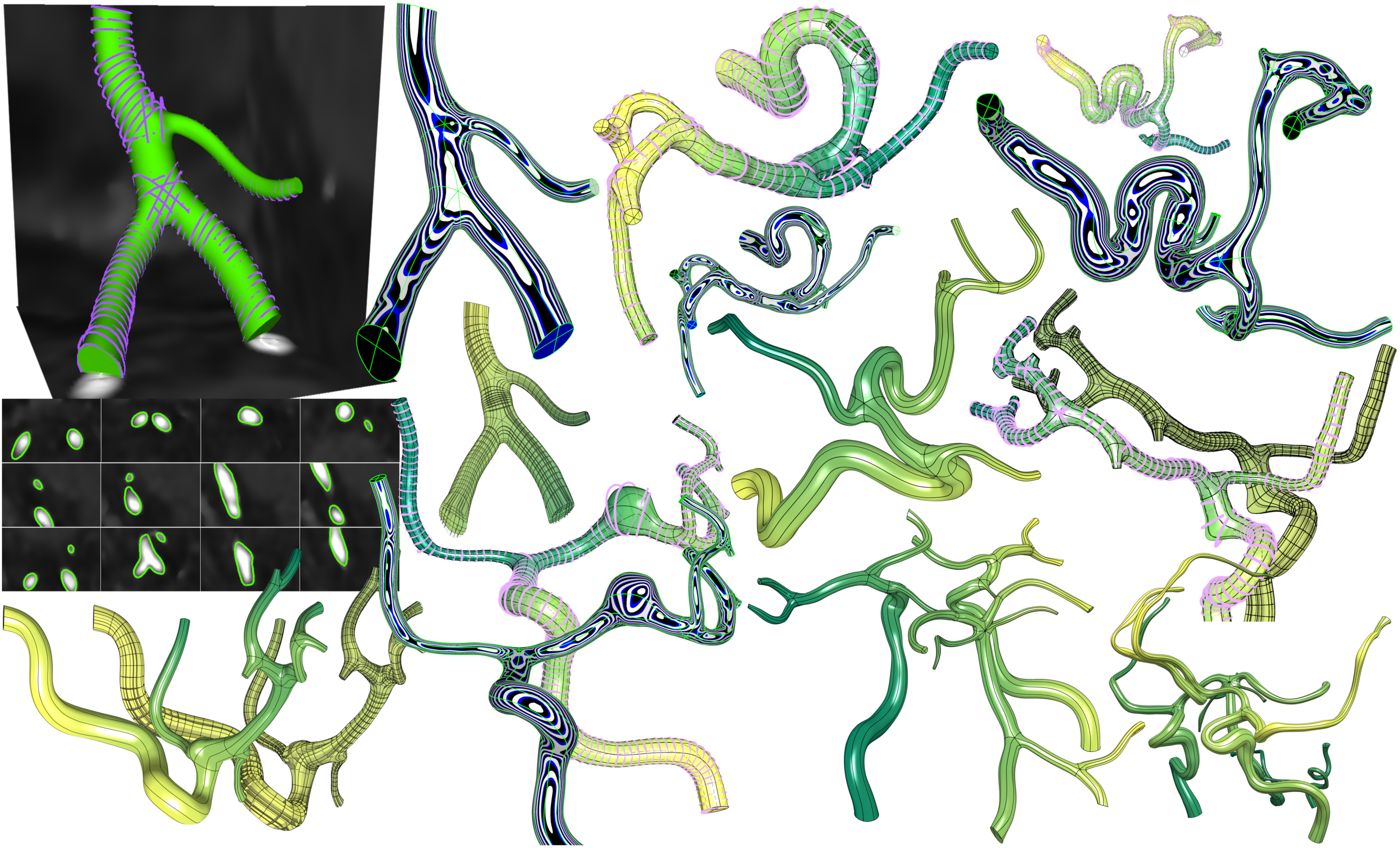}
    \caption{Gallery of Anatomical Structures Parametrised from Volumetric Scans. Contours, slice intersections, solid NURBS scaffolding with reflection lines and inner solid trabecular views. Scalable parametrisation to large anatomical structures.}
    \label{fig:ResultsAngio}
\end{figure}

\begin{table}[ht]
    \caption{Tracing anatomy from volumetric imaging: features of parametrised structures. Input scan: Imaging modality, $\mathbf{v}$ voxel size (mm). Solid NURBS Scaffolding: $|\mathrm{H_3}|$ total cubic-degree elements; $|\mathbf{P}|$ total control points in the multi-patch lattice; the computational time (s); geometrical parametrisation error (mm) is reported mean$\pm$sd. Hausdorff distance ($\varepsilon_{95}$).}
    \label{tab:ResultsAngio}
    \begin{minipage}{\columnwidth}
    \centering
    \begin{tabular}{cccccccc}
    %\toprule
        &\multicolumn{2}{c}{Input Scan}&\multicolumn{5}{c}{Solid NURBS Scaffolding}\\
        \toprule
        Structure & Modality & $\mathbf{v}$ (mm) & $|\mathrm{H_3}|$ & $|\mathbf{P}|$ & Time & $\varepsilon$ (mm) & $\varepsilon_{95}$ (mm) \\
        \midrule
        Angio 1 & CT & 0.26 isotropic & 52 & 10.1k & 42.7 & 0.26$\pm$0.15 & 0.55 \\
        Angio 2 & CT & 0.26 isotropic & 36 & 10.7k & 54.3 & 0.23$\pm$0.13 & 0.48 \\
        Angio 3 & CT & 0.35 isotropic & 52 & 11.5k & 70.1 & 0.26$\pm$0.14 & 0.49 \\
        Angio 4 & CT & 0.35 isotropic & 44 & 10.6k & 62.2 & 0.34$\pm$0.21 & 0.71 \\
        Angio 5 & CT & 0.35 isotropic & 52 & 11.4k & 70.7 & 0.25$\pm$0.15 & 0.54 \\
        Angio 6 & CT & 0.35 isotropic & 36 & 11.4k & 73.5 & 0.28$\pm$0.19 & 0.61 \\
        Brain Net 1 & MR & [0.36$\times$0.36$\times$0.5] & 100 & 19.6k & 122.2 & 0.34$\pm$0.21 & 0.72 \\
        Brain Net 2 & CT & [0.5$\times$0.5$\times$0.7] & 76 & 29.5k & 597.8 & 0.41$\pm$0.25 & 0.87 \\
        Phantom & MR & [0.36$\times$0.36$\times$0.5] & 20 & 3.5k & 12.5 & 0.27$\pm$0.15 & 0.53\\
        \bottomrule
    \end{tabular}
\end{minipage}
\end{table}

Anatomical structures from volumetric scans are parametrised using the introduced scaffolding (\cref{fig:ResultsAngio}).
The compact parametrisation produced a minimal luminal scaffolding comprising exactly 4 elements per branch.
Tubular structures with arbitrary branching and connectivity patterns are accurately recovered (\cref{tab:ResultsAngio}) for the elongation of the branches.
Also, irregular and convoluted tubular profiles as well as size-varying junctions are captured by the solid scaffolding with an organic outline using an optimal smoothing and fitting strategy.
Nearly G\textsuperscript{2} smoothness and continuity is achieved globally, by preserving local sharp features at the anatomical inlets and outlets allowing for boundary conditions in bio-mechanic computational simulations.
Similar tubular structures are parametrised in previous work \cite{zhang2007patient,urick2019review}, where elongated portions employ hexahedral meshes with dense cross-sectional subdivisions.
In these studies, neither data-driven, nor generic parametric form is given for bifurcations and busy junctions, which are conversely modelled with a dictionary of user-defined branching templates.
In \cite{zhang2007patient}, the parametrised anatomical structures reach circumferential G\textsuperscript{1} smoothness and continuity, whereas junctions locally exhibits sharp creases and longitudinal discontinuities.

\subsection{Isogeometric Analysis Simulations}
\label{IGASimul}

\begin{figure}
    \centering
    \begin{overpic}[width=.99\textwidth]{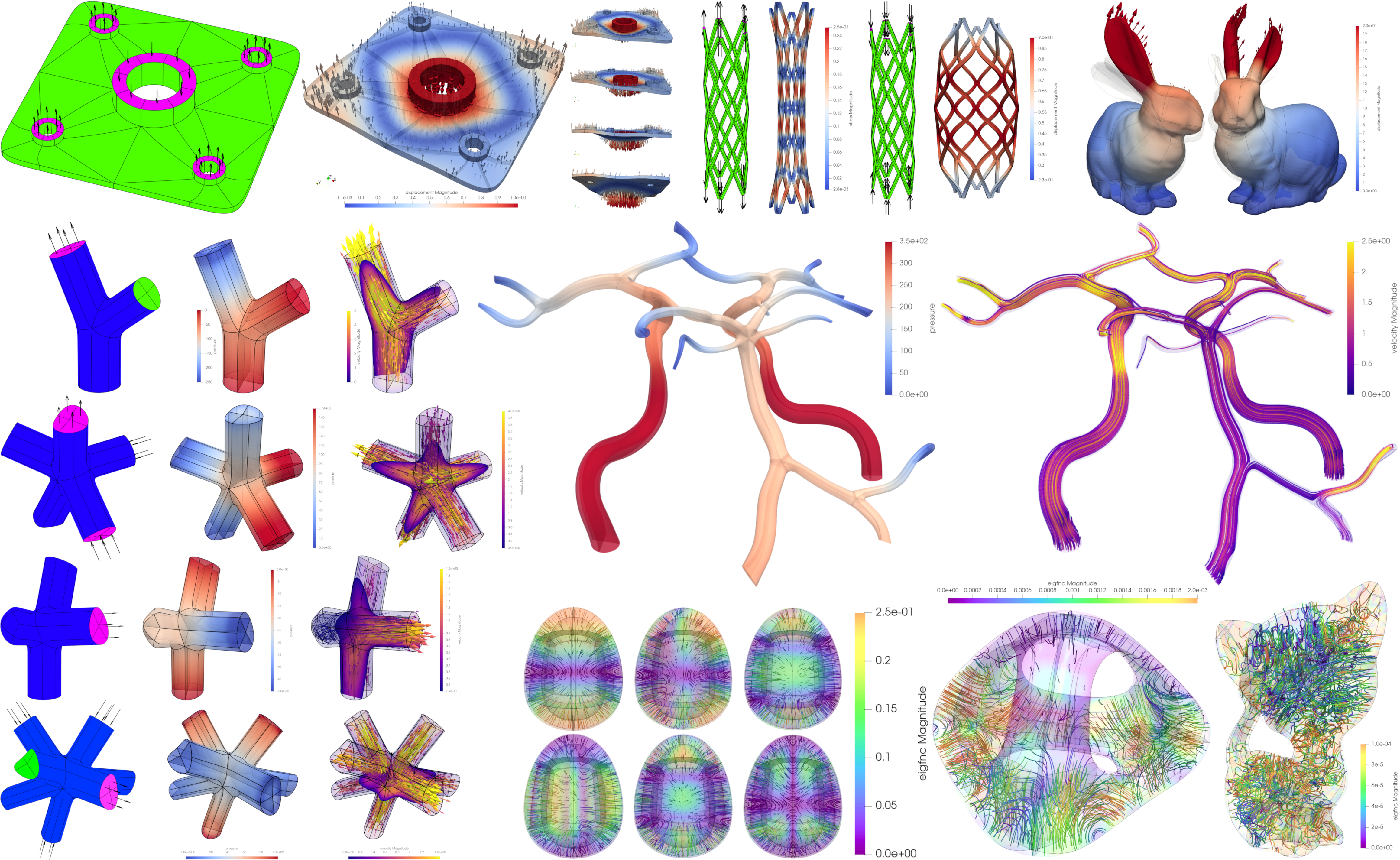}
        \put(0,60){\small{(a)}}
        \put(0,44){\small{(b)}}
        \put(35,18){\small{(c)}}
    \end{overpic}
    \caption{IGA Simulations using \cite{de2011geopdes}: (a) Linear-elastic deformation; (b) Fluid pressure and flow; (c) Set of Maxwell eigenfunctions.}
    \label{fig:ResultsIGAParaview}
\end{figure}

\begin{figure}
    \centering
    \begin{overpic}[width=.6\textwidth]{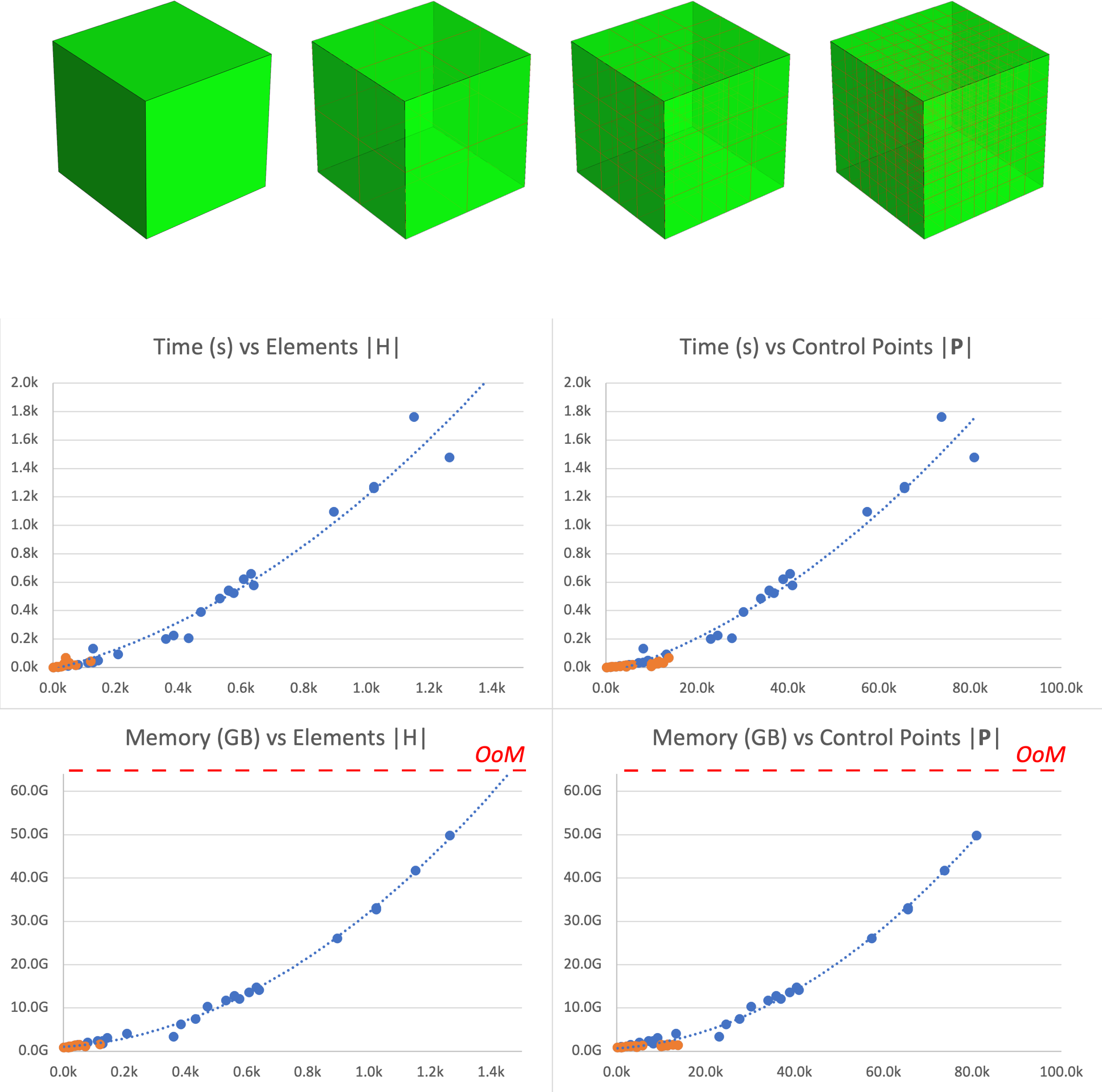}
        \put(2,98){\small{(a)}}
        \put(2,67){\small{(b)}}
        \put(13,75){\small{$|\mathrm{H}|: 1$}}
        \put(13,72){\small{$|\mathbf{P}|: 125$}}
        \put(37,75){\small{$|\mathrm{H}|: 8$}}
        \put(37,72){\small{$|\mathbf{P}|: 512$}}
        \put(61,75){\small{$|\mathrm{H}|: 64$}}
        \put(61,72){\small{$|\mathbf{P}|: 4$k}}
        \put(85,75){\small{$|\mathrm{H}|: 512$}}
        \put(85,72){\small{$|\mathbf{P}|: 32.8$k}}
    \end{overpic}
    \caption{IGA Simulations using \cite{de2011geopdes} - Benchmarking performance: (a) Cascade of conforming subdivisions; (b) Computational time and memory footprint as function of elements and control points density in the lattice; our scaffolding (orange dots) vs equivalent hexahedral meshes (blue dots). \textit{OoM}: Out of Memory threshold (red dashed line).}
    \label{fig:ResultsIGAPerformance}
\end{figure}

Representative computational problems are considered as proof-of-concept IGA applications to validate the parametrised scaffolding.
A linear elasticity problem \cite{auricchio2007fully} is solved considering the \textit{Plate} (CAD) and \textit{Stent} (CAD) structures, as well as the \textit{Bunny} (Mesh) geometry.
In this cases, boundary conditions and simulated solid deformations are shown in \cref{fig:ResultsIGAParaview}.
A steady-state fluid problem \cite{gomez2010isogeometric} is solved for representative luminal synthetic junctions, as well as for \textit{BrainNet 1} anatomical structure.
Boundary conditions and simulation solution profiles of pressure and flow distributions are shown in \cref{fig:ResultsIGAParaview}.
Lastly, the Maxwell eigenvalue problem \cite{vazquez2010isogeometric} is solved for \textit{Genus3} and \textit{Kitten} (Mesh), as well as for the \textit{Egg}$_W$ (Synthetic) structure.
In this case a set of eigenfunctions and the associated distributions are shown for both luminal and hollow wall geometries in \cref{fig:ResultsIGAParaview}.
All simulations are performed using \cite{de2011geopdes}, and the computational time as well as the memory footprint are provided in \cref{fig:ResultsIGAPerformance}.
The scalability of IGA simulations and associated performance is also evaluated on equivalent denser lattices.
An iterative cascade of conforming domain subdivisions (\cref{fig:ResultsIGAPerformance}) is progressively performed for the considered geometries, achieving comparable density as in hexahedral meshes \cite{LPPSC20,gao2017robust,gao2019feature,takayama2019dual}.

Successful simulations converged for all our scaffolding parametrisations (\cref{fig:ResultsIGAPerformance}), with smooth, continuous and consistent solution profiles.
The introduced framework exhibits low computational time and minimal memory footprint.
As in \cref{fig:ResultsIGAPerformance}, performance benchmarks shows a quadratic trend for increasing density of the lattice, suggesting improved scalability for structures parametrised with the introduced scaffolding.
Equivalent hexahedral meshes determined a progressive increase of elements per conforming subdivision, resulting in a parametric lattice of substantially higher density.
In this case, simulations only converged for meshes with less than 1.5k elements, as the memory footprint determined the limiting factor for the adopted IGA engine and solver.
All simulations were computed on a single-core CPU 3.60 GHz with 64GB RAM machine.

\section{Discussions and Conclusion}
\label{DiscsConcl}

This work introduced a novel scaffolding framework to trace 3D structures with solid NURBS elements for computational simulations using isogeometric analysis.

Results obtained form different input data showed the scaffolding framework is able to compactly generalise for the parametrisation of arbitrarily complex structures with minimal elements and low parametric density.
A novel interfacing configuration was introduced, i.e. the quadrilateral junction simplex, modelling the spatial partitioning of generic junctions for an underlying network of arbitrary connectivity patterns.
Conforming and data-driven fitting strategies were presented to determine the geometrical embedding of the scaffolding with (semi-)automatic pipelines for polygonal surface meshes and volumetric image-based data, respectively.
A novel solid smoothing paradigm is devised for the scaffolding lattice to obtain higher than positional geometric and parametric continuity at the interfaces and at the external boundaries of the structures, globally recovering an organic shape in the form of a continuum.
State-of-the-art tracing techniques employ raster (un-)structured polyhedral meshes to define the innermost lattice of 3D solid objects.
Despite these intrinsically differ from parametric NURBS geometries,
both qualitative and quantitative evaluations are proposed for the introduced framework based on analogies with the resulting scaffolding elements, lattice density, computational time and accuracy of fitted geometries.

To recover negligible geometrical deviations, tetrahedral meshes for conventional finite-element analyses \cite{si2015tetgen} required a dense tessellation of elements and vertices, which differs by several orders of magnitude to the proposed parametrisation.
%Compared to the proposed scaffolding, tetrahedral meshes for conventional finite-element simulations \cite{si2015tetgen} determined an unstructured plurality of denser elements by several orders of magnitude to recover fine details with negligible geometrical deviations.
Geometrical degradation is observed for iso-parametric optimised meshes, in particular for organic synthetic shapes and CAD designs with irregular and organic shapes.

The compactness of the introduced scaffolding was validated also on irregular geometries form 3D tessellated surfaces, where the resulting density of control points in the solid lattice was considerably lower than the input surface vertices.
By leveraging geodesic strategies and a conforming sampling, the proposed scaffolding parametrisation reported high geometrical accuracy with respect to the reference geometries, with minor deviations for sharp and high-curvature features.
Smoothness of the parametrised structures achieved nearly G\textsuperscript{2} continuity globally.

Similarly, convoluted anatomical structures were compactly parametrised from volumetric imaging data with a fully automatic pipeline.
Conforming fitting strategies based on geodesics and computer vision algorithms reported minor geometric deviations comparable to the input data resolution.
Smoothness and continuity of the parametrised biological structures achieved an organic profile underlying a homogeneous medium in the form of a continuum.

Several IGA computational simulations validated the compatibility of the parametrised scaffolding with benchmarking performance.
The simulations scalability was evaluated on equivalent hexahedral meshes obtained with cascades of conforming domain subdivisions to achieve similar element density as in state-of-the-art techniques\cite{LPPSC20}.
IGA simulations successfully converged for the proposed scaffolding framework in all cases, with smooth, continuous and consistent solution profiles.
The memory footprint, however, determined the limiting factor for simulations on equivalent  hexahedral tessellations, with a quadratic trend for increasing density of the meshes.

Previous studies on hexahedral meshes \cite{LPPSC20,gao2017robust,takayama2019dual}, sought a solid re-parametrisation from raster surface geometries as a viable option towards an IGA-compatible domain.
In similar CAD applications, the input tessellated geometry employed in finite-element analyses is often \textit{even} obtained from rasterising an original vectorial design.
For the sake of finite-element analyses, such re-parametrisation results in a convoluted and design-wise inefficient approach, specifically when IGA is used.
Although mesh degradation can be mitigated with denser lattices, the migration from a parametric continuous domain to finite-resolution meshes introduces geometrical approximations and downgrades higher degrees of continuity to a pure positional one. 
Moreover, generic tetrahedral or hexahedral-dominant meshes cannot be directly fed to IGA, since \textit{all} mesh elements require a conforming cuboid topology.
Any other cell type in the mesh, e.g. pyramidal or wedge-like elements, invalidate the computational domain when a Galerkin discretisation \cite{cottrell2009isogeometric} is considered for IGA.
Given the affinity of linear-element cuboid hexahedral meshes with tensors of splines, a parametric equivalent can be determined.
However, fitting NURBS cuboids \textit{a-posteriori} results in a dramatic increase of control points density for higher-degree spline elements.
By elevating a linear-degree cuboid (minimal lattice of 8 control points) to an equivalent of cubic-degree (minimal lattice of 64 control points) the lattice density increases by 700$\%$ for the individual cell.
Such increase of lattice density does not introduce, per se, higher degrees of continuity at the boundaries and interfaces of the solid mesh, instead the latter would simply remain positional (G$^0$) in the absence of a solid smoothing strategy.
From an IGA-conforming perspective, applying the required parametric constraints to an unstructured solid mesh may result in a complete dependency of all univariate components, i.e. along $u$, $v$, and $w$ in all elements, with limited parametrisation flexibility with respect to the introduced scaffolding.
Also, without an organised aggregation strategy, the scalability and simulation performance advantage observed on compact representations is dramatically diluted against dense tessellations, ultimately resulting in prohibitive or infeasible finite-element computations.

Based on the introduced framework, we argue that a streamlined solid design for finite-element analysis with IGA should rather begin with a consistent construction of a conforming scaffolding.
In the general form, this could leverage similar graphic user-interfaces, modelling routines and virtual environments employed for the construction of more complex composite CAD geometries.
This process would make use of guided and semi-automatic interactions for industrial and engineering designs, as in commercially available software, where modelling geometries rely on interactive user experiences.
Nonetheless, simpler geometries may benefit from a more streamlined and fully-automatic scaffolding pipeline (e.g. anatomical structures), which can dramatically impact on in-silico medicine and digital twins formulations on large scale, addressing complex systems, diverse physical scenarios and continuous solutions of PDEs.

\paragraph{Limitations and Future Work}

Current limitations of the introduced scaffolding framework stand in automatically mapping on reference surface meshes complex junction partitions for particularly busy branching patterns.
Although the existence of the junction simplex is guaranteed \cite{suarez2018scaffolding}, the overly compact geometrical embedding may result counter-intuitive and difficult to visualise, especially for the parametrisation of shape-varying structures as in complex polygonal tessellations.
As described in \cref{WallScaff,Othercaff}, alternative IGA-compatible scaffolding configurations are feasible, and may locally accommodate for geometrical simplifications and topological adjustments.
A simplified junction partition, e.g. the \textit{grafting scaffolding} in \cref{Othercaff}, can be developed for a non-minimal yet highly compact scaffolding, which locally decomposes the busy junctions with extra cuts and interleaving elements, diluting therefore the cascade of bisections in a more regular and distributed fashion.
Also, a more flexible formulation of the quadrilateral junction simplexes may locally relax the introduced geometrical and directional embedding to better model irregular and size-varying branching patterns.
On the one hand, this would improve the accuracy of the in-house fitting paradigm for polygonal structures with a minimal distortion parametrisation of the boundary surface, e.g. adopting standard UV-mapping and seam-cuts unwrapping techniques \cite{levy2002least}.
On the other hand, a more regular and distributed branching pattern of the junction simplex would reduce the valence of extraordinary vertices, therefore improving the geometrical smoothness and parametric continuity close to the singularities.
Although the introduced solid smoothing paradigm achieves nearly G$^2$ geometrical smoothness and parametric continuity for the entirety of the solid scaffolding, a minor degradation is observed at the medial locus and in the neighbourhood of extraordinary vertices.
This is however in line with the considerations of \cite{stam2001subdivision} for topological singularities.
The continuity observed in our parametrised geometries replicates the same experimental outcomes in \cite{stam2001subdivision,bajaj2002subdivision} for vertices of varying valence.
The devised smoothing strategy builds on the same averaging scheme, here with ad-hoc topological variations for multi-patch solid lattices.
Based on experimental observations, the introduced smoothing paradigm can be similarly applied also to solid multi-patch meshes which exhibit a different scaffolding configuration, comprising (un-)structured hexahedral cuboid elements.
%Alternative configurations and different non-minimal scaffoldings are indeed feasible starting from similar structure graphs, as shown in \cref{Othercaff}.
As shown in \cref{Othercaff}, both \textit{Cross} and \textit{Frame} geometries, exhibit a face-wise 6-connected scaffolding, similarly to hexahedral meshes, where each branch is extruded, without employing a quadratic junction simplex.

An organised structured and conformal aggregation strategy would potentially convert a cuboid hexahedral mesh into a compact scaffolding by means of an adaptive trade-off between splitting, refining and merging of cuboid elements.
This would represent a viable strategy for composite geometries requiring extensive sub-block decomposition \cite{chuang2000skeletonisation,lu2017evaluation,wang2017sheet,takayama2019dual}, without further geometric re-parametrisation.
%By extending such concept, the scaffolding can be seen as an equivalent cuboid hexahedral mesh in its least-compact representation.
%In this case, an adaptive trade-off between splitting, refining and conformally aggregating mesh elements into a conforming scaffolding would represent a viable strategy for composite geometries which require extensive local sub-block decomposition \cite{chuang2000skeletonisation,lu2017evaluation,wang2017sheet,takayama2019dual}.
An adaptive refinement may also address individual vectorial elements; recent IGA developments support hierarchical and T-NURBS elements \cite{hesch2016hierarchical,garau2018algorithms,bracco2018refinement,he2006manifold,bazilevs2010isogeometric}.
With this view, a local lattice refinement may be selectively employed in the neighbourhood of discontinuous regions and singular points (i.e. extraordinary vertices) with the aim of improving the local geometrical support, the smoothness and parametric continuity and towards adaptive degrees of freedom for computational simulations.
Adaptive and hierarchical local subdivisions of the vectorial elements \cite{garau2018algorithms,bracco2018refinement} could optimally circumvent the conforming propagation of the lattice refinement, as in a continuous Galerking method \cite{cockburn2009unified,chen2016nurbs}, without globally affecting the topological formulation and the parametric arrangement of the entire scaffolding.
Prospectively, the extension to hierarchical NURBS scaffolding elements could be further coupled with higher degrees splines and an optimal solid smoothing strategy along the lines of \cite{cashman2009nurbs,stam1998exact} where localised hierarchical refinements follow geometrical adjustments based on non-uniform knot vectors and on the diagonalisation of the lattice subdivision matrices.

\paragraph{Concluding Remarks}
\label{Conclusion}
The introduced scaffolding framework constitutes a fundamental advance towards bridging the gap between compactly tracing vectorial solid structures and continuous-domain computational simulations.
We believe our contributions could further stimulate and inspire novel developments with revolutionising applications on multi-disciplinary grounds for next-generation computational simulations.

% \begin{acks}
% Acknowledgements Here!
% \end{acks}

\bibliographystyle{ACM-Reference-Format}
\bibliography{GraphicsBiblio}


\begin{thebibliography}{84}

%%% the Web address.

\ifx \showCODEN    \undefined \def \showCODEN     #1{\unskip}     \fi
\ifx \showDOI      \undefined \def \showDOI       #1{#1}\fi
\ifx \showISBNx    \undefined \def \showISBNx     #1{\unskip}     \fi
\ifx \showISBNxiii \undefined \def \showISBNxiii  #1{\unskip}     \fi
\ifx \showISSN     \undefined \def \showISSN      #1{\unskip}     \fi
\ifx \showLCCN     \undefined \def \showLCCN      #1{\unskip}     \fi
\ifx \shownote     \undefined \def \shownote      #1{#1}          \fi
\ifx \showarticletitle \undefined \def \showarticletitle #1{#1}   \fi
\ifx \showURL      \undefined \def \showURL       {\relax}        \fi
% invisible to TeX
\providecommand\bibfield[2]{#2}
\providecommand\bibinfo[2]{#2}
\providecommand\natexlab[1]{#1}
\providecommand\showeprint[2][]{arXiv:#2}

\bibitem[\protect\citeauthoryear{Aspert, Santa-Cruz, and Ebrahimi}{Aspert
  et~al\mbox{.}}{2002}]%
        {Aspert2002}
\bibfield{author}{\bibinfo{person}{N. Aspert}, \bibinfo{person}{D. Santa-Cruz},
  {and} \bibinfo{person}{T. Ebrahimi}.} \bibinfo{year}{2002}\natexlab{}.
\newblock \showarticletitle{MESH: measuring errors between surfaces using the
  Hausdorff distance}. In \bibinfo{booktitle}{\emph{Proceedings. IEEE
  International Conference on Multimedia and Expo}}, Vol.~\bibinfo{volume}{1}.
  \bibinfo{pages}{705--708 vol.1}.
\newblock
\urldef\tempurl%
\url{https://doi.org/10.1109/ICME.2002.1035879}
\showDOI{\tempurl}


\bibitem[\protect\citeauthoryear{Attene, Cabiddu, Gagliardo, Giannini, and
  Monti}{Attene et~al\mbox{.}}{2016}]%
        {attene2016web}
\bibfield{author}{\bibinfo{person}{Marco Attene}, \bibinfo{person}{Daniela
  Cabiddu}, \bibinfo{person}{Stefano Gagliardo}, \bibinfo{person}{Franca
  Giannini}, {and} \bibinfo{person}{Marina Monti}.}
  \bibinfo{year}{2016}\natexlab{}.
\newblock \showarticletitle{A web repository to describe and execute shape
  oriented workflows}.
\newblock \bibinfo{journal}{\emph{Computer-Aided Design and Applications}}
  \bibinfo{volume}{13}, \bibinfo{number}{5} (\bibinfo{year}{2016}),
  \bibinfo{pages}{637--646}.
\newblock


\bibitem[\protect\citeauthoryear{Au, Tai, Chu, Cohen-Or, and Lee}{Au
  et~al\mbox{.}}{2008}]%
        {au2008skeleton}
\bibfield{author}{\bibinfo{person}{Oscar Kin-Chung Au},
  \bibinfo{person}{Chiew-Lan Tai}, \bibinfo{person}{Hung-Kuo Chu},
  \bibinfo{person}{Daniel Cohen-Or}, {and} \bibinfo{person}{Tong-Yee Lee}.}
  \bibinfo{year}{2008}\natexlab{}.
\newblock \showarticletitle{Skeleton extraction by mesh contraction}.
\newblock \bibinfo{journal}{\emph{ACM transactions on graphics (TOG)}}
  \bibinfo{volume}{27}, \bibinfo{number}{3} (\bibinfo{year}{2008}),
  \bibinfo{pages}{1--10}.
\newblock
\urldef\tempurl%
\url{https://doi.org/10.1145/1399504.1360643}
\showDOI{\tempurl}


\bibitem[\protect\citeauthoryear{Auricchio, da~Veiga, Buffa, Lovadina, Reali,
  and Sangalli}{Auricchio et~al\mbox{.}}{2007}]%
        {auricchio2007fully}
\bibfield{author}{\bibinfo{person}{F Auricchio}, \bibinfo{person}{L~Beirao da
  Veiga}, \bibinfo{person}{Annalisa Buffa}, \bibinfo{person}{C Lovadina},
  \bibinfo{person}{A Reali}, {and} \bibinfo{person}{G Sangalli}.}
  \bibinfo{year}{2007}\natexlab{}.
\newblock \showarticletitle{A fully “locking-free” isogeometric approach
  for plane linear elasticity problems: A stream function formulation}.
\newblock \bibinfo{journal}{\emph{Computer methods in applied mechanics and
  engineering}} \bibinfo{volume}{197}, \bibinfo{number}{1-4}
  (\bibinfo{year}{2007}), \bibinfo{pages}{160--172}.
\newblock
\urldef\tempurl%
\url{https://doi.org/10.1016/j.cma.2007.07.005}
\showDOI{\tempurl}


\bibitem[\protect\citeauthoryear{Bajaj, Schaefer, Warren, and Xu}{Bajaj
  et~al\mbox{.}}{2002}]%
        {bajaj2002subdivision}
\bibfield{author}{\bibinfo{person}{Chandrajit Bajaj}, \bibinfo{person}{Scott
  Schaefer}, \bibinfo{person}{Joe Warren}, {and} \bibinfo{person}{Guoliang
  Xu}.} \bibinfo{year}{2002}\natexlab{}.
\newblock \showarticletitle{A subdivision scheme for hexahedral meshes}.
\newblock \bibinfo{journal}{\emph{The visual computer}} \bibinfo{volume}{18},
  \bibinfo{number}{5-6} (\bibinfo{year}{2002}), \bibinfo{pages}{343--356}.
\newblock
\urldef\tempurl%
\url{https://doi.org/10.1007/s003710100150}
\showDOI{\tempurl}


\bibitem[\protect\citeauthoryear{Balzer and Werling}{Balzer and
  Werling}{2010}]%
        {balzer2010principles}
\bibfield{author}{\bibinfo{person}{Jonathan Balzer} {and}
  \bibinfo{person}{Stefan Werling}.} \bibinfo{year}{2010}\natexlab{}.
\newblock \showarticletitle{Principles of shape from specular reflection}.
\newblock \bibinfo{journal}{\emph{Measurement}} \bibinfo{volume}{43},
  \bibinfo{number}{10} (\bibinfo{year}{2010}), \bibinfo{pages}{1305--1317}.
\newblock
\urldef\tempurl%
\url{https://doi.org/10.1016/j.measurement.2010.07.013}
\showDOI{\tempurl}


\bibitem[\protect\citeauthoryear{Bazilevs, Beirao~da Veiga, Cottrell, Hughes,
  and Sangalli}{Bazilevs et~al\mbox{.}}{2006}]%
        {bazilevs2006isogeometric}
\bibfield{author}{\bibinfo{person}{Yuri Bazilevs}, \bibinfo{person}{L Beirao~da
  Veiga}, \bibinfo{person}{J~Austin Cottrell}, \bibinfo{person}{Thomas~JR
  Hughes}, {and} \bibinfo{person}{Giancarlo Sangalli}.}
  \bibinfo{year}{2006}\natexlab{}.
\newblock \showarticletitle{Isogeometric analysis: approximation, stability and
  error estimates for h-refined meshes}.
\newblock \bibinfo{journal}{\emph{Mathematical Models and Methods in Applied
  Sciences}} \bibinfo{volume}{16}, \bibinfo{number}{07} (\bibinfo{year}{2006}),
  \bibinfo{pages}{1031--1090}.
\newblock
\urldef\tempurl%
\url{https://doi.org/10.1142/s0218202506001455}
\showDOI{\tempurl}


\bibitem[\protect\citeauthoryear{Bazilevs, Calo, Cottrell, Evans, Hughes,
  Lipton, Scott, and Sederberg}{Bazilevs et~al\mbox{.}}{2010}]%
        {bazilevs2010isogeometric}
\bibfield{author}{\bibinfo{person}{Yuri Bazilevs}, \bibinfo{person}{Victor~M
  Calo}, \bibinfo{person}{John~A Cottrell}, \bibinfo{person}{John~A Evans},
  \bibinfo{person}{Thomas Jr~R Hughes}, \bibinfo{person}{S Lipton},
  \bibinfo{person}{Michael~A Scott}, {and} \bibinfo{person}{Thomas~W
  Sederberg}.} \bibinfo{year}{2010}\natexlab{}.
\newblock \showarticletitle{Isogeometric analysis using T-splines}.
\newblock \bibinfo{journal}{\emph{Computer Methods in Applied Mechanics and
  Engineering}} \bibinfo{volume}{199}, \bibinfo{number}{5-8}
  (\bibinfo{year}{2010}), \bibinfo{pages}{229--263}.
\newblock
\urldef\tempurl%
\url{https://doi.org/10.1016/j.cma.2009.02.036}
\showDOI{\tempurl}


\bibitem[\protect\citeauthoryear{Bazilevs, Calo, Hughes, and Zhang}{Bazilevs
  et~al\mbox{.}}{2008}]%
        {bazilevs2008isogeometric}
\bibfield{author}{\bibinfo{person}{Yuri Bazilevs}, \bibinfo{person}{Victor~M
  Calo}, \bibinfo{person}{Thomas~JR Hughes}, {and} \bibinfo{person}{Yongjie
  Zhang}.} \bibinfo{year}{2008}\natexlab{}.
\newblock \showarticletitle{Isogeometric fluid-structure interaction: theory,
  algorithms, and computations}.
\newblock \bibinfo{journal}{\emph{Computational mechanics}}
  \bibinfo{volume}{43}, \bibinfo{number}{1} (\bibinfo{year}{2008}),
  \bibinfo{pages}{3--37}.
\newblock
\urldef\tempurl%
\url{https://doi.org/10.1007/s00466-008-0315-x}
\showDOI{\tempurl}


\bibitem[\protect\citeauthoryear{Bingol and Krishnamurthy}{Bingol and
  Krishnamurthy}{2019}]%
        {bingol2019geomdl}
\bibfield{author}{\bibinfo{person}{Onur~Rauf Bingol} {and}
  \bibinfo{person}{Adarsh Krishnamurthy}.} \bibinfo{year}{2019}\natexlab{}.
\newblock \showarticletitle{{NURBS-Python}: An open-source object-oriented
  {NURBS} modeling framework in {Python}}.
\newblock \bibinfo{journal}{\emph{{SoftwareX}}}  \bibinfo{volume}{9}
  (\bibinfo{year}{2019}), \bibinfo{pages}{85--94}.
\newblock
\urldef\tempurl%
\url{https://doi.org/10.1016/j.softx.2018.12.005}
\showDOI{\tempurl}


\bibitem[\protect\citeauthoryear{Bracco, Giannelli, and V{\'a}zquez}{Bracco
  et~al\mbox{.}}{2018}]%
        {bracco2018refinement}
\bibfield{author}{\bibinfo{person}{Cesare Bracco}, \bibinfo{person}{Carlotta
  Giannelli}, {and} \bibinfo{person}{Rafael V{\'a}zquez}.}
  \bibinfo{year}{2018}\natexlab{}.
\newblock \showarticletitle{Refinement algorithms for adaptive isogeometric
  methods with hierarchical splines}.
\newblock \bibinfo{journal}{\emph{axioms}} \bibinfo{volume}{7},
  \bibinfo{number}{3} (\bibinfo{year}{2018}), \bibinfo{pages}{43}.
\newblock
\urldef\tempurl%
\url{https://doi.org/10.3390/axioms7030043}
\showDOI{\tempurl}


\bibitem[\protect\citeauthoryear{Bucelli, Salvador, Quarteroni,
  et~al\mbox{.}}{Bucelli et~al\mbox{.}}{2021}]%
        {bucelli2021multipatch}
\bibfield{author}{\bibinfo{person}{Michele Bucelli}, \bibinfo{person}{Matteo
  Salvador}, \bibinfo{person}{Alfio Quarteroni}, {et~al\mbox{.}}}
  \bibinfo{year}{2021}\natexlab{}.
\newblock \showarticletitle{Multipatch Isogeometric Analysis for
  electrophysiology: Simulation in a human heart}.
\newblock \bibinfo{journal}{\emph{Computer Methods in Applied Mechanics and
  Engineering}}  \bibinfo{volume}{376} (\bibinfo{year}{2021}),
  \bibinfo{pages}{113666}.
\newblock
\urldef\tempurl%
\url{https://doi.org/10.1016/j.cma.2021.113666}
\showDOI{\tempurl}


\bibitem[\protect\citeauthoryear{Bullitt, Zeng, Gerig, Aylward, Joshi, Smith,
  Lin, and Ewend}{Bullitt et~al\mbox{.}}{2005}]%
        {BULLITT20051232}
\bibfield{author}{\bibinfo{person}{Elizabeth Bullitt}, \bibinfo{person}{Donglin
  Zeng}, \bibinfo{person}{Guido Gerig}, \bibinfo{person}{Stephen Aylward},
  \bibinfo{person}{Sarang Joshi}, \bibinfo{person}{J.~Keith Smith},
  \bibinfo{person}{Weili Lin}, {and} \bibinfo{person}{Matthew~G. Ewend}.}
  \bibinfo{year}{2005}\natexlab{}.
\newblock \showarticletitle{Vessel Tortuosity and Brain Tumor Malignancy: A
  Blinded Study}.
\newblock \bibinfo{journal}{\emph{Academic Radiology}} \bibinfo{volume}{12},
  \bibinfo{number}{10} (\bibinfo{year}{2005}), \bibinfo{pages}{1232--1240}.
\newblock
\showISSN{1076-6332}
\urldef\tempurl%
\url{https://doi.org/10.1016/j.acra.2005.05.027}
\showDOI{\tempurl}


\bibitem[\protect\citeauthoryear{Burkhart, Hamann, and Umlauf}{Burkhart
  et~al\mbox{.}}{2010}]%
        {burkhart2010iso}
\bibfield{author}{\bibinfo{person}{Daniel Burkhart}, \bibinfo{person}{Bernd
  Hamann}, {and} \bibinfo{person}{Georg Umlauf}.}
  \bibinfo{year}{2010}\natexlab{}.
\newblock \showarticletitle{Iso-geometric Finite Element Analysis Based on
  Catmull-Clark: ubdivision Solids}. In \bibinfo{booktitle}{\emph{Computer
  Graphics Forum}}, Vol.~\bibinfo{volume}{29}. Wiley Online Library,
  \bibinfo{pages}{1575--1584}.
\newblock
\urldef\tempurl%
\url{https://doi.org/10.1111/j.1467-8659.2010.01766.x}
\showDOI{\tempurl}


\bibitem[\protect\citeauthoryear{Carraturo, Giannelli, Reali, and
  V{\'a}zquez}{Carraturo et~al\mbox{.}}{2019}]%
        {carraturo2019suitably}
\bibfield{author}{\bibinfo{person}{Massimo Carraturo},
  \bibinfo{person}{Carlotta Giannelli}, \bibinfo{person}{Alessandro Reali},
  {and} \bibinfo{person}{Rafael V{\'a}zquez}.} \bibinfo{year}{2019}\natexlab{}.
\newblock \showarticletitle{Suitably graded THB-spline refinement and
  coarsening: Towards an adaptive isogeometric analysis of additive
  manufacturing processes}.
\newblock \bibinfo{journal}{\emph{Computer Methods in Applied Mechanics and
  Engineering}}  \bibinfo{volume}{348} (\bibinfo{year}{2019}),
  \bibinfo{pages}{660--679}.
\newblock
\urldef\tempurl%
\url{https://doi.org/10.1016/j.cma.2019.01.044}
\showDOI{\tempurl}


\bibitem[\protect\citeauthoryear{Cashman, Augsd{\"o}rfer, Dodgson, and
  Sabin}{Cashman et~al\mbox{.}}{2009}]%
        {cashman2009nurbs}
\bibfield{author}{\bibinfo{person}{Thomas~J Cashman}, \bibinfo{person}{Ursula~H
  Augsd{\"o}rfer}, \bibinfo{person}{Neil~A Dodgson}, {and}
  \bibinfo{person}{Malcolm~A Sabin}.} \bibinfo{year}{2009}\natexlab{}.
\newblock \showarticletitle{NURBS with extraordinary points: high-degree,
  non-uniform, rational subdivision schemes}.
\newblock In \bibinfo{booktitle}{\emph{ACM SIGGRAPH 2009 papers}}.
  \bibinfo{pages}{1--9}.
\newblock
\urldef\tempurl%
\url{https://doi.org/10.1145/1576246.1531352}
\showDOI{\tempurl}


\bibitem[\protect\citeauthoryear{Catmull and Clark}{Catmull and Clark}{1978}]%
        {catmull1978recursively}
\bibfield{author}{\bibinfo{person}{Edwin Catmull} {and} \bibinfo{person}{James
  Clark}.} \bibinfo{year}{1978}\natexlab{}.
\newblock \showarticletitle{Recursively generated B-spline surfaces on
  arbitrary topological meshes}.
\newblock \bibinfo{journal}{\emph{Computer-aided design}} \bibinfo{volume}{10},
  \bibinfo{number}{6} (\bibinfo{year}{1978}), \bibinfo{pages}{350--355}.
\newblock
\urldef\tempurl%
\url{https://doi.org/10.1145/280811.280992}
\showDOI{\tempurl}


\bibitem[\protect\citeauthoryear{Chen, Simeon, and Klinkel}{Chen
  et~al\mbox{.}}{2016}]%
        {chen2016nurbs}
\bibfield{author}{\bibinfo{person}{L Chen}, \bibinfo{person}{B Simeon}, {and}
  \bibinfo{person}{S Klinkel}.} \bibinfo{year}{2016}\natexlab{}.
\newblock \showarticletitle{A NURBS based Galerkin approach for the analysis of
  solids in boundary representation}.
\newblock \bibinfo{journal}{\emph{Computer Methods in Applied Mechanics and
  Engineering}}  \bibinfo{volume}{305} (\bibinfo{year}{2016}),
  \bibinfo{pages}{777--805}.
\newblock
\urldef\tempurl%
\url{https://doi.org/10.1016/j.cma.2016.03.019}
\showDOI{\tempurl}


\bibitem[\protect\citeauthoryear{Cheng, Hu, Wang, Wang, and Tamura}{Cheng
  et~al\mbox{.}}{2015}]%
        {cheng2015accurate}
\bibfield{author}{\bibinfo{person}{Yuanzhi Cheng}, \bibinfo{person}{Xin Hu},
  \bibinfo{person}{Ji Wang}, \bibinfo{person}{Yadong Wang}, {and}
  \bibinfo{person}{Shinichi Tamura}.} \bibinfo{year}{2015}\natexlab{}.
\newblock \showarticletitle{Accurate vessel segmentation with constrained
  B-snake}.
\newblock \bibinfo{journal}{\emph{IEEE Transactions on Image Processing}}
  \bibinfo{volume}{24}, \bibinfo{number}{8} (\bibinfo{year}{2015}),
  \bibinfo{pages}{2440--2455}.
\newblock
\urldef\tempurl%
\url{https://doi.org/10.1109/tip.2015.2417683}
\showDOI{\tempurl}


\bibitem[\protect\citeauthoryear{Chuang, Tsai, and Ko}{Chuang
  et~al\mbox{.}}{2000}]%
        {chuang2000skeletonisation}
\bibfield{author}{\bibinfo{person}{Jen-Hui Chuang}, \bibinfo{person}{Chi-Hao
  Tsai}, {and} \bibinfo{person}{Min-Chi Ko}.} \bibinfo{year}{2000}\natexlab{}.
\newblock \showarticletitle{Skeletonisation of three-dimensional object using
  generalized potential field}.
\newblock \bibinfo{journal}{\emph{IEEE Transactions on Pattern Analysis and
  Machine Intelligence}} \bibinfo{volume}{22}, \bibinfo{number}{11}
  (\bibinfo{year}{2000}), \bibinfo{pages}{1241--1251}.
\newblock
\urldef\tempurl%
\url{https://doi.org/10.1109/34.888709}
\showDOI{\tempurl}


\bibitem[\protect\citeauthoryear{Cockburn, Gopalakrishnan, and
  Lazarov}{Cockburn et~al\mbox{.}}{2009}]%
        {cockburn2009unified}
\bibfield{author}{\bibinfo{person}{Bernardo Cockburn},
  \bibinfo{person}{Jayadeep Gopalakrishnan}, {and} \bibinfo{person}{Raytcho
  Lazarov}.} \bibinfo{year}{2009}\natexlab{}.
\newblock \showarticletitle{Unified hybridization of discontinuous Galerkin,
  mixed, and continuous Galerkin methods for second order elliptic problems}.
\newblock \bibinfo{journal}{\emph{SIAM J. Numer. Anal.}} \bibinfo{volume}{47},
  \bibinfo{number}{2} (\bibinfo{year}{2009}), \bibinfo{pages}{1319--1365}.
\newblock
\urldef\tempurl%
\url{https://doi.org/10.1137/070706616}
\showDOI{\tempurl}


\bibitem[\protect\citeauthoryear{Cottrell, Hughes, and Bazilevs}{Cottrell
  et~al\mbox{.}}{2009}]%
        {cottrell2009isogeometric}
\bibfield{author}{\bibinfo{person}{J~Austin Cottrell},
  \bibinfo{person}{Thomas~JR Hughes}, {and} \bibinfo{person}{Yuri Bazilevs}.}
  \bibinfo{year}{2009}\natexlab{}.
\newblock \bibinfo{booktitle}{\emph{Isogeometric analysis: toward integration
  of CAD and FEA}}.
\newblock \bibinfo{publisher}{John Wiley \& Sons}.
\newblock


\bibitem[\protect\citeauthoryear{Crane, Livesu, Puppo, and Qin}{Crane
  et~al\mbox{.}}{2020}]%
        {crane2020survey}
\bibfield{author}{\bibinfo{person}{Keenan Crane}, \bibinfo{person}{Marco
  Livesu}, \bibinfo{person}{Enrico Puppo}, {and} \bibinfo{person}{Yipeng Qin}.}
  \bibinfo{year}{2020}\natexlab{}.
\newblock \showarticletitle{A Survey of Algorithms for Geodesic Paths and
  Distances}.
\newblock \bibinfo{journal}{\emph{arXiv preprint arXiv:2007.10430}}
  (\bibinfo{year}{2020}).
\newblock


\bibitem[\protect\citeauthoryear{Dalcin, Collier, Vignal, Côrtes, and
  Calo}{Dalcin et~al\mbox{.}}{2016}]%
        {PetIGA}
\bibfield{author}{\bibinfo{person}{L. Dalcin}, \bibinfo{person}{N. Collier},
  \bibinfo{person}{P. Vignal}, \bibinfo{person}{A.M.A. Côrtes}, {and}
  \bibinfo{person}{V.M. Calo}.} \bibinfo{year}{2016}\natexlab{}.
\newblock \showarticletitle{PetIGA: A framework for high-performance
  isogeometric analysis}.
\newblock \bibinfo{journal}{\emph{Computer Methods in Applied Mechanics and
  Engineering}}  \bibinfo{volume}{308} (\bibinfo{year}{2016}),
  \bibinfo{pages}{151--181}.
\newblock
\showISSN{0045-7825}
\urldef\tempurl%
\url{https://doi.org/10.1016/j.cma.2016.05.011}
\showDOI{\tempurl}


\bibitem[\protect\citeauthoryear{De~Falco, Reali, and V{\'a}zquez}{De~Falco
  et~al\mbox{.}}{2011}]%
        {de2011geopdes}
\bibfield{author}{\bibinfo{person}{Carlo De~Falco}, \bibinfo{person}{Alessandro
  Reali}, {and} \bibinfo{person}{R V{\'a}zquez}.}
  \bibinfo{year}{2011}\natexlab{}.
\newblock \showarticletitle{GeoPDEs: a research tool for isogeometric analysis
  of PDEs}.
\newblock \bibinfo{journal}{\emph{Advances in Engineering Software}}
  \bibinfo{volume}{42}, \bibinfo{number}{12} (\bibinfo{year}{2011}),
  \bibinfo{pages}{1020--1034}.
\newblock
\urldef\tempurl%
\url{https://doi.org/10.1016/j.advengsoft.2011.06.010}
\showDOI{\tempurl}


\bibitem[\protect\citeauthoryear{Doo and Sabin}{Doo and Sabin}{1978}]%
        {doo1978behaviour}
\bibfield{author}{\bibinfo{person}{Daniel Doo} {and} \bibinfo{person}{Malcolm
  Sabin}.} \bibinfo{year}{1978}\natexlab{}.
\newblock \showarticletitle{Behaviour of recursive division surfaces near
  extraordinary points}.
\newblock \bibinfo{journal}{\emph{Computer-Aided Design}} \bibinfo{volume}{10},
  \bibinfo{number}{6} (\bibinfo{year}{1978}), \bibinfo{pages}{356--360}.
\newblock
\urldef\tempurl%
\url{https://doi.org/10.1145/280811.280991}
\showDOI{\tempurl}


\bibitem[\protect\citeauthoryear{Farin and Hansford}{Farin and
  Hansford}{1999}]%
        {farin1999discrete}
\bibfield{author}{\bibinfo{person}{Gerald Farin} {and} \bibinfo{person}{Dianne
  Hansford}.} \bibinfo{year}{1999}\natexlab{}.
\newblock \showarticletitle{Discrete coons patches}.
\newblock \bibinfo{journal}{\emph{Computer aided geometric design}}
  \bibinfo{volume}{16}, \bibinfo{number}{7} (\bibinfo{year}{1999}),
  \bibinfo{pages}{691--700}.
\newblock
\urldef\tempurl%
\url{https://doi.org/10.1016/s0167-8396(99)00031-x}
\showDOI{\tempurl}


\bibitem[\protect\citeauthoryear{Gao, Jakob, Tarini, and Panozzo}{Gao
  et~al\mbox{.}}{2017}]%
        {gao2017robust}
\bibfield{author}{\bibinfo{person}{Xifeng Gao}, \bibinfo{person}{Wenzel Jakob},
  \bibinfo{person}{Marco Tarini}, {and} \bibinfo{person}{Daniele Panozzo}.}
  \bibinfo{year}{2017}\natexlab{}.
\newblock \showarticletitle{Robust hex-dominant mesh generation using
  field-guided polyhedral agglomeration}.
\newblock \bibinfo{journal}{\emph{ACM Transactions on Graphics (TOG)}}
  \bibinfo{volume}{36}, \bibinfo{number}{4} (\bibinfo{year}{2017}),
  \bibinfo{pages}{1--13}.
\newblock
\urldef\tempurl%
\url{https://doi.org/10.1145/3072959.3073676}
\showDOI{\tempurl}


\bibitem[\protect\citeauthoryear{Gao, Shen, and Panozzo}{Gao
  et~al\mbox{.}}{2019}]%
        {gao2019feature}
\bibfield{author}{\bibinfo{person}{Xifeng Gao}, \bibinfo{person}{Hanxiao Shen},
  {and} \bibinfo{person}{Daniele Panozzo}.} \bibinfo{year}{2019}\natexlab{}.
\newblock \showarticletitle{Feature Preserving Octree-Based Hexahedral
  Meshing}. In \bibinfo{booktitle}{\emph{Computer Graphics Forum}},
  Vol.~\bibinfo{volume}{38}. Wiley Online Library, \bibinfo{pages}{135--149}.
\newblock
\urldef\tempurl%
\url{https://doi.org/10.1111/cgf.13795}
\showDOI{\tempurl}


\bibitem[\protect\citeauthoryear{Garau and V{\'a}zquez}{Garau and
  V{\'a}zquez}{2018}]%
        {garau2018algorithms}
\bibfield{author}{\bibinfo{person}{Eduardo~M Garau} {and}
  \bibinfo{person}{Rafael V{\'a}zquez}.} \bibinfo{year}{2018}\natexlab{}.
\newblock \showarticletitle{Algorithms for the implementation of adaptive
  isogeometric methods using hierarchical B-splines}.
\newblock \bibinfo{journal}{\emph{Applied Numerical Mathematics}}
  \bibinfo{volume}{123} (\bibinfo{year}{2018}), \bibinfo{pages}{58--87}.
\newblock
\urldef\tempurl%
\url{https://doi.org/10.1016/j.apnum.2017.08.006}
\showDOI{\tempurl}


\bibitem[\protect\citeauthoryear{Gomez, Hughes, Nogueira, and Calo}{Gomez
  et~al\mbox{.}}{2010}]%
        {gomez2010isogeometric}
\bibfield{author}{\bibinfo{person}{Hector Gomez}, \bibinfo{person}{Thomas~JR
  Hughes}, \bibinfo{person}{Xes{\'u}s Nogueira}, {and}
  \bibinfo{person}{Victor~M Calo}.} \bibinfo{year}{2010}\natexlab{}.
\newblock \showarticletitle{Isogeometric analysis of the isothermal
  Navier--Stokes--Korteweg equations}.
\newblock \bibinfo{journal}{\emph{Computer Methods in Applied Mechanics and
  Engineering}} \bibinfo{volume}{199}, \bibinfo{number}{25-28}
  (\bibinfo{year}{2010}), \bibinfo{pages}{1828--1840}.
\newblock
\urldef\tempurl%
\url{https://doi.org/10.1016/j.cma.2010.02.010}
\showDOI{\tempurl}


\bibitem[\protect\citeauthoryear{Gregson, Sheffer, and Zhang}{Gregson
  et~al\mbox{.}}{2011}]%
        {gregson2011all}
\bibfield{author}{\bibinfo{person}{James Gregson}, \bibinfo{person}{Alla
  Sheffer}, {and} \bibinfo{person}{Eugene Zhang}.}
  \bibinfo{year}{2011}\natexlab{}.
\newblock \showarticletitle{All-hex mesh generation via volumetric polycube
  deformation}. In \bibinfo{booktitle}{\emph{Computer graphics forum}},
  Vol.~\bibinfo{volume}{30}. Wiley Online Library, \bibinfo{pages}{1407--1416}.
\newblock
\urldef\tempurl%
\url{https://doi.org/10.1111/j.1467-8659.2011.02015.x}
\showDOI{\tempurl}


\bibitem[\protect\citeauthoryear{Hanson}{Hanson}{1994}]%
        {hanson1994quaternion}
\bibfield{author}{\bibinfo{person}{Andrew~J Hanson}.}
  \bibinfo{year}{1994}\natexlab{}.
\newblock \showarticletitle{Quaternion Frenet frames: Making optimal tubes and
  ribbons from curves}.
\newblock \bibinfo{journal}{\emph{Computer Science Department, Indiana
  University Bloomington, In}}  \bibinfo{volume}{47405} (\bibinfo{year}{1994}).
\newblock


\bibitem[\protect\citeauthoryear{He, Wang, Wang, Gu, and Qin}{He
  et~al\mbox{.}}{2006}]%
        {he2006manifold}
\bibfield{author}{\bibinfo{person}{Ying He}, \bibinfo{person}{Kexiang Wang},
  \bibinfo{person}{Hongyu Wang}, \bibinfo{person}{Xianfeng Gu}, {and}
  \bibinfo{person}{Hong Qin}.} \bibinfo{year}{2006}\natexlab{}.
\newblock \showarticletitle{Manifold T-spline}. In
  \bibinfo{booktitle}{\emph{International conference on geometric modeling and
  processing}}. Springer, \bibinfo{pages}{409--422}.
\newblock
\urldef\tempurl%
\url{https://doi.org/10.1007/11802914_29}
\showDOI{\tempurl}


\bibitem[\protect\citeauthoryear{Hesch, Franke, Dittmann, and Temizer}{Hesch
  et~al\mbox{.}}{2016}]%
        {hesch2016hierarchical}
\bibfield{author}{\bibinfo{person}{C Hesch}, \bibinfo{person}{M Franke},
  \bibinfo{person}{M Dittmann}, {and} \bibinfo{person}{I Temizer}.}
  \bibinfo{year}{2016}\natexlab{}.
\newblock \showarticletitle{Hierarchical NURBS and a higher-order phase-field
  approach to fracture for finite-deformation contact problems}.
\newblock \bibinfo{journal}{\emph{Computer Methods in Applied Mechanics and
  Engineering}}  \bibinfo{volume}{301} (\bibinfo{year}{2016}),
  \bibinfo{pages}{242--258}.
\newblock
\urldef\tempurl%
\url{https://doi.org/10.1016/j.cma.2015.12.011}
\showDOI{\tempurl}


\bibitem[\protect\citeauthoryear{Ho-Le}{Ho-Le}{1988}]%
        {ho1988finite}
\bibfield{author}{\bibinfo{person}{K Ho-Le}.} \bibinfo{year}{1988}\natexlab{}.
\newblock \showarticletitle{Finite element mesh generation methods: a review
  and classification}.
\newblock \bibinfo{journal}{\emph{Computer-aided design}} \bibinfo{volume}{20},
  \bibinfo{number}{1} (\bibinfo{year}{1988}), \bibinfo{pages}{27--38}.
\newblock
\urldef\tempurl%
\url{https://doi.org/10.1016/0010-4485(88)90138-8}
\showDOI{\tempurl}


\bibitem[\protect\citeauthoryear{Hughes, Cottrell, and Bazilevs}{Hughes
  et~al\mbox{.}}{2005}]%
        {hughes2005isogeometric}
\bibfield{author}{\bibinfo{person}{Thomas~JR Hughes}, \bibinfo{person}{John~A
  Cottrell}, {and} \bibinfo{person}{Yuri Bazilevs}.}
  \bibinfo{year}{2005}\natexlab{}.
\newblock \showarticletitle{Isogeometric analysis: CAD, finite elements, NURBS,
  exact geometry and mesh refinement}.
\newblock \bibinfo{journal}{\emph{Computer methods in applied mechanics and
  engineering}} \bibinfo{volume}{194}, \bibinfo{number}{39-41}
  (\bibinfo{year}{2005}), \bibinfo{pages}{4135--4195}.
\newblock
\urldef\tempurl%
\url{https://doi.org/10.1016/j.cma.2004.10.008}
\showDOI{\tempurl}


\bibitem[\protect\citeauthoryear{J{\"u}ttler, Langer, Mantzaflaris, Moore, and
  Zulehner}{J{\"u}ttler et~al\mbox{.}}{2014}]%
        {juttler2014geometry}
\bibfield{author}{\bibinfo{person}{Bert J{\"u}ttler}, \bibinfo{person}{Ulrich
  Langer}, \bibinfo{person}{Angelos Mantzaflaris}, \bibinfo{person}{Stephen~E
  Moore}, {and} \bibinfo{person}{Walter Zulehner}.}
  \bibinfo{year}{2014}\natexlab{}.
\newblock \showarticletitle{Geometry+ simulation modules: Implementing
  isogeometric analysis}.
\newblock \bibinfo{journal}{\emph{PAMM}} \bibinfo{volume}{14},
  \bibinfo{number}{1} (\bibinfo{year}{2014}), \bibinfo{pages}{961--962}.
\newblock
\urldef\tempurl%
\url{https://doi.org/10.1002/pamm.201410461}
\showDOI{\tempurl}


\bibitem[\protect\citeauthoryear{Kanitsar, Fleischmann, Wegenkittl, Felkel, and
  Groller}{Kanitsar et~al\mbox{.}}{2002}]%
        {Kanitsar2002}
\bibfield{author}{\bibinfo{person}{A. Kanitsar}, \bibinfo{person}{D.
  Fleischmann}, \bibinfo{person}{R. Wegenkittl}, \bibinfo{person}{P. Felkel},
  {and} \bibinfo{person}{E. Groller}.} \bibinfo{year}{2002}\natexlab{}.
\newblock \showarticletitle{CPR - curved planar reformation}. In
  \bibinfo{booktitle}{\emph{IEEE Visualization, 2002. VIS 2002.}}
  \bibinfo{pages}{37--44}.
\newblock
\urldef\tempurl%
\url{https://doi.org/10.1109/VISUAL.2002.1183754}
\showDOI{\tempurl}


\bibitem[\protect\citeauthoryear{Kimmel and Sethian}{Kimmel and
  Sethian}{1998}]%
        {kimmel1998computing}
\bibfield{author}{\bibinfo{person}{Ron Kimmel} {and} \bibinfo{person}{James~A
  Sethian}.} \bibinfo{year}{1998}\natexlab{}.
\newblock \showarticletitle{Computing geodesic paths on manifolds}.
\newblock \bibinfo{journal}{\emph{Proceedings of the national academy of
  Sciences}} \bibinfo{volume}{95}, \bibinfo{number}{15} (\bibinfo{year}{1998}),
  \bibinfo{pages}{8431--8435}.
\newblock
\urldef\tempurl%
\url{https://doi.org/10.1073/pnas.95.15.8431}
\showDOI{\tempurl}


\bibitem[\protect\citeauthoryear{Krishnamurthy and Levoy}{Krishnamurthy and
  Levoy}{1996}]%
        {krishnamurthy1996fitting}
\bibfield{author}{\bibinfo{person}{Venkat Krishnamurthy} {and}
  \bibinfo{person}{Marc Levoy}.} \bibinfo{year}{1996}\natexlab{}.
\newblock \showarticletitle{Fitting smooth surfaces to dense polygon meshes}.
  In \bibinfo{booktitle}{\emph{Proceedings of the 23rd annual conference on
  Computer graphics and interactive techniques}}. \bibinfo{pages}{313--324}.
\newblock
\urldef\tempurl%
\url{https://doi.org/10.1145/237170.237270}
\showDOI{\tempurl}


\bibitem[\protect\citeauthoryear{L{\'e}vy, Petitjean, Ray, and
  Maillot}{L{\'e}vy et~al\mbox{.}}{2002}]%
        {levy2002least}
\bibfield{author}{\bibinfo{person}{Bruno L{\'e}vy}, \bibinfo{person}{Sylvain
  Petitjean}, \bibinfo{person}{Nicolas Ray}, {and} \bibinfo{person}{J{\'e}rome
  Maillot}.} \bibinfo{year}{2002}\natexlab{}.
\newblock \showarticletitle{Least squares conformal maps for automatic texture
  atlas generation}.
\newblock \bibinfo{journal}{\emph{ACM transactions on graphics (TOG)}}
  \bibinfo{volume}{21}, \bibinfo{number}{3} (\bibinfo{year}{2002}),
  \bibinfo{pages}{362--371}.
\newblock
\urldef\tempurl%
\url{https://doi.org/10.1145/566654.566590}
\showDOI{\tempurl}


\bibitem[\protect\citeauthoryear{Livesu, Muntoni, Puppo, and Scateni}{Livesu
  et~al\mbox{.}}{2016}]%
        {livesu2016skeleton}
\bibfield{author}{\bibinfo{person}{Marco Livesu}, \bibinfo{person}{Alessandro
  Muntoni}, \bibinfo{person}{Enrico Puppo}, {and} \bibinfo{person}{Riccardo
  Scateni}.} \bibinfo{year}{2016}\natexlab{}.
\newblock \showarticletitle{Skeleton-driven adaptive hexahedral meshing of
  tubular shapes}. In \bibinfo{booktitle}{\emph{Computer Graphics Forum}},
  Vol.~\bibinfo{volume}{35}. Wiley Online Library, \bibinfo{pages}{237--246}.
\newblock
\urldef\tempurl%
\url{https://doi.org/10.1111/cgf.13021}
\showDOI{\tempurl}


\bibitem[\protect\citeauthoryear{Livesu, Pietroni, Puppo, Sheffer, and
  Cignoni}{Livesu et~al\mbox{.}}{2020}]%
        {LPPSC20}
\bibfield{author}{\bibinfo{person}{Marco Livesu}, \bibinfo{person}{Nico
  Pietroni}, \bibinfo{person}{Enrico Puppo}, \bibinfo{person}{Alla Sheffer},
  {and} \bibinfo{person}{Paolo Cignoni}.} \bibinfo{year}{2020}\natexlab{}.
\newblock \showarticletitle{LoopyCuts: Practical Feature-Preserving Block
  Decomposition for Strongly Hex-Dominant Meshing}.
\newblock \bibinfo{journal}{\emph{ACM Transactions on Graphics (SIGGRAPH)}}
  \bibinfo{volume}{39}, \bibinfo{number}{4} (\bibinfo{year}{2020}).
\newblock
\urldef\tempurl%
\url{https://doi.org/10.1145/3386569.3392472}
\showDOI{\tempurl}


\bibitem[\protect\citeauthoryear{Livesu, Sheffer, Vining, and Tarini}{Livesu
  et~al\mbox{.}}{2015}]%
        {livesu2015practical}
\bibfield{author}{\bibinfo{person}{Marco Livesu}, \bibinfo{person}{Alla
  Sheffer}, \bibinfo{person}{Nicholas Vining}, {and} \bibinfo{person}{Marco
  Tarini}.} \bibinfo{year}{2015}\natexlab{}.
\newblock \showarticletitle{Practical hex-mesh optimization via edge-cone
  rectification}.
\newblock \bibinfo{journal}{\emph{ACM Transactions on Graphics (TOG)}}
  \bibinfo{volume}{34}, \bibinfo{number}{4} (\bibinfo{year}{2015}),
  \bibinfo{pages}{1--11}.
\newblock
\urldef\tempurl%
\url{https://doi.org/10.1145/2766905}
\showDOI{\tempurl}


\bibitem[\protect\citeauthoryear{Livesu, Vining, Sheffer, Gregson, and
  Scateni}{Livesu et~al\mbox{.}}{2013}]%
        {livesu2013polycut}
\bibfield{author}{\bibinfo{person}{Marco Livesu}, \bibinfo{person}{Nicholas
  Vining}, \bibinfo{person}{Alla Sheffer}, \bibinfo{person}{James Gregson},
  {and} \bibinfo{person}{Riccardo Scateni}.} \bibinfo{year}{2013}\natexlab{}.
\newblock \showarticletitle{Polycut: Monotone graph-cuts for polycube
  base-complex construction}.
\newblock \bibinfo{journal}{\emph{ACM Transactions on Graphics (TOG)}}
  \bibinfo{volume}{32}, \bibinfo{number}{6} (\bibinfo{year}{2013}),
  \bibinfo{pages}{1--12}.
\newblock
\urldef\tempurl%
\url{https://doi.org/10.1145/2508363.2508388}
\showDOI{\tempurl}


\bibitem[\protect\citeauthoryear{Loop}{Loop}{1987}]%
        {loop1987smooth}
\bibfield{author}{\bibinfo{person}{Charles Loop}.}
  \bibinfo{year}{1987}\natexlab{}.
\newblock \showarticletitle{Smooth subdivision surfaces based on triangles}.
\newblock \bibinfo{journal}{\emph{Master's thesis, University of Utah,
  Department of Mathematics}} (\bibinfo{year}{1987}).
\newblock


\bibitem[\protect\citeauthoryear{Lu, Quadros, and Shimada}{Lu
  et~al\mbox{.}}{2017}]%
        {lu2017evaluation}
\bibfield{author}{\bibinfo{person}{Jean Hsiang-Chun Lu},
  \bibinfo{person}{William~Roshan Quadros}, {and} \bibinfo{person}{Kenji
  Shimada}.} \bibinfo{year}{2017}\natexlab{}.
\newblock \showarticletitle{Evaluation of user-guided semi-automatic
  decomposition tool for hexahedral mesh generation}.
\newblock \bibinfo{journal}{\emph{Journal of Computational Design and
  Engineering}} \bibinfo{volume}{4}, \bibinfo{number}{4}
  (\bibinfo{year}{2017}), \bibinfo{pages}{330--338}.
\newblock
\urldef\tempurl%
\url{https://doi.org/10.1016/j.jcde.2017.05.001}
\showDOI{\tempurl}


\bibitem[\protect\citeauthoryear{Ma and Kruth}{Ma and Kruth}{1998}]%
        {ma1998nurbs}
\bibfield{author}{\bibinfo{person}{Weiyin Ma} {and} \bibinfo{person}{J-P
  Kruth}.} \bibinfo{year}{1998}\natexlab{}.
\newblock \showarticletitle{NURBS curve and surface fitting for reverse
  engineering}.
\newblock \bibinfo{journal}{\emph{The International Journal of Advanced
  Manufacturing Technology}} \bibinfo{volume}{14}, \bibinfo{number}{12}
  (\bibinfo{year}{1998}), \bibinfo{pages}{918--927}.
\newblock
\urldef\tempurl%
\url{https://doi.org/10.1007/bf01179082}
\showDOI{\tempurl}


\bibitem[\protect\citeauthoryear{Martinez, Velho, and Carvalho}{Martinez
  et~al\mbox{.}}{2004}]%
        {martinez2004geodesic}
\bibfield{author}{\bibinfo{person}{Dimas Martinez}, \bibinfo{person}{Luiz
  Velho}, {and} \bibinfo{person}{P~Cezar Carvalho}.}
  \bibinfo{year}{2004}\natexlab{}.
\newblock \showarticletitle{Geodesic paths on triangular meshes}. In
  \bibinfo{booktitle}{\emph{Proceedings. 17th Brazilian Symposium on Computer
  Graphics and Image Processing}}. IEEE, \bibinfo{pages}{210--217}.
\newblock


\bibitem[\protect\citeauthoryear{Mortenson}{Mortenson}{1997}]%
        {mortenson1997geometric}
\bibfield{author}{\bibinfo{person}{Michael~E Mortenson}.}
  \bibinfo{year}{1997}\natexlab{}.
\newblock \bibinfo{booktitle}{\emph{Geometric modeling}}.
\newblock \bibinfo{publisher}{John Wiley \& Sons, Inc.}
\newblock


\bibitem[\protect\citeauthoryear{Pagani and Scott}{Pagani and Scott}{2018}]%
        {pagani2018curvature}
\bibfield{author}{\bibinfo{person}{Luca Pagani} {and} \bibinfo{person}{Paul~J
  Scott}.} \bibinfo{year}{2018}\natexlab{}.
\newblock \showarticletitle{Curvature based sampling of curves and surfaces}.
\newblock \bibinfo{journal}{\emph{Computer Aided Geometric Design}}
  \bibinfo{volume}{59} (\bibinfo{year}{2018}), \bibinfo{pages}{32--48}.
\newblock
\urldef\tempurl%
\url{https://doi.org/10.1016/j.cagd.2017.11.004}
\showDOI{\tempurl}


\bibitem[\protect\citeauthoryear{Pan, Xu, Xu, and Zhang}{Pan
  et~al\mbox{.}}{2016}]%
        {pan2016isogeometric}
\bibfield{author}{\bibinfo{person}{Qing Pan}, \bibinfo{person}{Guoliang Xu},
  \bibinfo{person}{Gang Xu}, {and} \bibinfo{person}{Yongjie Zhang}.}
  \bibinfo{year}{2016}\natexlab{}.
\newblock \showarticletitle{Isogeometric analysis based on extended
  Catmull--Clark subdivision}.
\newblock \bibinfo{journal}{\emph{Computers \& Mathematics with Applications}}
  \bibinfo{volume}{71}, \bibinfo{number}{1} (\bibinfo{year}{2016}),
  \bibinfo{pages}{105--119}.
\newblock
\urldef\tempurl%
\url{https://doi.org/10.1016/j.camwa.2015.11.012}
\showDOI{\tempurl}


\bibitem[\protect\citeauthoryear{Panotopoulou, Ross, Welker, Hubert, and
  Morin}{Panotopoulou et~al\mbox{.}}{2018}]%
        {panotopoulou2018scaffolding}
\bibfield{author}{\bibinfo{person}{Athina Panotopoulou},
  \bibinfo{person}{Elissa Ross}, \bibinfo{person}{Kathrin Welker},
  \bibinfo{person}{Evelyne Hubert}, {and} \bibinfo{person}{G{\'e}raldine
  Morin}.} \bibinfo{year}{2018}\natexlab{}.
\newblock \showarticletitle{Scaffolding a skeleton}.
\newblock In \bibinfo{booktitle}{\emph{Research in Shape Analysis}}.
  \bibinfo{publisher}{Springer}, \bibinfo{pages}{17--35}.
\newblock
\urldef\tempurl%
\url{https://doi.org/10.1007/978-3-319-77066-6_2}
\showDOI{\tempurl}


\bibitem[\protect\citeauthoryear{Panozzo, Puppo, Tarini, and
  Sorkine-Hornung}{Panozzo et~al\mbox{.}}{2014}]%
        {panozzo2014frame}
\bibfield{author}{\bibinfo{person}{Daniele Panozzo}, \bibinfo{person}{Enrico
  Puppo}, \bibinfo{person}{Marco Tarini}, {and} \bibinfo{person}{Olga
  Sorkine-Hornung}.} \bibinfo{year}{2014}\natexlab{}.
\newblock \showarticletitle{Frame fields: Anisotropic and non-orthogonal cross
  fields}.
\newblock \bibinfo{journal}{\emph{ACM Transactions on Graphics (TOG)}}
  \bibinfo{volume}{33}, \bibinfo{number}{4} (\bibinfo{year}{2014}),
  \bibinfo{pages}{1--11}.
\newblock
\urldef\tempurl%
\url{https://doi.org/10.1145/2601097.2601179}
\showDOI{\tempurl}


\bibitem[\protect\citeauthoryear{Petera and Pittman}{Petera and
  Pittman}{1994}]%
        {petera1994isoparametric}
\bibfield{author}{\bibinfo{person}{J. Petera} {and} \bibinfo{person}{J.~F.~T.
  Pittman}.} \bibinfo{year}{1994}\natexlab{}.
\newblock \showarticletitle{Isoparametric Hermite Elements}.
\newblock \bibinfo{journal}{\emph{Internat. J. Numer. Methods Engrg.}}
  \bibinfo{volume}{37}, \bibinfo{number}{20} (\bibinfo{year}{1994}),
  \bibinfo{pages}{3489--3519}.
\newblock
\urldef\tempurl%
\url{https://doi.org/10.1002/nme.1620372006}
\showDOI{\tempurl}


\bibitem[\protect\citeauthoryear{Piccinelli, Veneziani, Steinman, Remuzzi, and
  Antiga}{Piccinelli et~al\mbox{.}}{2009}]%
        {piccinelli2009framework}
\bibfield{author}{\bibinfo{person}{Marina Piccinelli},
  \bibinfo{person}{Alessandro Veneziani}, \bibinfo{person}{David~A Steinman},
  \bibinfo{person}{Andrea Remuzzi}, {and} \bibinfo{person}{Luca Antiga}.}
  \bibinfo{year}{2009}\natexlab{}.
\newblock \showarticletitle{A framework for geometric analysis of vascular
  structures: application to cerebral aneurysms}.
\newblock \bibinfo{journal}{\emph{IEEE transactions on medical imaging}}
  \bibinfo{volume}{28}, \bibinfo{number}{8} (\bibinfo{year}{2009}),
  \bibinfo{pages}{1141--1155}.
\newblock
\urldef\tempurl%
\url{https://doi.org/10.1109/tmi.2009.2021652}
\showDOI{\tempurl}


\bibitem[\protect\citeauthoryear{Piegl}{Piegl}{1991}]%
        {Piegl1991}
\bibfield{author}{\bibinfo{person}{L. Piegl}.} \bibinfo{year}{1991}\natexlab{}.
\newblock \showarticletitle{On NURBS: a Survey}.
\newblock \bibinfo{journal}{\emph{IEEE Computer Graphics and Applications}}
  \bibinfo{volume}{11}, \bibinfo{number}{01} (\bibinfo{date}{jan}
  \bibinfo{year}{1991}), \bibinfo{pages}{55--71}.
\newblock
\showISSN{1558-1756}
\urldef\tempurl%
\url{https://doi.org/10.1109/38.67702}
\showDOI{\tempurl}


\bibitem[\protect\citeauthoryear{Piegl and Tiller}{Piegl and Tiller}{1996}]%
        {piegl1996nurbs}
\bibfield{author}{\bibinfo{person}{Les Piegl} {and} \bibinfo{person}{Wayne
  Tiller}.} \bibinfo{year}{1996}\natexlab{}.
\newblock \bibinfo{booktitle}{\emph{The NURBS book}}.
\newblock \bibinfo{publisher}{Springer Science \& Business Media}.
\newblock
\urldef\tempurl%
\url{https://doi.org/10.1007/978-3-642-59223-2}
\showDOI{\tempurl}


\bibitem[\protect\citeauthoryear{Pietroni, Puppo, Marcias, Scopigno, and
  Cignoni}{Pietroni et~al\mbox{.}}{2016}]%
        {pietroni2016tracing}
\bibfield{author}{\bibinfo{person}{Nico Pietroni}, \bibinfo{person}{Enrico
  Puppo}, \bibinfo{person}{Giorgio Marcias}, \bibinfo{person}{Roberto
  Scopigno}, {and} \bibinfo{person}{Paolo Cignoni}.}
  \bibinfo{year}{2016}\natexlab{}.
\newblock \showarticletitle{Tracing field-coherent quad layouts}. In
  \bibinfo{booktitle}{\emph{Computer Graphics Forum}},
  Vol.~\bibinfo{volume}{35}. Wiley Online Library, \bibinfo{pages}{485--496}.
\newblock
\urldef\tempurl%
\url{https://doi.org/10.1111/cgf.13045}
\showDOI{\tempurl}


\bibitem[\protect\citeauthoryear{Rao}{Rao}{2017}]%
        {rao2017finite}
\bibfield{author}{\bibinfo{person}{Singiresu~S Rao}.}
  \bibinfo{year}{2017}\natexlab{}.
\newblock \bibinfo{booktitle}{\emph{The finite element method in engineering}}.
\newblock \bibinfo{publisher}{Butterworth-heinemann}.
\newblock


\bibitem[\protect\citeauthoryear{Russo}{Russo}{2006}]%
        {russo2006polygonal}
\bibfield{author}{\bibinfo{person}{Mario Russo}.}
  \bibinfo{year}{2006}\natexlab{}.
\newblock \bibinfo{booktitle}{\emph{Polygonal modeling: basic and advanced
  techniques}}.
\newblock \bibinfo{publisher}{Jones \& Bartlett Learning}.
\newblock


\bibitem[\protect\citeauthoryear{Saha, Borgefors, and di~Baja}{Saha
  et~al\mbox{.}}{2016}]%
        {saha2016survey}
\bibfield{author}{\bibinfo{person}{Punam~K Saha}, \bibinfo{person}{Gunilla
  Borgefors}, {and} \bibinfo{person}{Gabriella~Sanniti di Baja}.}
  \bibinfo{year}{2016}\natexlab{}.
\newblock \showarticletitle{A survey on skeletonization algorithms and their
  applications}.
\newblock \bibinfo{journal}{\emph{Pattern recognition letters}}
  \bibinfo{volume}{76} (\bibinfo{year}{2016}), \bibinfo{pages}{3--12}.
\newblock
\urldef\tempurl%
\url{https://doi.org/10.1016/j.patrec.2015.04.006}
\showDOI{\tempurl}


\bibitem[\protect\citeauthoryear{Saha, Borgefors, and di~Baja}{Saha
  et~al\mbox{.}}{2017}]%
        {saha2017skeletonization}
\bibfield{author}{\bibinfo{person}{Punam~K Saha}, \bibinfo{person}{Gunilla
  Borgefors}, {and} \bibinfo{person}{Gabriella~Sanniti di Baja}.}
  \bibinfo{year}{2017}\natexlab{}.
\newblock \bibinfo{booktitle}{\emph{Skeletonization: Theory, methods and
  applications}}.
\newblock \bibinfo{publisher}{Academic Press}.
\newblock


\bibitem[\protect\citeauthoryear{Sethian}{Sethian}{1996}]%
        {sethian1996fast}
\bibfield{author}{\bibinfo{person}{James~A Sethian}.}
  \bibinfo{year}{1996}\natexlab{}.
\newblock \showarticletitle{A fast marching level set method for monotonically
  advancing fronts}.
\newblock \bibinfo{journal}{\emph{Proceedings of the National Academy of
  Sciences}} \bibinfo{volume}{93}, \bibinfo{number}{4} (\bibinfo{year}{1996}),
  \bibinfo{pages}{1591--1595}.
\newblock
\urldef\tempurl%
\url{https://doi.org/10.1073/pnas.93.4.1591}
\showDOI{\tempurl}


\bibitem[\protect\citeauthoryear{Sethian}{Sethian}{1999}]%
        {sethian1999level}
\bibfield{author}{\bibinfo{person}{James~Albert Sethian}.}
  \bibinfo{year}{1999}\natexlab{}.
\newblock \bibinfo{booktitle}{\emph{Level set methods and fast marching
  methods: evolving interfaces in computational geometry, fluid mechanics,
  computer vision, and materials science}}. Vol.~\bibinfo{volume}{3}.
\newblock \bibinfo{publisher}{Cambridge university press}.
\newblock


\bibitem[\protect\citeauthoryear{Si}{Si}{2015}]%
        {si2015tetgen}
\bibfield{author}{\bibinfo{person}{Hang Si}.} \bibinfo{year}{2015}\natexlab{}.
\newblock \showarticletitle{TetGen, a Delaunay-based quality tetrahedral mesh
  generator}.
\newblock \bibinfo{journal}{\emph{ACM Transactions on Mathematical Software
  (TOMS)}} \bibinfo{volume}{41}, \bibinfo{number}{2} (\bibinfo{year}{2015}),
  \bibinfo{pages}{1--36}.
\newblock
\urldef\tempurl%
\url{https://doi.org/10.1145/2629697}
\showDOI{\tempurl}


\bibitem[\protect\citeauthoryear{Stam}{Stam}{1998a}]%
        {stam1998evaluation}
\bibfield{author}{\bibinfo{person}{Jos Stam}.}
  \bibinfo{year}{1998}\natexlab{a}.
\newblock \showarticletitle{Evaluation of loop subdivision surfaces}. In
  \bibinfo{booktitle}{\emph{SIGGRAPH’98 CDROM Proceedings}}. Citeseer.
\newblock


\bibitem[\protect\citeauthoryear{Stam}{Stam}{1998b}]%
        {stam1998exact}
\bibfield{author}{\bibinfo{person}{Jos Stam}.}
  \bibinfo{year}{1998}\natexlab{b}.
\newblock \showarticletitle{Exact evaluation of Catmull-Clark subdivision
  surfaces at arbitrary parameter values}. In
  \bibinfo{booktitle}{\emph{Proceedings of the 25th annual conference on
  Computer graphics and interactive techniques}}. \bibinfo{pages}{395--404}.
\newblock
\urldef\tempurl%
\url{https://doi.org/10.1145/280814.280945}
\showDOI{\tempurl}


\bibitem[\protect\citeauthoryear{Stam}{Stam}{2001}]%
        {stam2001subdivision}
\bibfield{author}{\bibinfo{person}{Jos Stam}.} \bibinfo{year}{2001}\natexlab{}.
\newblock \showarticletitle{On subdivision schemes generalizing uniform
  B-spline surfaces of arbitrary degree}.
\newblock \bibinfo{journal}{\emph{Computer Aided Geometric Design}}
  \bibinfo{volume}{18}, \bibinfo{number}{5} (\bibinfo{year}{2001}),
  \bibinfo{pages}{383--396}.
\newblock
\urldef\tempurl%
\url{https://doi.org/10.1016/s0167-8396(01)00038-3}
\showDOI{\tempurl}


\bibitem[\protect\citeauthoryear{Stolpner, Kry, and Siddiqi}{Stolpner
  et~al\mbox{.}}{2011}]%
        {stolpner2011medial}
\bibfield{author}{\bibinfo{person}{Svetlana Stolpner}, \bibinfo{person}{Paul
  Kry}, {and} \bibinfo{person}{Kaleem Siddiqi}.}
  \bibinfo{year}{2011}\natexlab{}.
\newblock \showarticletitle{Medial spheres for shape approximation}.
\newblock \bibinfo{journal}{\emph{IEEE transactions on pattern analysis and
  machine intelligence}} \bibinfo{volume}{34}, \bibinfo{number}{6}
  (\bibinfo{year}{2011}), \bibinfo{pages}{1234--1240}.
\newblock
\urldef\tempurl%
\url{https://doi.org/10.1109/tpami.2011.254}
\showDOI{\tempurl}


\bibitem[\protect\citeauthoryear{Su{\'a}rez and Hubert}{Su{\'a}rez and
  Hubert}{2017}]%
        {suarez2017scaffolding}
\bibfield{author}{\bibinfo{person}{AJ~Fuentes Su{\'a}rez} {and}
  \bibinfo{person}{Evelyne Hubert}.} \bibinfo{year}{2017}\natexlab{}.
\newblock \showarticletitle{Scaffolding skeletons using spherical Voronoi
  diagrams}.
\newblock \bibinfo{journal}{\emph{Electronic Notes in Discrete Mathematics}}
  \bibinfo{volume}{62} (\bibinfo{year}{2017}), \bibinfo{pages}{45--50}.
\newblock
\urldef\tempurl%
\url{https://doi.org/10.1016/j.endm.2017.10.009}
\showDOI{\tempurl}


\bibitem[\protect\citeauthoryear{Su{\'a}rez and Hubert}{Su{\'a}rez and
  Hubert}{2018}]%
        {suarez2018scaffolding}
\bibfield{author}{\bibinfo{person}{AJ~Fuentes Su{\'a}rez} {and}
  \bibinfo{person}{Evelyne Hubert}.} \bibinfo{year}{2018}\natexlab{}.
\newblock \showarticletitle{Scaffolding skeletons using spherical Voronoi
  diagrams: Feasibility, regularity and symmetry}.
\newblock \bibinfo{journal}{\emph{Computer-Aided Design}}
  \bibinfo{volume}{102} (\bibinfo{year}{2018}), \bibinfo{pages}{83--93}.
\newblock
\urldef\tempurl%
\url{https://doi.org/10.1016/j.cad.2018.04.016}
\showDOI{\tempurl}


\bibitem[\protect\citeauthoryear{Surazhsky, Surazhsky, Kirsanov, Gortler, and
  Hoppe}{Surazhsky et~al\mbox{.}}{2005}]%
        {surazhsky2005fast}
\bibfield{author}{\bibinfo{person}{Vitaly Surazhsky}, \bibinfo{person}{Tatiana
  Surazhsky}, \bibinfo{person}{Danil Kirsanov}, \bibinfo{person}{Steven~J
  Gortler}, {and} \bibinfo{person}{Hugues Hoppe}.}
  \bibinfo{year}{2005}\natexlab{}.
\newblock \showarticletitle{Fast exact and approximate geodesics on meshes}.
\newblock \bibinfo{journal}{\emph{ACM transactions on graphics (TOG)}}
  \bibinfo{volume}{24}, \bibinfo{number}{3} (\bibinfo{year}{2005}),
  \bibinfo{pages}{553--560}.
\newblock
\urldef\tempurl%
\url{https://doi.org/10.1145/1186822.1073228}
\showDOI{\tempurl}


\bibitem[\protect\citeauthoryear{Takayama}{Takayama}{2019}]%
        {takayama2019dual}
\bibfield{author}{\bibinfo{person}{Kenshi Takayama}.}
  \bibinfo{year}{2019}\natexlab{}.
\newblock \showarticletitle{Dual sheet meshing: An interactive approach to
  robust hexahedralization}. In \bibinfo{booktitle}{\emph{Computer Graphics
  Forum}}, Vol.~\bibinfo{volume}{38}. Wiley Online Library,
  \bibinfo{pages}{37--48}.
\newblock
\urldef\tempurl%
\url{https://doi.org/10.1111/cgf.13617}
\showDOI{\tempurl}


\bibitem[\protect\citeauthoryear{Tarini, Hormann, Cignoni, and Montani}{Tarini
  et~al\mbox{.}}{2004}]%
        {tarini2004polycube}
\bibfield{author}{\bibinfo{person}{Marco Tarini}, \bibinfo{person}{Kai
  Hormann}, \bibinfo{person}{Paolo Cignoni}, {and} \bibinfo{person}{Claudio
  Montani}.} \bibinfo{year}{2004}\natexlab{}.
\newblock \showarticletitle{Polycube-maps}.
\newblock \bibinfo{journal}{\emph{ACM transactions on graphics (TOG)}}
  \bibinfo{volume}{23}, \bibinfo{number}{3} (\bibinfo{year}{2004}),
  \bibinfo{pages}{853--860}.
\newblock
\urldef\tempurl%
\url{https://doi.org/10.1145/1186562.1015810}
\showDOI{\tempurl}


\bibitem[\protect\citeauthoryear{Urick, Sanders, Hossain, Zhang, and
  Hughes}{Urick et~al\mbox{.}}{2019}]%
        {urick2019review}
\bibfield{author}{\bibinfo{person}{Benjamin Urick}, \bibinfo{person}{Travis~M
  Sanders}, \bibinfo{person}{Shaolie~S Hossain}, \bibinfo{person}{Yongjie~J
  Zhang}, {and} \bibinfo{person}{Thomas~JR Hughes}.}
  \bibinfo{year}{2019}\natexlab{}.
\newblock \showarticletitle{Review of patient-specific vascular modeling:
  template-based isogeometric framework and the case for CAD}.
\newblock \bibinfo{journal}{\emph{Archives of Computational Methods in
  Engineering}} \bibinfo{volume}{26}, \bibinfo{number}{2}
  (\bibinfo{year}{2019}), \bibinfo{pages}{381--404}.
\newblock
\urldef\tempurl%
\url{https://doi.org/10.1007/s11831-017-9246-z}
\showDOI{\tempurl}


\bibitem[\protect\citeauthoryear{V{\'a}zquez}{V{\'a}zquez}{2016}]%
        {vazquez2016new}
\bibfield{author}{\bibinfo{person}{Rafael V{\'a}zquez}.}
  \bibinfo{year}{2016}\natexlab{}.
\newblock \showarticletitle{A new design for the implementation of isogeometric
  analysis in Octave and Matlab: GeoPDEs 3.0}.
\newblock \bibinfo{journal}{\emph{Computers \& Mathematics with Applications}}
  \bibinfo{volume}{72}, \bibinfo{number}{3} (\bibinfo{year}{2016}),
  \bibinfo{pages}{523--554}.
\newblock
\urldef\tempurl%
\url{https://doi.org/10.1016/j.camwa.2016.05.010}
\showDOI{\tempurl}


\bibitem[\protect\citeauthoryear{V{\'a}zquez and Buffa}{V{\'a}zquez and
  Buffa}{2010}]%
        {vazquez2010isogeometric}
\bibfield{author}{\bibinfo{person}{Rafael V{\'a}zquez} {and}
  \bibinfo{person}{Annalisa Buffa}.} \bibinfo{year}{2010}\natexlab{}.
\newblock \showarticletitle{Isogeometric analysis for electromagnetic
  problems}.
\newblock \bibinfo{journal}{\emph{IEEE Transactions on Magnetics}}
  \bibinfo{volume}{46}, \bibinfo{number}{8} (\bibinfo{year}{2010}),
  \bibinfo{pages}{3305--3308}.
\newblock
\urldef\tempurl%
\url{https://doi.org/10.1109/tmag.2010.2044563}
\showDOI{\tempurl}


\bibitem[\protect\citeauthoryear{Wang, He, Li, Gu, and Qin}{Wang
  et~al\mbox{.}}{2008}]%
        {wang2008polycube}
\bibfield{author}{\bibinfo{person}{Hongyu Wang}, \bibinfo{person}{Ying He},
  \bibinfo{person}{Xin Li}, \bibinfo{person}{Xianfeng Gu}, {and}
  \bibinfo{person}{Hong Qin}.} \bibinfo{year}{2008}\natexlab{}.
\newblock \showarticletitle{Polycube splines}.
\newblock \bibinfo{journal}{\emph{Computer-Aided Design}} \bibinfo{volume}{40},
  \bibinfo{number}{6} (\bibinfo{year}{2008}), \bibinfo{pages}{721--733}.
\newblock
\urldef\tempurl%
\url{https://doi.org/10.1016/j.cad.2008.01.012}
\showDOI{\tempurl}


\bibitem[\protect\citeauthoryear{Wang, Shen, Chen, Wu, and Gao}{Wang
  et~al\mbox{.}}{2017}]%
        {wang2017sheet}
\bibfield{author}{\bibinfo{person}{Rui Wang}, \bibinfo{person}{Chun Shen},
  \bibinfo{person}{Jinming Chen}, \bibinfo{person}{Haiyan Wu}, {and}
  \bibinfo{person}{Shuming Gao}.} \bibinfo{year}{2017}\natexlab{}.
\newblock \showarticletitle{Sheet operation based block decomposition of solid
  models for hex meshing}.
\newblock \bibinfo{journal}{\emph{Computer-Aided Design}}  \bibinfo{volume}{85}
  (\bibinfo{year}{2017}), \bibinfo{pages}{123--137}.
\newblock
\urldef\tempurl%
\url{https://doi.org/10.1016/j.cad.2016.07.016}
\showDOI{\tempurl}


\bibitem[\protect\citeauthoryear{Xu, Mourrain, Galligo, and Rabczuk}{Xu
  et~al\mbox{.}}{2014}]%
        {xu2014high}
\bibfield{author}{\bibinfo{person}{Gang Xu}, \bibinfo{person}{Bernard
  Mourrain}, \bibinfo{person}{Andr{\'e} Galligo}, {and} \bibinfo{person}{Timon
  Rabczuk}.} \bibinfo{year}{2014}\natexlab{}.
\newblock \showarticletitle{High-quality construction of analysis-suitable
  trivariate NURBS solids by reparameterization methods}.
\newblock \bibinfo{journal}{\emph{Computational Mechanics}}
  \bibinfo{volume}{54}, \bibinfo{number}{5} (\bibinfo{year}{2014}),
  \bibinfo{pages}{1303--1313}.
\newblock
\urldef\tempurl%
\url{https://doi.org/10.1007/s00466-014-1060-y}
\showDOI{\tempurl}


\bibitem[\protect\citeauthoryear{Zhang, Bazilevs, Goswami, Bajaj, and
  Hughes}{Zhang et~al\mbox{.}}{2007}]%
        {zhang2007patient}
\bibfield{author}{\bibinfo{person}{Yongjie Zhang}, \bibinfo{person}{Yuri
  Bazilevs}, \bibinfo{person}{Samrat Goswami}, \bibinfo{person}{Chandrajit~L
  Bajaj}, {and} \bibinfo{person}{Thomas~JR Hughes}.}
  \bibinfo{year}{2007}\natexlab{}.
\newblock \showarticletitle{Patient-specific vascular NURBS modeling for
  isogeometric analysis of blood flow}.
\newblock \bibinfo{journal}{\emph{Computer methods in applied mechanics and
  engineering}} \bibinfo{volume}{196}, \bibinfo{number}{29-30}
  (\bibinfo{year}{2007}), \bibinfo{pages}{2943--2959}.
\newblock
\urldef\tempurl%
\url{https://doi.org/10.1016/j.cma.2007.02.009}
\showDOI{\tempurl}


\bibitem[\protect\citeauthoryear{Zhou and Jacobson}{Zhou and Jacobson}{2016}]%
        {zhou2016thingi10k}
\bibfield{author}{\bibinfo{person}{Qingnan Zhou} {and} \bibinfo{person}{Alec
  Jacobson}.} \bibinfo{year}{2016}\natexlab{}.
\newblock \showarticletitle{Thingi10k: A dataset of 10,000 3d-printing models}.
\newblock \bibinfo{journal}{\emph{arXiv preprint arXiv:1605.04797}}
  (\bibinfo{year}{2016}).
\newblock


\end{thebibliography}

\appendix

\section{Wall Scaffolding Construction}
\label{WallScaff}

\begin{figure}
    \centering
    \begin{overpic}[width=.99\textwidth]{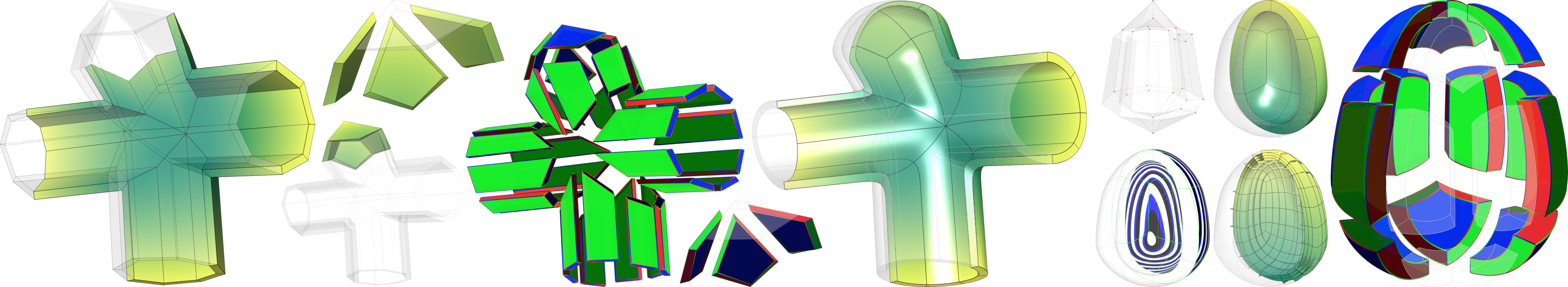}
        \put(1,17){\small{(a)}}
        \put(19,17){\small{(b)}}
        \put(40,17){\small{(c)}}
        \put(50,17){\small{(d)}}
        \put(69,17){\small{(e)}}
    \end{overpic}
    \caption{Wall Scaffolding Construction: (a) 4-way-junction: 8 elements per elongated branch forming an octagonal ring; (b) Additional capping elements per terminal branch. (c) Boundaries and conforming interfaces of adjacent elements facing the parametric directions; (d) Smooth limit geometry; (e) Egg$_W$ thin wall as closed hollow chamber: scaffolding comprising two sets of capping elements.}
    \label{fig:WallScaff}
\end{figure}

The wall scaffolding is modelled as an extension of the luminal one (\cref{LumScaff}).
It comprises the organised set of vectorial elements occupying the hollow volumetric space enclosing a luminal region, e.g. a set of connecting hollow pipes or hollow chambers with thin or thick walls, in the form of a solid shell (\cref{fig:WallScaff}).
The wall scaffolding can be considered either separately, or additionally coupled with a luminal scaffolding.

\paragraph{Minimal Elements} The minimal number of elements depends on the complexity of the underlying structure.
Building on concepts introduced in \cref{LumScaff}, the wall scaffolding requires 8 elements longitudinally surrounding each branch.
Additional 4 elements are required when one end of any branch is closed, i.e. the wall covers and closes the terminal portion of the hollow structure.
In general, the elements of the wall scaffolding are  proportional to the boundary sides of a virtually nested luminal scaffolding, and the construction of the wall resembles a conforming extrusion along the luminal interfaces.

\paragraph{Arrangement and Adjacency} In general the wall scaffolding cross-sectionally configures the elements of each elongated branch in an octagonal ring (\cref{fig:WallScaff}).
As in the luminal region, these elements are elongated along the parametric direction $w$, mapping the longitudinal direction of a branch.
The parametric directions $u$ and $v$ of the wall scaffolding independently map a pseudo-circumferential direction, and a pseudo-radial direction, respectively, relative to the elongation of the branch.
Each adjacent element of the wall scaffolding has interfaces facing the pseudo-circumferential direction $u$.
In the absence of a coupled luminal scaffolding, two sets of boundaries are determined, both facing the pseudo-radial direction $v$; one facing the interior hollow luminal region, another facing the exterior of the wall structure.
However, when the wall is coupled with a luminal scaffolding, both scaffoldings share a mutual set of interfaces, constituting a complementary and multi-compartmental nested structure.

For a closed terminal branch the wall scaffolding additionally introduces 4 capping elements (\cref{fig:WallScaff}).
These form a quadrant-like configuration and their parametric arrangement is consistent with a luminal scaffolding element, i.e. $u$ and $v$ no longer map independently the pseudo-circumferential and the pseudo-radial directions, respectively.
%Each capping element is also adjacent with 2 elongated elements of the octagonal ring configuration.
Each capping element determines at least 4 interfaces.
Two contiguous sides (one facing $u$, and the other facing $v$) are shared with 2 adjacent capping elements; whereas other two contiguous sides (one facing $u$, and the other facing $v$, on opposite directions) are shared with 2 adjacent elongated elements.

Without a coupled luminal region, both sides of the capping element facing $w$ are referred as boundaries, each belonging to the interior and to the exterior wall boundaries, respectively.
Conversely, with a coupled luminal scaffolding, each capping element shares an extra interface between the nested scaffoldings. %The remaining side of the capping element constitutes the exterior boundary of the wall scaffolding.

The formulation of a wall scaffolding configuration at the interface of a \mbox{$n$-way-junction} is a straightforward extension of the luminal quadrilateral junction simplex $\mathcal{Q}$ in \cref{QuadJuncSimplex}.
Both interior ($\mathcal{Q}_{\textrm{int}}$) and exterior ($\mathcal{Q}_{\textrm{ext}}$) simplexes of the of the wall scaffolding replicate the same configuration and differ in size, accounting for the pseudo-radial thickness of the wall.
In this case, the resulting set of interfaces facing $w$ is determined by connecting one-to-one each vertex of $\mathcal{Q}_{\textrm{int}}$ with the respective vertex of  $\mathcal{Q}_{\textrm{ext}}$, and the cross-sectional octagonal configuration is kept for all the branches. %joining the \mbox{$n$-way-junction}.
% Also, the subset of the vertices of both $\mathcal{Q}_{\textrm{int}}$ and $\mathcal{Q}_{\textrm{ext}}$ coincides with the a subset of control points as in \cref{eq_NURBScuboid}.

\paragraph{Conforming Constraints}
\label{WallScaffConformConstr}
The conforming conditions depend on coupling the wall with a nested luminal scaffolding.

For an uncoupled open wall scaffolding, no mutual dependency is determined for the parametric components of each element along the different parametric directions. 
Conversely, for an uncoupled closed wall scaffolding, i.e. in presence of capping elements, the parametric components of each element along $u_{(\textrm{wall~branch})}$ and $v_{(\textrm{wall~branch})}$ are equivalent, whereas only the parametric component along $w_{(\textrm{wall~branch})}$ is independent. The capping elements of each terminal branch, however, make exception: all the parametric components are dependent and mutually equivalent along $(u,v,w)_{(\textrm{wall~cap})}$ to $(u,v)_{(\textrm{wall~branch})}$.

When a wall scaffolding is coupled with a nested luminal one, the conforming constraints are mutually related between scaffoldings.
In particular, for a coupled closed wall scaffolding, the parametric components of each element along $(u,v)_{(\textrm{wall~branch})}$ coincide with those of $(u,v)_{(\textrm{luminal})}$.
Also, the parametric components along $w_{(\textrm{wall~brach})}$ coincide with those along $w_{(\textrm{luminal})}$.
All the capping elements of the closed wall scaffolding are mutually dependent, i.e. $(u,v,w)_{(\textrm{wall~cap})}$ coincide to $(u,v)_{(\textrm{wall~branch})}$ as well as to $(u,v)_{(\textrm{luminal})}$.
Lastly, for a coupled open wall scaffolding, the parametric components of each element along $u_{(\textrm{wall~branch})}$ coincide with those of $(u,v)_{(\textrm{luminal})}$, whereas $v_{(\textrm{wall~branch})}$ and $w_{(\textrm{wall~branch})}$ are independent.

\section{Other Conforming Scaffolding Configurations}
\label{Othercaff}

\begin{figure}
    \centering
    \begin{overpic}[width=.99\textwidth]{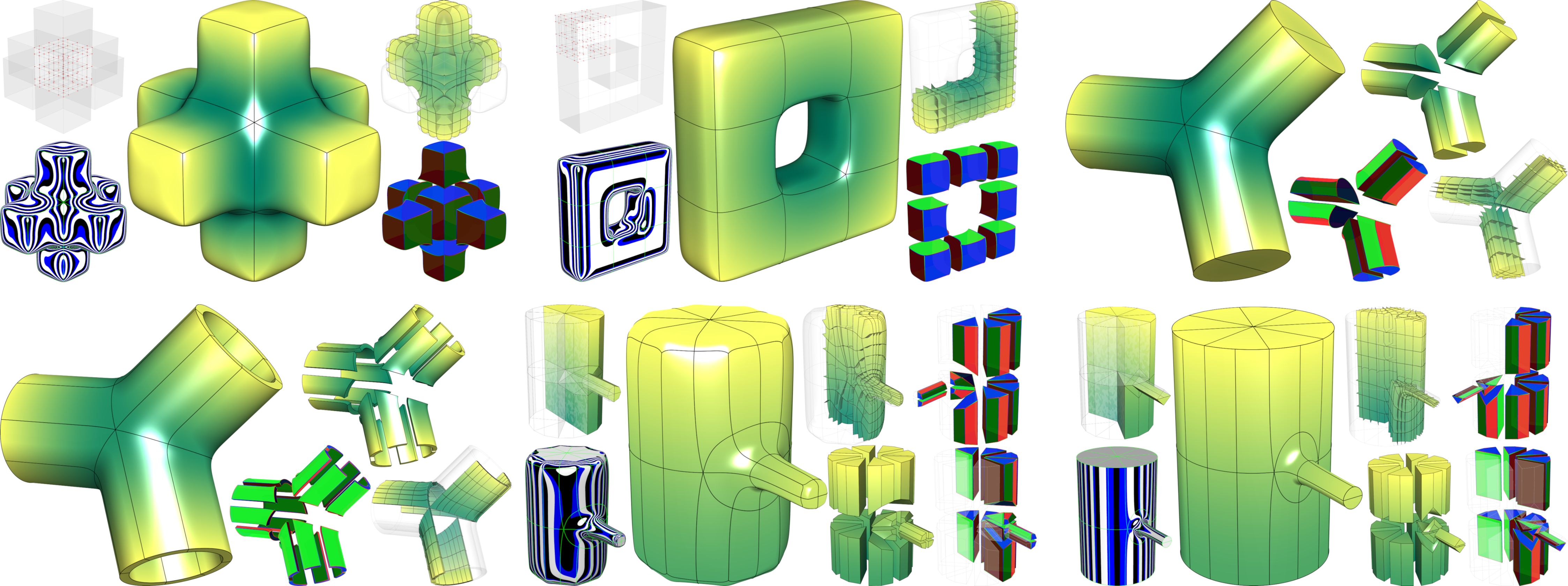}
        \put(-1,36){\small{(a)}}
        \put(33,36){\small{(b)}}
        \put(67,36){\small{(c)}}
        \put(0,17){\small{(d)}}
        \put(31,17){\small{(e)}}
        \put(66,17){\small{(f)}}
    \end{overpic}
    \caption{Other Scaffolding Construction: (a) \textit{Cross} and (b) \textit{Frame} geometries with 6-face-connectivity scaffolding. (c) Exceptionally reduced planar scaffolding for bifurcating structures; (d) Associated configuration of the wall scaffolding. (e) Grafting scaffolding: \textit{Pinocchio} geometry with primary and secondary branches; (f) \textit{Pinocchio} topological variation, arrangement and adjacency.}
    \label{fig:OtherScaff}
\end{figure}

For completeness, alternative conforming scaffolding configurations are simply mentioned below.
These may result in local deviations from the introduced framework with minor topological and parametric variations affecting the density of vectorial elements, their arrangement and the associated conforming constraints.

By way of example, a \textit{6-face-connectivity scaffolding} is shown in \cref{fig:OtherScaff}, for the synthetic \textit{Cross} and \textit{Frame} geometries.
Starting from an innermost cube, such configuration consist of an independent conformal extrusion along each boundary face into a set of adjacent vectorial elements.
Note that the extension of such configuration includes structured (e.g. voxel-based) or unstructured hexahedral meshing approaches.

Based on \cref{LumScaff}, a \textit{planar scaffolding} exceptionally halves the number of vectorial elements per branch.
This is suitable for isolated 3-way-junctions as in \cref{fig:OtherScaff}.
Alternatively, in case of a network, this configuration is applicable only if \textit{all} the junctions are either bifurcations, or they exhibit a planar arrangement.
In such case, note that the structural torsional twist along connecting branches of the luminal scaffolding may increase up to $\theta_{\text{max}} = \frac{\pi}{2}$ by adopting the same geometrical embedding in \cref{GeomEmbedding}. Note that such exceptional reduction can affect the configuration of a coupled wall scaffolding, or symmetrical and modular structures (e.g. \textit{Gear}).

Lastly, a \textit{grafting scaffolding} defines a hierarchical branching relationship and labels branches of the skeleton as primary, secondary, and so on (e.g. based on size), these being interleaved by cross-sectional cuts in \cref{fig:OtherScaff}.
This configuration is suitable for particularly busy or spatially close junctions by locally simplifying the interfacing simplexes.
Note that, the grafting configuration reduces the scaffolding elements of the secondary branch relative to the primary one, with potential limitations for downstream branching and connectivity patterns.





\end{document}